\documentclass[final,3p,times]{elsarticle}
\usepackage{amssymb}
\usepackage{amsthm}
\usepackage[]{amsmath}
\usepackage{graphicx}
\usepackage{graphics}
\usepackage{subfigure}
\usepackage{latexsym}
\usepackage[]{amsmath}
\usepackage{amssymb}
\usepackage{indentfirst}
\usepackage{epsfig}
\usepackage{epstopdf}
\usepackage{float}
\usepackage{setspace}
\usepackage{booktabs}
\usepackage{times}
\usepackage{anysize}
\usepackage{endnotes}
\usepackage{amsmath}
\usepackage{mathrsfs}
\usepackage{enumitem}
\numberwithin{equation}{section}

\biboptions{numbers,sort&compress}
\usepackage[colorlinks, citecolor=red]{hyperref}
\usepackage{amsmath}
\usepackage{times}
\usepackage{anysize}
\marginsize{1.5cm}{1.5cm}{0.7cm}{1.5cm}
\selectfont
\journal {}

\makeatletter
\def\ps@pprintTitle{%
	\let\@oddhead\@empty
	\let\@evenhead\@empty
	\let\@oddfoot\@empty
	\let\@evenfoot\@oddfoot}
\makeatother

\begin{document}

\begin{frontmatter}

\title{Inverse scattering transform for the focusing two-component Hirota equation with nonzero boundary conditions}

\author{ Feng Zhang\fnref{label1} }
\author{ Pengfei Han\fnref{label1} }
\author{ Yi Zhang\fnref{label1}$^*$  }
\cortext[cor1]{Corresponding author}
\ead{zy2836@163.com}

\address[label1]{ School of Mathematical Sciences, Zhejiang Normal University, Jinhua 321004, PR China}

\begin{abstract}
Applying the inverse scattering transform to study a focusing two-component Hirota equation with nonzero boundary conditions at infinity. Through the spectral problem and the adjoint spectral problem, the analyticity properties and symmetry relations of the eigenfunctions and scattering coefficients are systematically investigated. Based on these properties, the discrete spectrum is comprehensively characterized, producing three types of discrete eigenvalues. The asymptotic behavior of the eigenfunctions and scattering coefficients is explored in detail. Furthermore, the inverse scattering problem is formulated through a suitable matrix Riemann-Hilbert problem, and the trace formula along with the reconstruction formula for the potential function are derived. The reflectionless condition facilitates the successful construction of $N$-soliton solution. Finally, various types of solitons are obtained by considering different types of discrete eigenvalues and their combinations, and the interactions of these solitons are demonstrated.
\end{abstract}

\begin{keyword}
focusing two-component Hirota equation; inverse scattering transform; $N$-soliton solution.

\end{keyword}
\end{frontmatter}

\section{Introduction}

The nonlinear Schr\"odinger (NLS) equation has garnered significant attention due to its theoretical significance and broad practical applications. In most dispersion energy preserving systems, the scalar form of the NLS equation can be derived by taking appropriate limits, making it as a widely recognized framework for describing the dynamics of weakly nonlinear dispersive wave trains \cite{N1,N2}.  It finds broad applications in fields such as deep water waves, Bose-Einstein condensates, plasma physics, nonlinear fiber optics, and magnetization spin waves \cite{N3,N4,A1}. As a quintessential example of integrable systems, it serves not only as a foundation for understanding nonlinear wave phenomena but also as a driving force for advancing the study of integrable systems. While the NLS equation has achieved remarkable success in numerous scenarios, higher-order extensions are necessary to account for complex nonlinear wave phenomena, such as oceanic waves and ultrashort optical pulses propagation. In nonlinear optics, as optical pulses shorten, nonlinear effects and higher-order dispersive, including self-frequency shift, self-steepening and third-order dispersion induced by stimulated Raman scattering, become pronounced and cannot be ignored \cite{N5,N7}. These intricate phenomena have spurred extensive research into developing higher-order extensions of the NLS equation \cite{N8,N9,N10,N12}.

This study investigates a focusing two-component Hirota equation, expressed as
\begin{equation}\label{eq1}
i{\tilde {\boldsymbol{u}}_t} + {\alpha _1}\left( {{{\tilde {\boldsymbol{u}}}_{xx}} + 2\tilde {\boldsymbol{u}}{{\left\| {\tilde {\boldsymbol{u}}} \right\|}^2}} \right) + i{\alpha _2}\left( {{{\tilde {\boldsymbol{u}}}_{xxx}} + 3{{\tilde {\boldsymbol{u}}}_x}{{\left\| {\tilde {\boldsymbol{u}}} \right\|}^2} + 3\tilde {\boldsymbol{u}}{{\tilde {\boldsymbol{u}}}^\dag }{{\tilde {\boldsymbol{u}}}_x}} \right) = 0,
\end{equation}
where $\tilde {\boldsymbol{u}} = {\left( {{{\tilde u}_1},{{\tilde u}_2}} \right)^T}$ denotes a two-component vector function, ${\left\| {\tilde {\boldsymbol{u}}} \right\|^2} = {\left| {{{\tilde u}_1}} \right|^2} + {\left| {{{\tilde u}_1}} \right|^2}$, $\dag$ represents the Hermitian transpose, and ${\alpha _j}{\kern 2pt}(j = 1,2)$ are arbitrary real constants associated with second and third order dispersions, respectively \cite{N13,N14,N15}. Using the variable transformation $\tilde {\boldsymbol{u}} = {\boldsymbol{u}}{e^{ - 2iu_o^2t}}$ to this equation yieds the focusing two-component Hirota equation with nonzero boundary conditions (NZBCs) at infinity as
\begin{equation}\label{eq2}
\begin{split}
i{{ {\boldsymbol{u}}}_t} + {\alpha _1}\left[ {{{ {\boldsymbol{u}}}_{xx}} + 2 {\boldsymbol{u}}\left( {{{\left\| { {\boldsymbol{u}}} \right\|}^2} - u_o^2} \right)} \right] + i{\alpha _2}\left[ {{{ {\boldsymbol{u}}}_{xxx}} + 3{{ {\boldsymbol{u}}}_x}{{\left\| { {\boldsymbol{u}}} \right\|}^2} + 3 {\boldsymbol{u}}{{ {\boldsymbol{u}}}^\dag }{{ {\boldsymbol{u}}}_x}} \right] = 0, \\
{\kern 0pt}  \mathop {\lim }\limits_{x \to  \pm \infty } {\boldsymbol{u}}\left( {x,t} \right) = {{\boldsymbol{u}}_ \pm } = {{\boldsymbol{u}}_o}{e^{i{h _ \pm }}},{\kern 163pt}
\end{split}
\end{equation}
where $ {\boldsymbol{u}} = {\left( {{{ u}_1},{{ u}_2}} \right)^T}$ is a two-component vector function, ${h _ \pm }$ are arbitrary real numbers, ${\boldsymbol{u}}_o$ is a constant vector and ${u_o} = \left\| {{{\boldsymbol{u}}_o}} \right\|$.

It is straightforward to see that when ${\alpha _1=1}$ and  ${\alpha _2=0}$, Eq.(\ref{eq1}) simplifies to the classical focusing Manakov system; when ${\alpha _1=0}$ and  ${\alpha _2=1}$, it degenerates into the coupled modified
Korteweg-de Vries equation; and when ${{\tilde u}_2}=0$, it transforms into the scalar Hirota equation. The Hirota equation has widespread applications in various physical systems. For example, the equation can be used to describe nonlinear effects in optical fibers, such as self-interaction and dispersion, which significantly affect the propagation characteristics and waveform evolution of optical pulses \cite{N17,N18,N19,N20}. Additionally, it can be applied to study wave behavior in cold atomic gases, especially in superfluids or Bose-Einstein condensates, where the nonlinear properties of the wave often lead to the emergence of localized wave phenomena such as soliton solutions \cite{N21,N23,N24}.

The inverse scattering transform (IST) is a powerful method for studying nonlinear evolution equations, particularly effective in solving the initial value problems associated with integrable systems \cite{M2,N35,A2,N38,A3,A4}. Originally introduced through the work of Gardner et al. for the KdV equation \cite{N25}, IST was subsequently extended by Zakharov and Shabat to the NLS equation \cite{N26}, establishing the foundational framework for later developments. The method was further adapted to integrable systems with NZBCs in \cite{N27}, with significant refinements made in \cite{N28,N29}. Over time, the application of IST has evolved beyond simple scalar equations to encompass more complex multi-component and matrix equations, such as Manakov system \cite{N30,N31}, multi-component Gerdjikov-Ivanov equation \cite{N32,A5,A6}, coupled Lakshmanan-Porsezian-Daniel equation \cite{N33}, and matrix NLS equation \cite{N34}. While these extensions are mathematically more challenging, they allow for the exploration of richer physical phenomena.

Although the IST has been effectively applied to the defocusing two-component Hirota equation \cite{N40} and the scalar Hirota equation \cite{N41}, constructing the IST for the focusing two-component Hirota equation remains essential. This necessity stems from the increased complexity in handling the analyticity, asymptotic behavior and symmetries of the scattering coefficients and Jost eigenfunctions, along with the intricate process of formulating the associated Riemann-Hilbert problem. Moreover, the discrete spectrum of the focusing system is more diverse, encompassing a broader range of eigenvalue types. The complexity of these discrete eigenvalues directly impacts the construction and interaction of soliton solutions. From a practical perspective, the focusing two-component Hirota equation is widely used in nonlinear optics, fluctuation physics, and other fields, demonstrating significant value in describing multi-component weak nonlinear fluctuations. Employing the IST to analyze this equation not only lays the groundwork for understanding the long-time asymptotic behavior of complex fluctuation phenomena, but also leads to the novel soliton solutions and their interactions. This provides invaluable insights into the theoretical study of integrable systems.

The structure of this paper is organized as follows: In Section 2, we construct the direct scattering problem and investigate the analyticity and symmetry relations of the Jost eigenfunctions and scattering coefficients. In Section 3, a characterization of the discrete spectrum is provided, and the asymptotic behavior of the modified Jost eigenfunctions and scattering coefficients is analyzed. In Section 4, we formulate the inverse scattering problem by constructing an appropriate Riemann-Hilbert problem, including the derivation of trace and reconstruction formulas. In Section 5, the $N$-soliton solutions are derived, and various solitons are obtained based on the discrete eigenvalues. In Section 6, we present a brief summary and discussion.

\section{ Direct scattering }

The focusing two-component Hirota equation (\ref{eq2}) corresponds to the $3 \times 3$ Lax pair \cite{N13}, written by
\begin{equation}\label{eq3}
\begin{array}{l}
{\boldsymbol{\varphi} _x} = \boldsymbol{X}\left( {k,x,t} \right)\boldsymbol{\varphi} ,\\{\boldsymbol{\varphi} _t} = \boldsymbol{T}\left( {k,x,t} \right)\boldsymbol{\varphi} ,
\end{array}
\end{equation}
where
$$\boldsymbol{X} =  - ik\boldsymbol{\sigma} + \boldsymbol{U},$$
$$\boldsymbol{T} = {\alpha _1}\left[ {2i{k^2}\boldsymbol{\sigma} - 2k\boldsymbol{U} + i\boldsymbol{\sigma}\left( {{\boldsymbol{U}^2}{ + u_o^2} - {\boldsymbol{U}_x}} \right)} \right] + {\alpha _2}\left[ { - 4i{k^3}\boldsymbol{{\sigma}} + 4{k^2}\boldsymbol{U} + 2ik\boldsymbol{{\sigma}}\left( {{\boldsymbol{U}_x} - {\boldsymbol{U}^2}} \right) - \boldsymbol{U}{\boldsymbol{U}_x} + {\boldsymbol{U}_x}\boldsymbol{U} - {\boldsymbol{U}_{xx}} + 2{\boldsymbol{U}^3}} \right],$$
$$\boldsymbol{U} = \left( {\begin{array}{*{20}{c}}
0&{ - {\boldsymbol{u}^\dag }}\\
\boldsymbol{u}&{\tilde {\boldsymbol{0}}}
\end{array}} \right),{\kern 5pt}\boldsymbol{{\sigma}} = \left( {\begin{array}{*{20}{c}}
1&{{\boldsymbol{0}^T}}\\
\boldsymbol{0}&-\boldsymbol{I}
\end{array}} \right).$$
Here, $\boldsymbol{I}$ is the $2 \times 2$ identity matrix, $k$ denotes the spectral parameter, ${ {\boldsymbol{0}}}={\left( {0,0} \right)^T}$, and ${\tilde {\boldsymbol{0}}}$ is a $2 \times 2$ zero matrix. By calculating the zero curvature condition ${\boldsymbol{X}_t} - {\boldsymbol{T}_x} + \left[ {\boldsymbol{X},\boldsymbol{T}} \right] = \boldsymbol{0}$, we can obtain Eq.(\ref{eq2}).

\subsection{Uniformization}

To ensure the proper definition of the Jost eigenfunctions, it is necessary to first examine the asymptotic spectral problem as $x \to \pm \infty $, which is written as
\begin{equation}\label{eq5}
\begin{split}
{\boldsymbol{\varphi} _x} = {\boldsymbol{X}_ \pm }\boldsymbol{\varphi},{\kern 5pt} {\boldsymbol{X}_ \pm } = \mathop {\lim }\limits_{x \to  \pm \infty } \boldsymbol{X} =  - ik\boldsymbol{\sigma}  + {\boldsymbol{U}_ \pm },{\kern 125pt}\\
{\boldsymbol{\varphi} _t} = {\boldsymbol{T}_ \pm }\boldsymbol{\varphi} ,{\kern 5pt} {\boldsymbol{T}_ \pm } = \mathop {\lim }\limits_{x \to  \pm \infty } \boldsymbol{T} = {\alpha _1}\left[ {2i{k^2}\boldsymbol{\sigma}  - 2k{\boldsymbol{U}_ \pm } + i\boldsymbol{\sigma} \left( {\boldsymbol{U}_ \pm ^2 + u_o^2} \right)} \right] + 2{\alpha _2}\left[ { - 2i{k^3}\boldsymbol{\sigma}  + 2{k^2}{\boldsymbol{U}_ \pm } - ik\boldsymbol{\sigma U}_ \pm ^2 + \boldsymbol{U}_ \pm ^3} \right],
\end{split}
\end{equation}
where
$${\boldsymbol{U}_ \pm } = \mathop {\lim }\limits_{x \to  \pm \infty } \boldsymbol{U} = \left( {\begin{array}{*{20}{c}}
0&{ - \boldsymbol{u}_ \pm ^\dag }\\
{{\boldsymbol{u}_ \pm }}&{\tilde {\boldsymbol{0}}}
\end{array}} \right).$$
According to the expressions (\ref{eq5}) of the matrices ${\boldsymbol{X}_ \pm }$ and ${\boldsymbol{T}_ \pm }$, the eigenvalue matrix of ${\boldsymbol{X}_ \pm }$ is given by $i{\boldsymbol{\varpi} _1}$, and that of ${\boldsymbol{T}_ \pm }$ is given by  $-i{\boldsymbol{\varpi} _2}$

\begin{equation}\label{eq7}
\begin{split}
i{\boldsymbol{\varpi} _1} = \mathrm{diag}\left[ { - i\lambda ,ik,i\lambda } \right],{\kern 145pt}\\
 - i{\boldsymbol{\varpi} _2} = \mathrm{diag}\left[ {2i\lambda \left( {{\alpha _1}k + {\alpha _2}\left( { - 2{k^2} + u_o^2} \right)} \right),4i{\alpha _2}{k^3} - i{\alpha _1}\left( {{k^2} + {\lambda ^2}} \right), - 2i\lambda \left( {{\alpha _1}k + {\alpha _2}\left( { - 2{k^2} + u_o^2} \right)} \right)} \right],
\end{split}
\end{equation}
where ${\lambda ^2} = u_o^2 + {k^2}$. It is evident that $\lambda \left( k \right)$ is a multivalued function. To uniformize it, we introduce a two-sheeted Riemann surface with branch points at $ \pm i{u_o}$ and a branch cut along the segment $\left[ { - i{u_o},i{u_o}} \right]$. For further analysis, we employ the conformal mapping $z = \lambda  + k$ and define $z$ as a uniformization variable. The corresponding inverse mappings are expressed as
\begin{equation}\label{eq8}
\lambda  = \frac{1}{2}\left( {z + \frac{{u_o^2}}{z}} \right),{\kern 5pt}k = \frac{1}{2}\left( {z - \frac{{u_o^2}}{z}} \right).
\end{equation}
The first equation of the above formulas is known as the Joukowsky transformation, which maps the regions ${\mathop{\rm Im}\nolimits} \lambda > 0$ and ${\mathop{\rm Im}\nolimits} \lambda < 0$ to the following sets, respectively:
\begin{equation}\label{eq9}
{\Xi ^ + } = \left\{ {z \in \mathbb{C}:  \left( {{{\left| z \right|}^2} - u_o^2} \right){\mathop{\rm Im}\nolimits} z > 0} \right\},{\kern 5pt}{\Xi ^ - } = \left\{ {z \in \mathbb{C} :\left( {{{\left| z \right|}^2} - u_o^2} \right){\mathop{\rm Im}\nolimits} z < 0} \right\}.
\end{equation}
We define a circle ${C_o}$ centered at the origin with radius ${u_o}$.  Then, by setting ${\mathop{\rm Im}\nolimits} \lambda  = 0$, the continuous spectrum is determined to be $\Sigma  =  \mathbb{R} \cup {C_o}$. Notably, in the defocusing case, the continuous spectrum is only on the real axis $\mathbb{R}$.

\subsection{Jost eigenfunctions and analyticity}

To simplify the representation of the eigenvector matrix of $\boldsymbol{X}_ \pm$ and $\boldsymbol{T}_ \pm$, we introduce two vector symbols: $\boldsymbol{\nu}  = {\left( {{\nu _1},{\nu _2}} \right)^T},{\boldsymbol{\nu} ^ \bot } = {\left( {{\nu _2}, - {\nu _1}} \right)^\dag }$. Using these notations, the eigenvector matrix of $\boldsymbol{X}_ \pm$ and $\boldsymbol{T}_ \pm$ in the asymptotic scattering problem (\ref{eq5}) is given by
\renewcommand{\arraycolsep}{2pt}
\renewcommand{\arraystretch}{1.8}
$${\boldsymbol{G}_ \pm } = \left( {\begin{array}{*{20}{c}}
i\\
{ - \dfrac{{{\boldsymbol{u}_ \pm }}}{z}}
\end{array}\begin{array}{*{20}{c}}
0\\
{\dfrac{{\boldsymbol{u}_ \pm ^ \bot }}{{{\boldsymbol{u}_o}}}}
\end{array}\begin{array}{*{20}{c}}
{\dfrac{{{u_o}}}{z}}\\
{-\dfrac{{i{\boldsymbol{u}_ \pm }}}{{{\boldsymbol{u}_o}}}}
\end{array}} \right),$$
and it satisfies the following diagonalization relations
$${\boldsymbol{X}_ \pm } = i{\boldsymbol{G}_ \pm }{\boldsymbol{\varpi} _1}\boldsymbol{G}_ \pm ^{ - 1},{\kern 5pt}{\boldsymbol{T}_ \pm } =  - i{\boldsymbol{G}_ \pm }{\boldsymbol{\varpi} _2}\boldsymbol{G}_ \pm ^{ - 1},$$
where ${\boldsymbol{\varpi} _j{\kern 2pt}(j=1,2)}$ are defined in the Eqs.(\ref{eq7}). Furthermore, the determinant and inverse of $\boldsymbol{G}_\pm$ are given by
\begin{equation}\label{eq12}
\mu : = \det {\boldsymbol{G}_ \pm } = 1 + \frac{{u_o^2}}{z},{\kern 5pt}\boldsymbol{G}_ \pm ^{ - 1} = \frac{1}{\mu }\left( {\begin{array}{*{20}{c}}
{ - i}\\
0\\
{ - \dfrac{{i{u_o}}}{z}}
\end{array}\begin{array}{*{20}{c}}
{ - \dfrac{{\boldsymbol{u}_ \pm ^\dag }}{z}}\\
{\mu \dfrac{{{{\left( {\boldsymbol{u}_ \pm ^ \bot } \right)}^\dag }}}{{{u_o}}}}\\
{\dfrac{{\boldsymbol{u}_ \pm ^\dag }}{{{u_o}}}}
\end{array}} \right).
\end{equation}

Based on the asymptotic spectral problem (\ref{eq5}), the Jost solutions ${\boldsymbol{\varphi} _ \pm }\left( {z,x,t} \right)$ are introduced, with corresponding boundary conditions at $x \to \pm\infty$ given by
\begin{equation}\label{eq13}
{\boldsymbol{\varphi} _ \pm } = {\boldsymbol{G}_ \pm }{e^{i\boldsymbol{H} \left( {z} \right)}} + o\left( 1 \right),
\end{equation}
where
$$\boldsymbol{H} \left( {z,x,t} \right) = {\boldsymbol{\varpi} _1}x - {\boldsymbol{\varpi} _2}t = \mathrm{diag}\left( {{h _1},{h _2},-{h _1}} \right),$$
$${h _1} =  - \lambda x + 2\lambda \left[ {{\alpha _1}k + {\alpha _2}\left( { - 2{k^2} + u_o^2} \right)} \right],{\kern 5pt}{h _2} = kx - \left[ {{\alpha _1}\left( {{k^2} + {\lambda ^2}} \right) - 4{\alpha _2}{k^3}} \right].$$

Below, the modified Jost eigenfunctions ${\boldsymbol{\phi} _ \pm }\left( {z,x,t} \right)$ are introduced to eliminate the exponential oscillations in  ${\boldsymbol{\varphi}_ \pm}\left( {z,x,t} \right)$, which are defined as
\begin{equation}\label{eq15}
{\boldsymbol{\phi} _ \pm } = {\boldsymbol{\varphi} _ \pm }{e^{ - i\boldsymbol{H} \left( {z,x,t} \right)}},
\end{equation}
with $\mathop {\lim }\limits_{x \to  \pm \infty } {\boldsymbol{\phi} _ \pm } = {\boldsymbol{G}_ \pm }$. Using Eqs.(\ref{eq15}), the asymptotic spectral problem (\ref{eq5}) becomes equivalent to
\begin{equation}\label{eq16}
{\left( {\boldsymbol{G}_ \pm ^{ - 1}{\boldsymbol{\phi} _ \pm }} \right)_x} = \left[ {i{\boldsymbol{\varpi} _1},\boldsymbol{G}_ \pm ^{ - 1}{\boldsymbol{\phi} _ \pm }} \right] + \boldsymbol{G}_ \pm ^{ - 1}\Delta {\boldsymbol{U}_ \pm }{\boldsymbol{\phi} _ \pm },{\kern 5pt}{\left( {\boldsymbol{G}_ \pm ^{ - 1}{\boldsymbol{\phi} _ \pm }} \right)_t} = \left[ { - i{\boldsymbol{\varpi} _2},\boldsymbol{G}_ \pm ^{ - 1}{\boldsymbol{\phi} _ \pm }} \right] + \boldsymbol{G}_ \pm ^{ - 1}\Delta {\boldsymbol{T}_ \pm }{\boldsymbol{\phi} _ \pm },
\end{equation}
where $\Delta {\boldsymbol{U}_ \pm } = \boldsymbol{U} - {\boldsymbol{U}_ \pm },\Delta {\boldsymbol{T}_ \pm } = \boldsymbol{T} - {\boldsymbol{T}_ \pm }$. Integrating the first equation above with respect to $x$, we obtain the Volterra integral equations for ${\boldsymbol{\phi} _ \pm }$ as follows
\begin{equation}\label{eq17}
{\boldsymbol{\phi} _ \pm }\left( {z,x,t} \right) = {\boldsymbol{G}_ \pm } + \int_{ \pm \infty }^x {{\boldsymbol{G}_ \pm }{e^{i\left( {x -\gamma } \right){\boldsymbol{\varpi} _1}}}} \boldsymbol{G}_ \pm ^{ - 1}\Delta {\boldsymbol{U}_ \pm }{\boldsymbol{\phi} _ \pm }{e^{ - i\left( {x - \gamma } \right){\boldsymbol{\varpi} _1}}}\mathrm{d}\gamma.
\end{equation}
With the help of the integral equations (\ref{eq17}), the analyticity for the modified Jost eigenfunctions ${\boldsymbol{\phi} _ \pm }$ in the complex $z$-plane is established. Referring to \cite{N31}, assuming $\left(\boldsymbol{u}\left( { \mathbin{\vcenter{\hbox{\scalebox{0.5}{$\bullet$}}}}  ,t} \right) - {\boldsymbol{u}_ - } \right)\in {L^1}\left( {-\infty,a } \right)$ and $\left(\boldsymbol{u}\left( { \mathbin{\vcenter{\hbox{\scalebox{0.5}{$\bullet$}}}}  ,t} \right) - {\boldsymbol{u}_ + }\right) \in {L^1}\left( {a,+\infty } \right)$ for any real number $a$, the following proposition can be established.

\newtheorem{proposition}{Proposition}
\begin{proposition}
    The columns ${\boldsymbol{\phi} _{ \pm, l}}{\kern 2pt}(l = 1, 2 ,3)$ of ${\boldsymbol{\phi} _ \pm }$ can be analytically extended to the appropriate regions:
$$\begin{array}{l}
{\boldsymbol{\phi} _{ - ,1}}:z \in {\Xi _1},{\kern 5pt}{\boldsymbol{\phi} _{ - ,2}}:{\mathop{\rm Im}\nolimits} z < 0,{\kern 5pt}{\boldsymbol{\phi} _{ - ,3}}:z \in {\Xi _4},\\
{\boldsymbol{\phi} _{ + ,1}}:z \in {\Xi _2},{\kern 5pt}{\boldsymbol{\phi} _{ + ,2}}:{\mathop{\rm Im}\nolimits} z > 0,{\kern 5pt}{\boldsymbol{\phi} _{ + ,3}}:z \in {\Xi _3},
\end{array}$$
where ${\Xi _j}{\kern 2pt}(j = 1, \mkern-2mu\cdots\mkern-2mu ,4)$ are
$$\begin{array}{l}
{\Xi _1} \equiv \left\{ {z:\left| z \right| > {u_o} \wedge {\mathop{\rm Im}\nolimits} z > 0} \right\},{\kern 5pt}{\Xi _2} \equiv \left\{ {z:\left| z \right| > {u_o} \wedge {\mathop{\rm Im}\nolimits} z < 0} \right\},\\
{\Xi _3} \equiv \left\{ {z:\left| z \right| < {u_o} \wedge {\mathop{\rm Im}\nolimits} z < 0} \right\},{\kern 5pt}{\Xi _4} \equiv \left\{ {z:\left| z \right| < {u_o} \wedge {\mathop{\rm Im}\nolimits} z > 0} \right\},
\end{array}$$
and $\bigcup\nolimits_{j = 1}^4 {{\Xi _j}}  = \mathbb{C}$, ${\Xi ^ + } = {\Xi _1} \cup {\Xi _3},{\Xi ^ - } = {\Xi _2} \cup {\Xi _4}$.
    \label{prop:1}
\end{proposition}

It can be observed that there are four fundamental regions ${\Xi _j}{\kern 2pt}(j = 1, \mkern-2mu\cdots\mkern-2mu ,4)$ in this case, compared to only two in \cite{N40}, highlighting the increased complexity of the focusing case relative to the defocusing case.

From the definition (\ref{eq15}) of ${\boldsymbol{\phi} _ \pm }$, it follows that the columns ${\boldsymbol{\varphi} _{ \pm ,l}}$ of ${\boldsymbol{\varphi} _ \pm }$ align with the boundedness and analyticity properties of ${\boldsymbol{\phi} _{ \pm ,l}}$. Applying Abel's formula, if ${\boldsymbol{\varphi}}$ is a solution to the spectral problem (\ref{eq3}), then the following holds
$$\begin{array}{l}
{\partial x}\left( {\det {\boldsymbol{\varphi} }} \right) = \mathrm{Tr}\boldsymbol{X}\det {\boldsymbol{\varphi}},{\kern 5pt}\partial x\left( {\det {\boldsymbol{\phi} _ \pm }} \right) = \partial x\left( {\det {\boldsymbol{\varphi} _ \pm }{e^{ - i\boldsymbol{H}}}} \right) = 0,\\
{\partial t}\left( {\det {\boldsymbol{\varphi}}} \right) = \mathrm{Tr}\boldsymbol{T}\det {\boldsymbol{\varphi}},{\kern 8pt}\partial t\left( {\det {\boldsymbol{\phi} _ \pm }} \right) = \partial t\left( {\det {\boldsymbol{\varphi} _ \pm }{e^{ - i\boldsymbol{H}}}} \right) = 0,
\end{array}$$
where $\mathrm{Tr}\boldsymbol{X} = ik$, $\mathrm{Tr}\boldsymbol{T} =  - i\left[ {{\alpha _1}\left( {{\lambda ^2}+{k^2} } \right) - 4{\alpha _2}{k^3}} \right]$. Thus, for $z \in \mathop \Sigma \limits^ \circ   = \Sigma \backslash \left\{ { \pm i{u_o}} \right\}$, we derive from (\ref{eq15}) that
$$\det {\boldsymbol{\varphi} _ \pm } = \mu {e^{i{h _2}\left( {z,x,t} \right)}}.$$

Below, we introduce the scattering matrix $\boldsymbol{S}\left( z \right) = \left( {{s_{jl}}} \right){\kern 2pt}(j,l = 1, 2,3$), and the components of the this matrix are referred to as the scattering coefficients. As two fundamental solution matrices for the spectral problem (\ref{eq5}), ${\boldsymbol{\varphi} _ \pm }$ are related by constant matrix $\boldsymbol{S}\left( z \right)$, in the following manner
\begin{equation}\label{eq20}
{\boldsymbol{\varphi} _ - } = {\boldsymbol{\varphi} _ + }\boldsymbol{S}\left( z \right),{\kern 5pt}{\boldsymbol{\varphi} _ + } = {\boldsymbol{\varphi} _ - }\boldsymbol{B}\left( z \right),{\kern 5pt}z \in \mathop \Sigma \limits^ \circ,
\end{equation}
where it is clear that $\det \boldsymbol{S}\left( z \right) = 1$, and $\boldsymbol{B}\left( z \right) = {\boldsymbol{S}^{ - 1}}\left( z \right)= \left( {{b_{jl}}} \right)$.

For any $z$ within the domains where they are analytic, the modified eigenfunctions ${\boldsymbol{\phi} _ \pm }$ remain bounded when $x \in \mathbb{R}$. Moreover, for any $z$ on the continuous spectrum $\Sigma $, the asymptotic behavior of ${\boldsymbol{\phi} _ \pm }$ as $x \to  \mp \infty $ is given by
$$\begin{array}{l}
{\boldsymbol{\phi} _ + } = {\boldsymbol{G}_ - }{e^{i\boldsymbol{H}}}\boldsymbol{B}{e^{ - i\boldsymbol{H}}} + o\left( 1 \right),{\kern 5pt}x \to  - \infty,     {\kern 10pt}
{\boldsymbol{\phi} _ - } = {\boldsymbol{G}_ + }{e^{i\boldsymbol{H}}}\boldsymbol{S}{e^{ - i\boldsymbol{H}}} + o\left( 1 \right),{\kern 5pt}x \to  + \infty.
\end{array}$$
Using Eqs.(\ref{eq20}) and Proposition \ref{prop:1}, for $l = 1, 2, 3$, the analyticity of ${{s_{ll}}}$ and ${{b_{ll}}}$ is summarized in the following proposition.

\begin{proposition}
The scattering coefficients ${b_{ll}}$ and ${s_{ll}}$ can be analytically extended beyond $\Sigma $ into the following regions:
\begin{equation}\label{eq22}
\begin{array}{l}
{b_{11}}:z \in {\Xi _2},{\kern 5pt}{b_{22}}:{\mathop{\rm Im}\nolimits} z > 0,{\kern 5pt}{b_{33}}:z \in {\Xi _3},\\
{s_{11}}:z \in {\Xi _1},{\kern 5pt}{s_{22}}:{\mathop{\rm Im}\nolimits} z < 0,{\kern 5pt}{s_{33}}:z \in {\Xi _4}.
\end{array}
\end{equation}
\label{prop:2}
\end{proposition}

Through Propositions \ref{prop:1} and \ref{prop:2}, we find that the columns ${\boldsymbol{\varphi} _{ \pm ,l}}$ of the Jost eigenfunctions and the scattering coefficients ${b_{ll}}$ and ${s_{ll}}$ are analytic in their corresponding regions of the complex $z$-plane.  In contrast, in the defocusing case, the second column of the Jost eigenfunctions, as well as the scattering coefficients ${b_{22}}$ and ${s_{22}}$ are not analytic.

\subsection{Adjoint problem}

It follows from Proposition \ref{prop:1} that in an arbitrary region ${\Xi _j}$, only two columns of ${\boldsymbol{\varphi} _ \pm }$ are analytic. This requires us to introduce an adjoint spectral problem to construct the fully analytic eigenfunctions and then to address the inverse scattering problem \cite{N42}. The adjoint Lax pair is
\renewcommand{\arraystretch}{1.3}
\begin{equation}\label{eq23}
\begin{array}{l}
{\tilde {\boldsymbol{\varphi}} _x} = \tilde {\boldsymbol{X}}\left( {k,x,t} \right)\tilde {\boldsymbol{\varphi}} ,\\{\tilde {\boldsymbol{\varphi}} _t} = \tilde {\boldsymbol{T}}\left( {k,x,t} \right)\tilde {\boldsymbol{\varphi}} ,
\end{array}
\end{equation}
with
$$\tilde{ \boldsymbol{X}} = ik\boldsymbol{\sigma}  + {\boldsymbol{U}^*},{\kern 5pt} {\boldsymbol{U}^T} =  - {\boldsymbol{U}^*},{\kern 5pt}{\boldsymbol{U}^\dag } =  - \boldsymbol{U},$$
$$\tilde {\boldsymbol{T}} = {\alpha _1}\left[ { - 2i{k^2}\boldsymbol{\sigma}  - 2k{\boldsymbol{U}^*} - i\boldsymbol{\sigma} \left( {{{\left( {{\boldsymbol{U}^*}} \right)}^2} + u_o^2 - \boldsymbol{U}_x^*} \right)} \right] + {\alpha _2}\left[ {4i{k^3}\boldsymbol{\sigma}  + 4{k^2}{\boldsymbol{U}^*} - 2ik\boldsymbol{\sigma} \left( {\boldsymbol{U}_x^* - {{\left( {{\boldsymbol{U}^*}} \right)}^2}} \right) + \boldsymbol{U}_x^*{\boldsymbol{U}^*} - {\boldsymbol{U}^*}\boldsymbol{U}_x^*  - \boldsymbol{U}_{xx}^* + 2{{\left( {{\boldsymbol{U}^*}} \right)}^3}} \right].$$
Note that when $z \in \Sigma $, we have $\tilde {\boldsymbol{X}}\left( z \right) = {\boldsymbol{X}^*}\left( {{z^*}} \right)$ and $\tilde {\boldsymbol{T}}\left( z \right) = {\boldsymbol{T}^*}\left( {{z^*}} \right)$. Denoting the cross product by the symbol $' \times '$, we can establish the following proposition based on its fundamental properties \cite{N30,N40}.

\begin{proposition}
Assuming that ${\boldsymbol{\nu} _1}\left(z,x,t\right)$ and ${\boldsymbol{\nu} _2}\left( z,x,t\right)$ are two arbitrary solutions for the adjoint spectral problem (\ref{eq23}), it follows that
$$\boldsymbol{\nu}\left(z,x,t\right) = {e^{i{h_2}}}\left[ {{\boldsymbol{\nu} _1} \times {\boldsymbol{\nu} _2}} \right]$$
serves as a solution to the spectral problem (\ref{eq3}).
\label{prop:3}
 \end{proposition}

Based on this proposition, we generate four additional analytic eigenfunctions, each associated with a basic region ${\Xi _j}$, by formulating the Jost eigenfunctions of the adjoint spectral problem (\ref{eq23}). Similar to the previous process, the asymptotic spectral problem of Eqs.(\ref{eq23}) at $x \to  \pm \infty $ is given by
\begin{equation}\label{eq27}
\begin{array}{c}
{{\tilde {\boldsymbol{\varphi}} }_x} = {{\tilde {\boldsymbol{X}}}_ \pm }\tilde {\boldsymbol{\varphi}} ,{\kern 5pt}{{\tilde {\boldsymbol{X}}}_ \pm } = \mathop {\lim }\limits_{x \to  \pm \infty } \tilde {\boldsymbol{X}} = ik\boldsymbol{\sigma}  + \boldsymbol{U}_ \pm ^*,\\
{{\tilde {\boldsymbol{\varphi}} }_t} = {{\tilde {\boldsymbol{T}}}_ \pm }\tilde {\boldsymbol{\varphi}} ,{\kern 5pt}{{\tilde {\boldsymbol{T}}}_ \pm } = \mathop {\lim }\limits_{x \to  \pm \infty } \tilde {\boldsymbol{T}} = {\alpha _1}\left[ { - 2i{k^2}\boldsymbol{\sigma}  - 2k\boldsymbol{U}_ \pm ^* - i\boldsymbol{\sigma} \left( {{{\left( {\boldsymbol{U}_ \pm ^*} \right)}^2} + u_o^2} \right)} \right] + 2{\alpha _2}\left[ {2i{k^3}\boldsymbol{\sigma}  + 2{k^2}\boldsymbol{U}_ \pm ^* + ik\boldsymbol{\sigma} {{\left( {\boldsymbol{U}_ \pm ^*} \right)}^2} + {{\left( {\boldsymbol{U}_ \pm ^*} \right)}^3}} \right].
\end{array}
\end{equation}
The corresponding eigenvalue matrices of ${{\tilde {\boldsymbol{X}}}_ \pm }$ and ${{\tilde {\boldsymbol{T}}}_ \pm }$ are $ - i{\boldsymbol{\varpi} _1}$ and $ i{\boldsymbol{\varpi} _2}$, respectively, with the eigenvector matrix expressed as ${\tilde {\boldsymbol{G}}_ \pm }\left( z \right) = \boldsymbol{G}_ \pm ^*\left( {{z^*}} \right)$, satisfying $\det {\tilde {\boldsymbol{G}}_ \pm }\left( z \right) = \mu \left( z \right)$. These matrices satisfy the following relations
$${\tilde {\boldsymbol{X}}_ \pm } =  - i{\tilde {\boldsymbol{G}}_ \pm }\left( z \right){\boldsymbol{\varpi} _1}\tilde {\boldsymbol{G}}_ \pm ^{ - 1}\left( z \right),{\kern 5pt}{\tilde {\boldsymbol{T}}_ \pm } = i{\tilde {\boldsymbol{G}}_ \pm }\left( z \right){\boldsymbol{\varpi} _2}\tilde {\boldsymbol{G}}_ \pm ^{ - 1}\left( z \right).$$
As before, for $z \in \Sigma $, based on the Lax pair (\ref{eq23}),  the Jost solutions to the adjoint spectral problem (\ref{eq27}) are
$${{\tilde {\boldsymbol{\varphi }}}_ \pm }\left( {z} \right) = {{\tilde {\boldsymbol{G}}}_ \pm }{e^{ - i\boldsymbol{H}}} + o\left( 1 \right),{\kern 5pt}x \to  \pm \infty ,$$
and the modified adjoint eigenfunctions take the form
\begin{equation}\label{eq30}
{{\tilde {\boldsymbol{\phi}} }_ \pm }\left( {z} \right) = {{\tilde {\boldsymbol{\varphi }}}_ \pm }\left( z \right){e^{i\boldsymbol{H}}}.
\end{equation}
Similar to Proposition \ref{prop:1}, the columns ${\tilde{\boldsymbol{\phi}} _{ \pm, l}}{\kern 2pt}(l = 1, 2,3)$ of the modified adjoint eigenfunctions ${\tilde{\boldsymbol{\phi}} _ \pm }$ can be analytically extended to the appropriate regions:
$$\begin{array}{l}
{\tilde{\boldsymbol{\phi}} _{ + ,1}}:z \in {\Xi _1},{\kern 5pt}{\tilde{\boldsymbol{\phi}} _{ + ,2}}:{\mathop{\rm Im}\nolimits} z < 0,{\kern 5pt}{\tilde{\boldsymbol{\phi}} _{ + ,3}}:z \in {\Xi _4},\\
{\tilde{\boldsymbol{\phi}} _{ - ,1}}:z \in {\Xi _2},{\kern 5pt}{\tilde{\boldsymbol{\phi}} _{ - ,2}}:{\mathop{\rm Im}\nolimits} z > 0,{\kern 5pt}{\tilde{\boldsymbol{\phi}} _{ - ,3}}:z \in {\Xi _3}.
\end{array}$$

Similarly, the adjoint scattering matrices $\tilde {\boldsymbol{S}}$ and $\tilde {\boldsymbol{B}}$ correspond to
\begin{equation}\label{eq32}
{{\tilde {\boldsymbol{\varphi}} }_ - } = {{\tilde {\boldsymbol{\varphi}} }_ + }\tilde {\boldsymbol{S}},{\kern 5pt}{{\tilde {\boldsymbol{\varphi}} }_ + } = {{\tilde {\boldsymbol{\varphi}} }_ - }\tilde {\boldsymbol{B}},{\kern 5pt}z \in \mathop \Sigma \limits^ \circ  ,
\end{equation}
where $\tilde {\boldsymbol{B}} = {{\tilde {\boldsymbol{S}}}^{ - 1}}$, and the adjoint scattering coefficients ${\tilde{b}_{ll}}$ and ${\tilde{s}_{ll}}$ are analytically extendable beyond $\Sigma $ into the subsequent domains:
\begin{equation}\label{eq33}
\begin{array}{*{20}{l}}
{{{\tilde s}_{11}}:z \in {\Xi _2},{\kern 5pt}{{\tilde s}_{22}}:{\rm{Im}}z > 0,{\kern 5pt}{{\tilde s}_{33}}:z \in {\Xi _3},}\\
{{{\tilde b}_{11}}:z \in {\Xi _1},{\kern 5pt}{{\tilde b}_{22}}:{\rm{Im}}z < 0,{\kern 5pt}{{\tilde b}_{33}}:z \in {\Xi _4}.}
\end{array}
\end{equation}
Thus, based on the analyticity discussed above, using Proposition \ref{prop:3} one can define the four new solutions to the Eqs.(\ref{eq3}) as
\begin{equation}\label{eq34}
\begin{array}{l}
{\boldsymbol{\rho} _1}\left( {z} \right) = {e^{i{h_2}}}\left[ {{{\tilde {\boldsymbol{\varphi}} }_{ + ,1}}\left( {z} \right) \times {{\tilde {\boldsymbol{\varphi}} }_{ - ,2}}\left( {z} \right)} \right],{\kern 5pt}{\boldsymbol{\rho} _2}\left( {z} \right) = {e^{i{h_2}}}\left[ {{{\tilde {\boldsymbol{\varphi}} }_{ - ,1}}\left( {z} \right) \times {{\tilde {\boldsymbol{\varphi}} }_{ + ,2}}\left( {z} \right)} \right],\\
{\boldsymbol{\rho} _3}\left( {z} \right) = {e^{i{h_2}}}\left[ {{{\tilde {\boldsymbol{\varphi}} }_{ + ,2}}\left( {z} \right) \times {{\tilde {\boldsymbol{\varphi}} }_{ - ,3}}\left( {z} \right)} \right],{\kern 5pt}{\boldsymbol{\rho} _4}\left( {z} \right) = {e^{i{h_2}}}\left[ {{{\tilde {\boldsymbol{\varphi}} }_{ - ,2}}\left( {z} \right) \times {{\tilde {\boldsymbol{\varphi}} }_{ + ,3}}\left( {z} \right)} \right],
\end{array}
\end{equation}
which are known as auxiliary eigenfunctions. Moreover, according to the Eq.(\ref{eq30}) and the analyticity of ${\tilde{\boldsymbol{\phi}} _{ \pm, l}}$, it follows that ${\boldsymbol{\rho} _j}$ is completely analytic within the corresponding fundamental region ${\Xi _j}$. Notice that we construct four auxiliary eigenfunctions, which is two more than in the defocusing case. It is also important to note that there is a simple relationship between ${\boldsymbol{\varphi} _ \pm }$ and ${{\tilde {\boldsymbol{\varphi}} }_ \pm }$ when $z \in \Sigma $.

\newtheorem{corollary}{Corollary}
\begin{corollary}\label{cor:1}
Let $l$, $j$ and $n$ be arbitrary cyclic indices, when $z \in \Sigma $, it holds that
$$\begin{array}{l}
{\boldsymbol{\varphi} _{ \pm ,l}}\left( z \right) = \dfrac{{{e^{i{h_2}}}}}{{{\mu _l}}}\left[ {{{\tilde {\boldsymbol{\varphi}} }_{ \pm ,j}}\left( {z} \right) \times {{\tilde {\boldsymbol{\varphi}} }_{ \pm ,n}}\left( {z} \right)} \right],{\kern 5pt}
{{\tilde{\boldsymbol{ \varphi}} }_{ \pm ,l}}\left( z \right) = \dfrac{{{e^{ - i{h_2}}}}}{{{\mu _l}}}\left[ {{\boldsymbol{\varphi} _{ \pm ,j}}\left( {z} \right) \times {\boldsymbol{\varphi} _{ \pm ,n}}\left( {z} \right)} \right],
\end{array}$$
where ${\mu _1}\left( z \right) ={\mu _3}\left( z \right) = 1$ and ${\mu _2}\left( z \right) = \mu \left( z \right).$
\end{corollary}

From this, it can be derived that the relationship between the adjoint scattering matrix $\tilde {\boldsymbol{S}}$ and the scattering matrix $\boldsymbol{S}$ is given by
\begin{equation}\label{eq36}
\tilde {\boldsymbol{S}} = \hat {\boldsymbol{\sigma}} {\left( {{\boldsymbol{S}^{ - 1}}} \right)^T}{{\hat {\boldsymbol{\sigma}} }^{ - 1}},{\kern 5pt}\hat {\boldsymbol{\sigma}}  = \mathrm{diag}\left( {1,\mu ,1} \right),
\end{equation}
where $\mu $ is defined in Eqs.(\ref{eq12}). For $z \in \Sigma $, from Eqs.(\ref{eq32}), the definition (\ref{eq34}) of the auxiliary eigenfunction ${\boldsymbol{\rho} _j}$, and Corollary \ref{cor:1}, the decompositions for the Jost eigenfunctions ${\boldsymbol{\varphi} _ \pm }$ are obtained as
\renewcommand{\arraystretch}{2}
\begin{equation}\label{eq37}
\begin{array}{l}
{\boldsymbol{\varphi} _{ + ,1}} = \dfrac{{{\boldsymbol{\rho} _4} + {b_{21}}{\boldsymbol{\varphi} _{ + ,2}}}}{{{b_{22}}}} = \dfrac{{{b_{31}}{\boldsymbol{\varphi} _{ + ,3}} + {\boldsymbol{\rho} _3}}}{{{b_{33}}}},{\kern 5pt}
{\boldsymbol{\varphi} _{ - ,1}} = \dfrac{{{\boldsymbol{\rho} _3} + {s_{21}}{\boldsymbol{\varphi} _{ - ,2}}}}{{{s_{22}}}} = \dfrac{{{s_{31}}{\boldsymbol{\varphi} _{ - ,3}} + {\boldsymbol{\rho} _4}}}{{{s_{33}}}},\\
{\boldsymbol{\varphi} _{ + ,3}} = \dfrac{{{\boldsymbol{\rho} _1} + {b_{23}}{\boldsymbol{\varphi} _{ + ,2}}}}{{{b_{22}}}} = \dfrac{{{b_{13}}{\boldsymbol{\varphi} _{ + ,1}} + {\boldsymbol{\rho} _2}}}{{{b_{11}}}},{\kern 5pt}
{\boldsymbol{\varphi} _{ - ,3}} = \dfrac{{{\boldsymbol{\rho} _2} + {s_{23}}{\boldsymbol{\varphi} _{ - ,2}}}}{{{s_{22}}}} = \dfrac{{{s_{13}}{\boldsymbol{\varphi} _{ - ,1}} + {\boldsymbol{\rho} _1}}}{{{s_{11}}}}.
\end{array}
\end{equation}

Furthermore, in order to eliminate the exponential oscillations in ${\boldsymbol{\rho} _j}\left( z \right)$, the modified auxiliary eigenfunctions are introduced as
\begin{equation}\label{eq38}
{\boldsymbol{\chi} _1}\left( z \right) = {\boldsymbol{\rho} _1}\left( z \right){e^{i{h_1}}},{\kern 5pt}{\boldsymbol{\chi} _2}\left( z \right) = {\boldsymbol{\rho} _2}\left( z \right){e^{i{h_1}}},{\kern 5pt}{\boldsymbol{\chi} _3}\left( z \right) = {\boldsymbol{\rho} _3}\left( z \right){e^{ - i{h_1}}},{\kern 5pt}{\boldsymbol{\chi} _4}\left( z \right) = {\boldsymbol{\rho} _4}\left( z \right){e^{ - i{h_1}}}.
\end{equation}

\subsection{Symmetries}

In the case of NZBCs, the scattering problem (\ref{eq3}) exhibits two kinds of involutions: $\left( {\lambda, k } \right) \to \left( {{\lambda ^*}, {k^*}} \right)$ and $\left( {\lambda, k } \right) \to \left( {- \lambda, k } \right)$. These transformations correspond to $z \to {z^*}$ and $z \to  - {{u_o^2}/z}$, respectively. In other words, the scattering coefficients and eigenfunctions obey two sets of symmetry relations.

\begin{proposition}
If $\boldsymbol{\varphi} \left( z \right)$ is a nonsingular solution of the spectral problem (\ref{eq3}), then so are ${\left( {{\boldsymbol{\varphi} ^\dag }\left( {{z^*}} \right)} \right)^{ - 1}}$ and $\boldsymbol{\varphi} \left(  - {{u_o^2}/z} \right)$.
\label{prop:4}
\end{proposition}

The following is a detailed discussion of the two symmetries based on the above proposition.

\textbf{First symmetry:} Under the transformation $z \to {z^*}$, the Eqs.(\ref{eq8}) show that $\left( {\lambda, k } \right) \to \left( {{\lambda ^*}, {k^*}} \right)$. The property ${\boldsymbol{H}^\dag }\left( {{z^*}} \right) = \boldsymbol{H}\left( z \right)$ and Proposition \ref{prop:4} imply that ${\boldsymbol{\varphi} _ \pm }$ satisfy the following symmetry relations
\begin{equation}\label{eq39}
{\boldsymbol{\varphi} _ \pm }\left( z \right) = {\left( {\boldsymbol{\varphi} _ \pm ^\dag \left( {{z^*}} \right)} \right)^{ - 1}}\bar {\boldsymbol{\sigma}},{\kern 5pt}\bar {\boldsymbol{\sigma}}  = \mathrm{diag}\left( {\mu ,1,\mu } \right),{\kern 5pt}z \in \Sigma .
\end{equation}
Subsequently, substituting Eqs.(\ref{eq37}) into the expression (\ref{eq39}) yields the symmetry relations of the columns ${\boldsymbol{\varphi} _{ \pm ,l}}$ of the Jost eigenfunctions ${\boldsymbol{\varphi} _{ \pm }}$ as
\renewcommand{\arraystretch}{2}
$$\begin{array}{l}
\boldsymbol{\varphi} _{ - ,1}^*\left( {{z^*}} \right) = \dfrac{{{e^{ - i{h_2}\left( z \right)}}}}{{{s_{22}}\left( z \right)}}\left[ {{\boldsymbol{\varphi} _{ - ,2}}\left( z \right) \times {\boldsymbol{\rho} _2}}\left( z \right) \right],{\kern 8pt}\boldsymbol{\varphi} _{ + ,1}^*\left( {{z^*}} \right) = \dfrac{{{e^{ - i{h_2}\left( z \right)}}}}{{{b_{22}}\left( z \right)}}\left[ {{\boldsymbol{\varphi} _{ + ,2}}\left( z \right) \times {\boldsymbol{\rho} _1}}\left( z \right) \right],\\
\boldsymbol{\varphi} _{ - ,2}^*\left( {{z^*}} \right)  = \dfrac{{{e^{ - i{h_2}\left( z \right)}}}}{{\mu {s_{33}}\left( z \right)}}\left[ {{\boldsymbol{\varphi} _{ - ,3}}\left( z \right) \times {\boldsymbol{\rho} _4}}\left( z \right) \right]=\dfrac{{{e^{ - i{h_2}\left( z \right)}}}}{{\mu {s_{11}}\left( z \right)}}\left[ {{\boldsymbol{\rho} _1}\left( z \right) \times {\boldsymbol{\varphi} _{ - ,1}}}\left( z \right) \right],\\
\boldsymbol{\varphi} _{ + ,2}^*\left( {{z^*}} \right) = \dfrac{{{e^{ - i{h_2}\left( z \right)}}}}{{\mu {b_{33}}\left( z \right)}}\left[ {{\boldsymbol{\varphi} _{ + ,3}}\left( z \right) \times {\boldsymbol{\rho} _3}}\left( z \right) \right]=\dfrac{{{e^{ - i{h_2}\left( z \right)}}}}{{\mu {b_{11}}\left( z \right)}}\left[ {{\boldsymbol{\rho} _2}\left( z \right) \times {\boldsymbol{\varphi} _{ + ,1}}}\left( z \right) \right],\\
\boldsymbol{\varphi} _{ - ,3}^*\left( {{z^*}} \right) = \dfrac{{{e^{ - i{h_2}\left( z \right)}}}}{{{s_{22}}\left( z \right)}}\left[ {{\boldsymbol{\rho} _3}\left( z \right) \times {\boldsymbol{\varphi} _{ - ,2}}}\left( z \right) \right],{\kern 8pt}\boldsymbol{\varphi} _{ + ,3}^*\left( {{z^*}} \right) = \dfrac{{{e^{ - i{h_2}\left( z \right)}}}}{{{b_{22}}\left( z \right)}}\left[ {{\boldsymbol{\rho} _4}\left( z \right) \times {\boldsymbol{\varphi} _{ + ,2}}}\left( z \right) \right].
\end{array}$$

By direct calculation, we can find that scattering matrices $\boldsymbol{S}$, $\boldsymbol{B}$ satisfy the following relation
\begin{equation}\label{eq41}
{\boldsymbol{S}^\dag }\left( {{z^*}} \right) = \bar{\boldsymbol{\sigma}} \left( z \right)\boldsymbol{B}\left( z \right){\bar{\boldsymbol{\sigma}} ^{ - 1}}\left( z \right),{\kern 5pt}z \in \Sigma,
\end{equation}
i.e.
\begin{equation}\label{eq42}
\begin{array}{l}
{s_{11}}\left( z \right) = b_{11}^*\left( {{z^*}} \right),{\kern 10pt}{s_{12}}\left( z \right) = b_{21}^*\left( {{z^*}} \right)/{\mu ^*}\left( {{z^*}} \right),{\kern 10pt}{s_{13}}\left( z \right) = b_{31}^*\left( {{z^*}} \right),{\kern 10pt}{s_{21}}\left( z \right) = {\mu ^*}\left( {{z^*}} \right)b_{12}^*\left( {{z^*}} \right),\\
{s_{22}}\left( z \right) = b_{22}^*\left( {{z^*}} \right),{s_{23}}\left( z \right) = {\mu ^*}\left( {{z^*}} \right)b_{32}^*\left( {{z^*}} \right),{s_{31}}\left( z \right) = b_{13}^*\left( {{z^*}} \right),{s_{32}}\left( z \right) = \frac{{b_{23}^*\left( {{z^*}} \right)}}{{{\mu ^*}\left( {{z^*}} \right)}},{s_{33}}\left( z \right) = b_{33}^*\left( {{z^*}} \right).
\end{array}
\end{equation}
The Schwarz reflection principle implies the below symmetry relations
\begin{equation}\label{eq43}
\begin{array}{l}
{s_{11}}\left( z \right) = b_{11}^*\left( {{z^*}} \right),{\kern 3pt}z \in {\Xi _2},{\kern 8pt}
{s_{22}}\left( z \right) = b_{22}^*\left( {{z^*}} \right),{\kern 3pt}{\rm{Im}}z > 0,{\kern 8pt}
{s_{33}}\left( z \right) = b_{33}^*\left( {{z^*}} \right),{\kern 3pt}z \in {\Xi _3}.
\end{array}
\end{equation}

Similarly, the symmetry relations of ${\boldsymbol{\rho} _j}$ are
\begin{equation}\label{eq44}
\begin{array}{l}
\boldsymbol{\rho} _1^*\left( {{z^*}} \right) = {e^{ - i{h_2}}}\left[ {{\boldsymbol{\varphi} _{ + ,1}}\left( z \right) \times {\boldsymbol{\varphi} _{ - ,2}}\left( z \right)} \right],{\kern 5pt}z \in {\Xi _2},{\kern 10pt}
\boldsymbol{\rho} _2^*\left( {{z^*}} \right) = {e^{ - i{h_2}}}\left[ {{\boldsymbol{\varphi} _{ - ,1}}\left( z \right) \times {\boldsymbol{\varphi} _{ + ,2}}\left( z \right)} \right],{\kern 5pt}z \in {\Xi _1},\\
\boldsymbol{\rho} _3^*\left( {{z^*}} \right) = {e^{ - i{h_2}}}\left[ {{\boldsymbol{\varphi} _{ + ,2}}\left( z \right) \times {\boldsymbol{\varphi} _{ - ,3}}\left( z \right)} \right],{\kern 5pt}z \in {\Xi _4},{\kern 10pt}
\boldsymbol{\rho} _4^*\left( {{z^*}} \right) = {e^{ - i{h_2}}}\left[ {{\boldsymbol{\varphi} _{ - ,2}}\left( z \right) \times {\boldsymbol{\varphi} _{ + ,3}}\left( z \right)} \right],{\kern 5pt}z \in {\Xi _3}.
\end{array}
\end{equation}
In additional, Corollary \ref{cor:1} states that for cyclic indices $l$, $j$ and $n$, there are
$$\boldsymbol{\varphi} _{ \pm ,l}^*\left( {{z^*}} \right) = \frac{{{e^{ - i{h_2}}}}}{{{\mu _j}}}\left[ {{\boldsymbol{\varphi} _{ \pm ,j}}\left( z \right) \times {\boldsymbol{\varphi} _{ \pm ,n}}}\left( z \right) \right],{\kern 5pt}z \in \Sigma .$$

\textbf{Second symmetry:} Under the transformation $z \to  - {{u_o^2}/z}$, the Eqs.(\ref{eq8}) map $\left( {\lambda, k} \right) \to \left( {- \lambda, k } \right)$. Applying Proposition \ref{prop:4}, and using the relation $\boldsymbol{H}\left( { - {{u_o^2}/z}} \right) = \mathrm{diag}\left( { - 1,1, - 1} \right)\boldsymbol{H}\left( z \right)$, for $z \in \Sigma $, the Jost eigenfunctions ${\boldsymbol{\varphi} _ \pm }$ have the another symmetry relations
\renewcommand{\arraystretch}{1.2}
$${\boldsymbol{\varphi} _ \pm }\left( z \right) = {\boldsymbol{\varphi} _ \pm }\left( { - \frac{{u_o^2}}{z}} \right)\tilde {\boldsymbol{\sigma}} ,{\kern 5pt}\tilde {\boldsymbol{\sigma}}  = \left( {\begin{array}{*{20}{c}}
0\\
0\\
{{{i{u_o}} \mathord{\left/
 {\vphantom {{i{u_o}} z}} \right.
 \kern-\nulldelimiterspace} z}}
\end{array}\begin{array}{*{20}{c}}
0\\
1\\
0
\end{array}\begin{array}{*{20}{c}}
{{{i{u_o}} \mathord{\left/
 {\vphantom {{i{u_o}} z}} \right.
 \kern-\nulldelimiterspace} z}}\\
0\\
0
\end{array}} \right).$$
Following previous discussions, and considering the analytic regions of the eigenfunctions $\boldsymbol{\varphi}_ \pm$, the above relations can be reformulated as follows
\renewcommand{\arraystretch}{2.2}
\begin{equation}\label{eq47}
\begin{array}{l}
{\boldsymbol{\varphi} _{ \pm ,1}}\left( z \right) = \dfrac{{i{u_o}}}{z}{\boldsymbol{\varphi} _{ \pm ,3}}\left( { - \dfrac{{u_o^2}}{z}} \right),{\kern 8pt}z \in \left\{ {z:\left| z \right| > {u_o} \wedge {\mathop{\rm Im}\nolimits} z {\kern 3pt} {\scriptstyle{\lessgtr}} {\kern 3pt}0 } \right\},\\
{\boldsymbol{\varphi} _{ \pm ,2}}\left( z \right) = {\boldsymbol{\varphi} _{ \pm ,2}}\left( { - \dfrac{{u_o^2}}{z}} \right),{\kern 22pt}z \in \left\{ {z:{\mathop{\rm Im}\nolimits} z {\kern 3pt} {\scriptstyle{\lessgtr}} {\kern 3pt}0 } \right\},\\
{\boldsymbol{\varphi} _{ \pm ,3}}\left( z \right) = \dfrac{{i{u_o}}}{z}{\boldsymbol{\varphi} _{ \pm ,1}}\left( { - \dfrac{{u_o^2}}{z}} \right),{\kern 8pt}z \in \left\{ {z:\left| z \right| < {u_o} \wedge {\mathop{\rm Im}\nolimits} z  {\kern 3pt} {\scriptstyle{\lessgtr}} {\kern 3pt}0 } \right\}.
\end{array}
\end{equation}

As in Eq.({\ref{eq41}), we again apply the definition (\ref{eq20}) of the scattering matrix $\boldsymbol{S}$, which leads to
$$\boldsymbol{S}\left( { - \frac{{u_o^2}}{z}} \right) = \tilde {\boldsymbol{\sigma}} \boldsymbol{S}\left( z \right){{\tilde {\boldsymbol{\sigma}} }^{ - 1}},{\kern 5pt}z \in \Sigma ,$$
i.e.
\begin{equation}\label{eq49}
\begin{array}{l}
{s_{11}}\left( { - \dfrac{{u_o^2}}{z}} \right) = {s_{33}}\left( z \right),{\kern 15pt}{s_{12}}\left( { - \dfrac{{u_o^2}}{z}} \right) = \dfrac{{i{u_o}}}{z}{s_{32}}\left( z \right),{\kern 4pt}{s_{13}}\left( { - \dfrac{{u_o^2}}{z}} \right) = {s_{31}}\left( z \right),\\
{s_{21}}\left( { - \dfrac{{u_o^2}}{z}} \right) = \dfrac{z}{{i{u_o}}}{s_{23}}\left( z \right),{\kern 2pt}{s_{22}}\left( { - \dfrac{{u_o^2}}{z}} \right) = {s_{22}}\left( z \right),{\kern 18pt}{s_{23}}\left( { - \dfrac{{u_o^2}}{z}} \right) = \dfrac{z}{{i{u_o}}}{s_{21}}\left( z \right),\\
{s_{31}}\left( { - \dfrac{{u_o^2}}{z}} \right) = {s_{13}}\left( z \right),{\kern 16pt}{s_{32}}\left( { - \dfrac{{u_o^2}}{z}} \right) = \dfrac{{i{u_o}}}{z}{s_{12}}\left( z \right),{\kern 4pt}{s_{33}}\left( { - \dfrac{{u_o^2}}{z}} \right) = {s_{11}}\left( z \right).
\end{array}
\end{equation}
The scattering coefficients ${b_{lj}}$ are evident to follow the same relations. Based on the analyticity (\ref{eq33}) of the scattering coefficients ${a_{ll}}$ and ${b_{ll}}$, the following equations hold
\begin{equation}\label{eq50}
\begin{array}{l}
{s_{11}}\left( z \right) = {s_{33}}\left( { - \dfrac{{u_o^2}}{z}} \right),{\kern 5pt}z \in {\Xi _1},{\kern 15pt}{b_{11}}\left( z \right) = {b_{33}}\left( { - \dfrac{{u_o^2}}{z}} \right),{\kern 5pt}z \in {\Xi _2},\\
{b_{22}}\left( z \right) = {b_{22}}\left( { - \dfrac{{u_o^2}}{z}} \right),{\kern 5pt}{\mathop{\rm Im}\nolimits} z \ge 0,{\kern 8pt}{s_{22}}\left( z \right) = {s_{22}}\left( { - \dfrac{{u_o^2}}{z}} \right),{\kern 5pt}{\mathop{\rm Im}\nolimits} z \le 0.
\end{array}
\end{equation}

Using the symmetries (\ref{eq47}) of the eigenfunctions ${\boldsymbol{\varphi} _ \pm }$, the symmetries of ${\boldsymbol{\rho} _j}$ are derived as follows
$$\begin{array}{l}
{\boldsymbol{\rho} _1}\left( z \right) = \dfrac{{i{u_o}}}{z}{\boldsymbol{\rho} _4}\left( { - \dfrac{{u_o^2}}{z}} \right),{\kern 5pt}z \in {\Xi _1},{\kern 10pt}
{\boldsymbol{\rho} _2}\left( z \right) = \dfrac{{i{u_o}}}{z}{\boldsymbol{\rho} _3}\left( { - \dfrac{{u_o^2}}{z}} \right),{\kern 5pt}z \in {\Xi _2}.
\end{array}$$
We incorporate the symmetries (\ref{eq42}) and (\ref{eq49}) of the scattering coefficients to define the reflection coefficients ${\beta _l}$ as
\begin{equation}\label{eq52}
\begin{array}{l}
{\beta _1}\left( z \right) = \dfrac{{{s_{21}}\left( z \right)}}{{{s_{11}}\left( z \right)}} = \mu \dfrac{{b_{12}^*\left( {{z^*}} \right)}}{{b_{11}^*\left( {{z^*}} \right)}},{\kern 5pt}{\beta _1}\left( { - \dfrac{{u_o^2}}{z}} \right) = \dfrac{z}{{i{u_o}}}\dfrac{{{s_{23}}\left( z \right)}}{{{s_{33}}\left( z \right)}} = \dfrac{{z\mu }}{{i{u_o}}}\dfrac{{b_{32}^*\left( {{z^*}} \right)}}{{b_{33}^*\left( {{z^*}} \right)}},\\
{\beta _2}\left( z \right) = \dfrac{{{s_{31}}\left( z \right)}}{{{s_{11}}\left( z \right)}} =  \dfrac{{b_{13}^*\left( {{z^*}} \right)}}{{b_{11}^*\left( {{z^*}} \right)}},{\kern 9pt}{\beta _2}\left( { - \dfrac{{u_o^2}}{z}} \right) = \dfrac{{{s_{13}}\left( z \right)}}{{{s_{33}}\left( z \right)}} = \dfrac{{b_{31}^*\left( {{z^*}} \right)}}{{b_{33}^*\left( {{z^*}} \right)}},\\
{\beta _3}\left( z \right) = \dfrac{{{s_{32}}\left( z \right)}}{{{s_{22}}\left( z \right)}} = \dfrac{{b_{23}^*\left( {{z^*}} \right)}}{{\mu b_{22}^*\left( {{z^*}} \right)}},{\kern 5pt}{\beta _3}\left( { - \dfrac{{u_o^2}}{z}} \right) = \dfrac{{i{u_o}}}{z}\dfrac{{{s_{12}}\left( z \right)}}{{{s_{22}}\left( z \right)}} = \dfrac{{i{u_o}}}{{z\mu }}\dfrac{{b_{21}^*\left( {{z^*}} \right)}}{{b_{22}^*\left( {{z^*}} \right)}},
\end{array}
\end{equation}
which will aid in our construction for the inverse scattering problem.

\section{Discrete spectrum and asymptotic behavior}

\subsection{Discrete spectrum}

Because there are four fundamental regions in the focusing case, this implies that the discrete spectrum is richer than in the defocusing case. We introduce four matrices ${\boldsymbol{P}_j}$ to describe the discrete spectrum, which are analytic within the corresponding fundamental domains ${\Xi _j}{\kern 2pt} (j = 1, \mkern-2mu\cdots\mkern-2mu ,4)$. The matrices ${\boldsymbol{P}_j}$ are defined in terms of
\begin{equation}\label{eq53}
\begin{array}{l}
{\boldsymbol{P}_1}\left( z \right) = \left( {{\boldsymbol{\varphi} _{ - ,1}}\left( z \right),{\boldsymbol{\varphi} _{ + ,2}}\left( z \right),{\boldsymbol{\rho} _1}}\left( z \right) \right),{\kern 5pt}\det {\boldsymbol{P}_1} = {s_{11}}{b_{22}}\mu{e^{i{h_2}}},{\kern 5pt}z \in {\Xi _1},\\
{\boldsymbol{P}_2}\left( z \right) = \left( {{\boldsymbol{\varphi} _{ + ,1}}\left( z \right),{\boldsymbol{\varphi} _{ - ,2}}\left( z \right),{\boldsymbol{\rho} _2}}\left( z \right) \right),{\kern 5pt}\det {\boldsymbol{P}_2} = {s_{22}}{b_{11}}\mu{e^{i{h_2}}},{\kern 5pt}z \in {\Xi _2},\\
{\boldsymbol{P}_3}\left( z \right) = \left( {{\boldsymbol{\rho} _3}\left( z \right),{\boldsymbol{\varphi} _{ - ,2}}\left( z \right),{\boldsymbol{\varphi} _{ + ,3}}}\left( z \right) \right),{\kern 5pt}\det {\boldsymbol{P}_3} = {s_{22}}{b_{33}}\mu{e^{i{h_2}}},{\kern 5pt}z \in {\Xi _3},\\
{\boldsymbol{P}_4}\left( z \right) = \left( {{\boldsymbol{\rho} _4}\left( z \right),{\boldsymbol{\varphi} _{ + ,2}}\left( z \right),{\boldsymbol{\varphi} _{ - ,3}}}\left( z \right) \right),{\kern 5pt}\det {\boldsymbol{P}_4} = {s_{33}}{b_{22}}\mu{e^{i{h_2}}},{\kern 5pt}z \in {\Xi _4}.
\end{array}
\end{equation}
The determinant of ${\boldsymbol{P}_1}$ indicates that its columns are linearly dependent at the zeros of ${s_{11}}$, ${b_{22}}$ and $\mu$. In the same way, ${\boldsymbol{P}_l}{\kern 2pt} (l = 2, 3,4)$ exhibit the same property. Furthermore, according to the symmetry relations (\ref{eq43}) and (\ref{eq50}), the relationship between these zeros can be summarized in the following propositions.

\begin{proposition}
Let ${\mathop{\rm Im}\nolimits} {z_o} > 0$, so that
$${b_{22}}\left( {{z_o}} \right) = 0 \Leftrightarrow {s_{22}}\left( {z_o^*} \right) = 0 \Leftrightarrow {s_{22}}\left( { - \frac{{u_o^2}}{{z_o^*}}} \right) = 0 \Leftrightarrow {b_{22}}\left( { - \frac{{u_o^2}}{{{z_o}}}} \right) = 0.$$
\label{prop:5}
\end{proposition}

\begin{proposition}
Let ${\mathop{\rm Im}\nolimits} {z_o} < 0$ and $\left| {{z_o}} \right| \ge {u_o}$, so that
$${b_{11}}\left( {z_o^*} \right) = 0 \Leftrightarrow  {s_{11}}\left( {{z_o}} \right)= 0 \Leftrightarrow {b_{33}}\left( { - \frac{{u_o^2}}{{z_o^*}}} \right) = 0 \Leftrightarrow {s_{33}}\left( { - \frac{{u_o^2}}{{{z_o}}}} \right) = 0.$$
\label{prop:6}
\end{proposition}

The above propositions indicate that the discrete eigenvalues occur in the ${z_n},z_n^*,{{ - u_o^2}/{{z_n}}}$ and ${{ - u_o^2}/{z_n^*}}$. Consequently, for a specified ${z} \in {\Xi _1}$, there are three distinct categories of eigenvalues:
\begin{enumerate}[label=\Roman*., align=left, leftmargin=1.5cm, labelsep=0.0cm]
    \item $s_{11}\left( {{z}} \right) = 0 \wedge {b_{22}}\left( {{z}} \right) \ne 0$. This is classified as an eigenvalue of the first type.
    \item ${s_{11}}\left( {{z}} \right) \ne 0 \wedge {b_{22}}\left( {{z}} \right) = 0$. This is classified as an eigenvalue of the second type.
    \item ${s_{11}}\left( {{z}} \right) = {b_{22}}\left( {{z}} \right) = 0$. This is classified as an eigenvalue of the third type.
\end{enumerate}

In the following, we examine the properties of each of these three types for eigenvalues. For this purpose, based on the symmetries of the eigenfunctions and auxiliary eigenfunctions, the following propositions hold.

\begin{proposition}
Assuming that ${z_o} \in {\Xi _1}$, the following equivalent relations hold:
$${\boldsymbol{\rho} _1}\left( {{z_o}} \right) = \boldsymbol{0} \Leftrightarrow {\boldsymbol{\rho} _4}\left( { - \frac{{u_o^2}}{{{z_o}}}} \right) = \boldsymbol{0} \Leftrightarrow {\boldsymbol{\varphi} _{ - ,2}}\left( {z_o^*} \right) = {\varsigma _o}{\boldsymbol{\varphi} _{ + ,1}}\left( {z_o^*} \right) \Leftrightarrow {\boldsymbol{\varphi} _{ - ,2}}\left( { - \frac{{u_o^2}}{{z_o^*}}} \right) = {{\hat \varsigma }_o}{\boldsymbol{\varphi} _{ + ,3}}\left( { - \frac{{u_o^2}}{{z_o^*}}} \right),$$
where ${\varsigma _o}$ and ${{\hat \varsigma }_o}$ are arbitrary nonzero constants.
\label{prop:7}
\end{proposition}

\begin{proposition}
Assuming that ${z_o} \in {\Xi _1}$, the following equivalent relations hold:
$${\boldsymbol{\rho} _2}\left( {z_o^*} \right) = \boldsymbol{0} \Leftrightarrow {\boldsymbol{\rho} _3}\left( { - \frac{{u_o^2}}{{z_o^*}}} \right) = \boldsymbol{0} \Leftrightarrow {\boldsymbol{\varphi} _{ - ,1}}\left( {{z_o}} \right) = {{\tilde \varsigma }_o}{\boldsymbol{\varphi} _{ + ,2}}\left( {{z_o}} \right) \Leftrightarrow {\boldsymbol{\varphi} _{ - ,3}}\left( { - \frac{{u_o^2}}{{{z_o}}}} \right) = {\check{\varsigma} _o}{\boldsymbol{\varphi} _{ + ,2}}\left( { - \frac{{u_o^2}}{{{z_o}}}} \right),$$
where ${\tilde{\varsigma} _o}$ and ${\check{\varsigma} _o}$ are arbitrary nonzero constants.
\label{prop:8}
\end{proposition}

Assuming that all discrete eigenvalues are simple, based on the above results, we can derive the following conclusions, which provide a complete characterization of the discrete spectrum.

\begin{proposition}
Let ${z_o} \in {\Xi _1}$ be a simple discrete eigenvalue, and the following hold:
\begin{enumerate}[label=\roman*., align=left, leftmargin=0.9cm, labelsep=0.1cm]
    \item  Assuming ${z_o}$ is the first kind eigenvalue, there exist constants ${a_o},{{\hat a}_o},{ \check{a}_o}$ and ${{\bar a}_o}$ such that
   $$\begin{array}{l}
{\boldsymbol{\varphi} _{ - ,1}}\left( {{z_o}} \right) = \dfrac{{{a_o}}}{{{b_{22}}\left( {{z_o}} \right)}}{\boldsymbol{\rho} _1}\left( {{z_o}} \right),{\kern 5pt}{\boldsymbol{\rho} _2}\left( {z_o^*} \right) = {{\hat a}_o}{\boldsymbol{\varphi} _{ + ,1}}\left( {z_o^*} \right),{\kern 5pt}
{\boldsymbol{\rho} _3}\left( { - \dfrac{{u_o^2}}{{z_o^*}}} \right) = {\check{a}_o}{\boldsymbol{\varphi} _{ + ,3}}\left( { - \dfrac{{u_o^2}}{{z_o^*}}} \right),{\kern 5pt}{\boldsymbol{\varphi} _{ - ,3}}\left( { - \dfrac{{u_o^2}}{{{z_o}}}} \right) = {{\bar a}_o}{\boldsymbol{\rho} _4}\left( { - \dfrac{{u_o^2}}{{{z_o}}}} \right).
\end{array}$$
    \item Assuming ${z_o}$ is the second kind eigenvalue, there exist constants ${c_o},{{\hat c}_o},{ \check{c}_o}$ and ${{\bar c}_o}$ such that
   $$\begin{array}{l}
{\boldsymbol{\rho} _1}\left( {{z_o}} \right) = {c_o}{\boldsymbol{\varphi} _{ + ,2}}\left( {{z_o}} \right),{\kern 5pt}{\boldsymbol{\varphi} _{ - ,2}}\left( {z_o^*} \right) = {{\hat c}_o}{\boldsymbol{\rho} _2}\left( {z_o^*} \right),{\kern 5pt}
{\boldsymbol{\varphi} _{ - ,2}}\left( { - \dfrac{{u_o^2}}{{z_o^*}}} \right) = {\check{c}_o}{\boldsymbol{\rho} _3}\left( { - \dfrac{{u_o^2}}{{z_o^*}}} \right),{\kern 5pt}{\boldsymbol{\rho} _4}\left( { - \dfrac{{u_o^2}}{{{z_o}}}} \right) = {{\bar c}_o}{\boldsymbol{\varphi} _{ + ,2}}\left( { - \dfrac{{u_o^2}}{{{z_o}}}} \right).
\end{array}$$
    \item Assuming ${z_o}$ is the third kind eigenvalue, there exist constants ${d_o},{{\hat d}_o},{ \check{d}_o}$ and ${{\bar d}_o}$ such that
    $$\begin{array}{l}
{\boldsymbol{\varphi} _{ - ,1}}\left( {{z_o}} \right) = {d_o}{\boldsymbol{\varphi} _{ + ,2}}\left( {{z_o}} \right),{\kern 5pt}{\boldsymbol{\varphi} _{ - ,2}}\left( {z_o^*} \right) = {{\hat d}_o}{\boldsymbol{\varphi} _{ + ,1}}\left( {z_o^*} \right),{\kern 5pt}
{\boldsymbol{\varphi} _{ - ,2}}\left( { - \dfrac{{u_o^2}}{{z_o^*}}} \right) = {\check{d}_o}{\boldsymbol{\varphi} _{ + ,3}}\left( { - \dfrac{{u_o^2}}{{z_o^*}}} \right),{\kern 5pt}{\boldsymbol{\varphi} _{ - ,3}}\left( { - \dfrac{{u_o^2}}{{{z_o}}}} \right) = {{\bar d}_o}{\boldsymbol{\varphi} _{ + ,2}}\left( { - \dfrac{{u_o^2}}{{{z_o}}}} \right).
\end{array}$$
   \end{enumerate}
\label{prop:9}
\end{proposition}

In the following, we rewrite the above proposition using the modified Jost eigenfunctions $\boldsymbol{\phi}_{\pm}$ and the modified auxiliary eigenfunctions $\boldsymbol{\chi}_j$, aiming to computing the residual conditions more conveniently.

Let $\left\{ {{z_n}} \right\}_{n = 1}^{{N_1}}$ be the set of all first type eigenvalues, and then
$$\begin{array}{l}
{\boldsymbol{\phi} _{ - ,1}}\left( {{z_n}} \right) = \dfrac{{{a_n}}}{{{b_{22}}\left( {{z_n}} \right)}}{\boldsymbol{\chi} _1}\left( {{z_n}} \right){e^{ - 2i{h_1}\left( {{z_n}} \right)}},{\kern 4pt}{\boldsymbol{\phi} _{ - ,3}}\left( { - \dfrac{{u_o^2}}{{{z_n}}}} \right) = {{\bar a}_n}{\boldsymbol{\chi} _4}\left( { - \dfrac{{u_o^2}}{{{z_n}}}} \right){e^{ - 2i{h_1}\left( {{z_n}} \right)}},\\
{\boldsymbol{\chi} _3}\left( { - \dfrac{{u_o^2}}{{z_n^*}}} \right) = {\check{a}_n}{\boldsymbol{\phi} _{ + ,3}}\left( { - \dfrac{{u_o^2}}{{z_n^*}}} \right){e^{2i{h_1}\left( {z_n^*} \right)}},{\kern 8pt}{\boldsymbol{\chi} _2}\left( {z_n^*} \right) = {{\hat a}_n}{\boldsymbol{\phi} _{ + ,1}}\left( {z_n^*} \right){e^{2i{h_1}\left( {z_n^*} \right)}}.
\end{array}$$

Let $\left\{ {{\omega _n}} \right\}_{n = 1}^{{N_2}}$ be the set of all second type eigenvalues, and then
$$\begin{array}{l}
{\boldsymbol{\chi} _1}\left( {{\omega_n}} \right) = {c_n}{\boldsymbol{\phi} _{ + ,2}}\left( {{\omega_n}} \right){e^{i\left( {{h_1} + {h_2}} \right)\left( {{\omega_n}} \right)}},{\kern 30pt}{\boldsymbol{\phi} _{ - ,2}}\left( {\omega_n^*} \right) = {{\hat c}_n}{\boldsymbol{\chi} _2}\left( {\omega_n^*} \right){e^{ - i\left( {{h_1} + {h_2}} \right)\left( {\omega_n^*} \right)}},\\
{\boldsymbol{\phi} _{ - ,2}}\left( { - \dfrac{{u_o^2}}{{\omega_n^*}}} \right) = {\check{c}_n}{\boldsymbol{\chi} _3}\left( { - \dfrac{{u_o^2}}{{\omega_n^*}}} \right){e^{ - i\left( {{h_1} + {h_2}} \right)\left( {\omega_n^*} \right)}},{\kern 5pt}{\boldsymbol{\chi} _4}\left( { - \dfrac{{u_o^2}}{{{\omega_n}}}} \right) = {{\bar c}_n}{\boldsymbol{\phi} _{ + ,2}}\left( { - \dfrac{{u_o^2}}{{{\omega_n}}}} \right){e^{i\left( {{h_1} + {h_2}} \right)\left( {{\omega_n}} \right)}}.
\end{array}$$

Let $\left\{ {{\xi _n}} \right\}_{n = 1}^{{N_2}}$ be the set of all third type eigenvalues, and then
$$\begin{array}{l}
{\boldsymbol{\phi} _{ - ,1}}\left( {{\xi_n}} \right) = {d_n}{\boldsymbol{\phi} _{ + ,2}}\left( {{\xi_n}} \right){e^{ - i\left( {{h_1} - {h_2}} \right)\left( {{\xi_n}} \right)}},{\kern 24pt}{\boldsymbol{\phi} _{ - ,2}}\left( {\xi_n^*} \right) = {{\hat d}_n}{\boldsymbol{\phi} _{ + ,1}}\left( {\xi_n^*} \right){e^{i\left( {{h_1} - {h_2}} \right)\left( {\xi_n^*} \right)}},\\
{\boldsymbol{\phi} _{ - ,2}}\left( { - \dfrac{{u_o^2}}{{\xi_n^*}}} \right) = {\check{d}_n}{\boldsymbol{\phi} _{ + ,3}}\left( { - \dfrac{{u_o^2}}{{\xi_n^*}}} \right){e^{i\left( {{h_1} - {h_2}} \right)\left( {\xi_n^*} \right)}},{\kern 5pt}{\boldsymbol{\phi} _{ - ,3}}\left( { - \dfrac{{u_o^2}}{{{\xi_n}}}} \right) = {{\bar d}_n}{\boldsymbol{\phi} _{ + ,2}}\left( { - \dfrac{{u_o^2}}{{{\xi_n}}}} \right){e^{ - i\left( {{h_1} - {h_2}} \right)\left( {{\xi_n}} \right)}}.
\end{array}$$

Combining the symmetries of the modified auxiliary eigenfunctions $\boldsymbol{\chi}_j$  and the modified Jost eigenfunctions $\boldsymbol{\phi}_{\pm}$, we get the symmetry relations of the norming constants in Proposition \ref{prop:9} and express them as follows.

\begin{proposition}
The norming constants satisfy the following symmetry properties
$$\begin{array}{l}
{{\bar a}_n} = \dfrac{{{a_n}}}{{{b_{22}}\left( {{z_n}} \right)}},{\kern 5pt}{{\hat a}_n} = {\check{a}_n} =  - a_n^*,{\kern 5pt}{{\bar c}_n} = \dfrac{{{\omega _n}}}{{i{u_o}}}{c_n},{\kern 5pt}{{\hat c}_n} =  - \dfrac{{c_n^*}}{{\mu \left( {\omega _n^*} \right){b_{11}}\left( {\omega _n^*} \right)}},{\kern 5pt}{{\bar d}_n} = \dfrac{{{\xi _n}}}{{i{u_o}}}{d_n},\\
\end{array}$$
$$\begin{array}{l}
{\check{c}_n} = \dfrac{{{u_o}}}{{i\omega _n^*}}\dfrac{{c_n^*}}{{\mu \left( {\omega _n^*} \right){b_{11}}\left( {\omega _n^*} \right)}},{\kern 19pt}{{\hat d}_n} =  - \dfrac{{{s'_{22}} \left( {\xi _n^*} \right)}}{{{b'_{11}} \left( {\xi _n^*} \right)}}\dfrac{{d_n^*}}{{\mu \left( {\xi _n^*} \right)}},{\kern 19pt}{\check{d}_n} = \dfrac{{{u_o}}}{{i\xi _n^*}}\dfrac{{{s'_{22}} \left( {\xi _n^*} \right)}}{{{b'_{11}} \left( {\xi _n^*} \right)}}\dfrac{{d_n^*}}{{\mu \left( {\xi _n^*} \right)}}.
\end{array}$$
\label{prop:10}
\end{proposition}

\subsection{Asymptotic behavior}

The asymptotic behavior of the scattering coefficients and eigenfunctions at $k \to \infty $ are studied to ensure proper normalization in the subsequent construction of the Riemann-Hilbert problems. The uniformization variable $z = \lambda  + k$ implies that the asymptotic behavior at $z \to \infty $ and $z \to 0$ need to be examined simultaneously. Therefore, the asymptotic behavior of $\boldsymbol{\phi} _{ \pm }$ is derived using the differential equations (\ref{eq16}) and the standard Wentzel-Kramers-Brillouin expansion method \cite{N32} as follows

\renewcommand{\arraystretch}{1.2}
\begin{equation}\label{eq63}
\begin{array}{l}
{\boldsymbol{\phi} _{ \pm ,1}} = \left( {\begin{array}{*{20}{c}}
1\\
{i\boldsymbol{u}/z}
\end{array}} \right) + {\rm O}\left( {{z^{ - 2}}} \right),{\kern 25pt}z \to \infty ,{\kern 15pt}{\boldsymbol{\phi} _{ \pm ,1}} = \left( {\begin{array}{*{20}{c}}
{{\boldsymbol{u}^\dag }{\boldsymbol{u}_ \pm }/u_o^2}\\
{i{\boldsymbol{u}_ \pm }/z}
\end{array}} \right) + {\rm O}\left( z \right),{\kern 3pt}z \to 0,\\
{\boldsymbol{\phi} _{ \pm ,2}} = \left( {\begin{array}{*{20}{c}}
{i{\boldsymbol{u}^\dag }\boldsymbol{u}_ \pm ^ \bot /{u_o}z}\\
{\boldsymbol{u}_ \pm ^ \bot /{u_o}}
\end{array}} \right) + {\rm O}\left( {{z^{ - 2}}} \right),{\kern 3pt}z \to \infty ,{\kern 15pt}{\boldsymbol{\phi} _{ \pm ,2}} = \left( {\begin{array}{*{20}{c}}
0\\
{\boldsymbol{u}_ \pm ^ \bot /{u_o}}
\end{array}} \right) + {\rm O}\left( z \right),{\kern 10pt}z \to 0,\\
{\boldsymbol{\phi} _{ \pm ,3}} = \left( {\begin{array}{*{20}{c}}
{i{\boldsymbol{u}^\dag }{\boldsymbol{u}_ \pm }/{u_o}z}\\
{{\boldsymbol{u}_ \pm }/{u_o}}
\end{array}} \right) + {\rm O}\left( {{z^{ - 2}}} \right),{\kern 3pt}z \to \infty ,{\kern 15pt}{\boldsymbol{\phi} _{ \pm ,3}} = \left( {\begin{array}{*{20}{c}}
{i{u_o}/z}\\
{\boldsymbol{u}/{u_o}}
\end{array}} \right) + {\rm O}\left( z \right),{\kern 15pt}z \to 0.
\end{array}
\end{equation}
Based on the asymptotic behavior (\ref{eq63}) of the modified Jost eigenfunctions $\boldsymbol{\phi} _{ \pm}$, and incorporating the definitions (\ref{eq38}) of ${\boldsymbol{\chi} _j}$ along with (\ref{eq34}) and (\ref{eq44}), we derive the asymptotic behavior of ${\boldsymbol{\chi} _j}{\kern 2pt}(j = 1, \mkern-2mu\cdots\mkern-2mu ,4)$ as follows

$$\begin{array}{l}
{\boldsymbol{\chi} _1}\left( z \right) = \left( {\begin{array}{*{20}{c}}
{i{\boldsymbol{u}^\dag }{\boldsymbol{u}_ - }/{u_o}z}\\
{{\boldsymbol{u}_ - }/{u_o}}
\end{array}} \right) + {\rm O}\left( {{z^{ - 2}}} \right),{\kern 20pt}z \to \infty ,{\kern 15pt}{\boldsymbol{\chi} _1}\left( z \right) = \left( {\begin{array}{*{20}{c}}
{i\boldsymbol{u}_ + ^\dag {\boldsymbol{u}_ - }/{u_o}z}\\
{\left( {\boldsymbol{u}_ + ^\dag \boldsymbol{u}} \right){\boldsymbol{u}_ - }/u_o^3}
\end{array}} \right) + {\rm O}\left( z \right),{\kern 3pt}z \to 0,\\
{\boldsymbol{\chi} _2}\left( z \right) = \left( {\begin{array}{*{20}{c}}
{i{\boldsymbol{u}^\dag }{\boldsymbol{u}_ + }/{u_o}z}\\
{{\boldsymbol{u}_ + }/{u_o}}
\end{array}} \right) + {\rm O}\left( {{z^{ - 2}}} \right),{\kern 20pt}z \to \infty ,{\kern 15pt}{\boldsymbol{\chi} _2}\left( z \right) = \left( {\begin{array}{*{20}{c}}
{i\boldsymbol{u}_ - ^\dag {\boldsymbol{u}_ + }/{u_o}z}\\
{\left( {\boldsymbol{u}_ - ^\dag \boldsymbol{u}} \right){\boldsymbol{u}_ + }/u_o^3}
\end{array}} \right) + {\rm O}\left( z \right),{\kern 3pt}z \to 0,\\
{\boldsymbol{\chi} _3}\left( z \right) = \left( {\begin{array}{*{20}{c}}
{\boldsymbol{u}_ - ^\dag {\boldsymbol{u}_ + }/u_o^2}\\
{i\left( {\boldsymbol{u}_ - ^\dag {\boldsymbol{u}_ + }} \right)\boldsymbol{u}/u_o^2z}
\end{array}} \right) + {\rm O}\left( {{z^{ - 2}}} \right),{\kern 3pt}z \to \infty ,{\kern 15pt}{\boldsymbol{\chi} _3}\left( z \right) = \left( {\begin{array}{*{20}{c}}
{{\boldsymbol{u}^\dag }{\boldsymbol{u}_ + }/u_o^2}\\
{i{\boldsymbol{u}_ + }/z}
\end{array}} \right) + {\rm O}\left( z \right),{\kern 18pt}z \to 0,\\
{\boldsymbol{\chi} _4}\left( z \right) = \left( {\begin{array}{*{20}{c}}
{\boldsymbol{u}_ + ^\dag {\boldsymbol{u}_ - }/u_o^2}\\
{i\left( {\boldsymbol{u}_ + ^\dag {\boldsymbol{u}_ - }} \right)\boldsymbol{u}/u_o^2z}
\end{array}} \right) + {\rm O}\left( {{z^{ - 2}}} \right),{\kern 3pt}z \to \infty ,{\kern 15pt}{\boldsymbol{\chi} _4}\left( z \right) = \left( {\begin{array}{*{20}{c}}
{{\boldsymbol{u}^\dag }{\boldsymbol{u}_ - }/u_o^2}\\
{i{\boldsymbol{u}_ - }/z}
\end{array}} \right) + {\rm O}\left( z \right),{\kern 18pt}z \to 0.
\end{array}$$
Similarly, by combining the the definitions (\ref{eq20}) of the scattering matrices ${\boldsymbol{S}}$ and ${\boldsymbol{B}}$ and asymptotic behavior (\ref{eq63}) of modified Jost eigenfunctions $\boldsymbol{\phi} _{ \pm}$, we derive the asymptotic behavior for ${s_{jl}}$ and ${b_{jl}}$ :
$$\begin{array}{l}
{s_{11}} = 1 + {\rm O}\left( {{z^{ - 1}}} \right),{\kern 22pt}{b_{11}} = 1 + {\rm O}\left( {{z^{ - 1}}} \right),{\kern 25pt}z \to \infty ,{\kern 15pt}{s_{11}} = \dfrac{{\boldsymbol{u}_ + ^\dag {\boldsymbol{u}_ - }}}{{u_o^2}} + {\rm O}\left( z \right),{\kern 3pt}{b_{11}} = \dfrac{{\boldsymbol{u}_ - ^\dag {\boldsymbol{u}_ + }}}{{u_o^2}} + {\rm O}\left( z \right),{\kern 3pt}z \to 0,\\
{s_{22}} = \dfrac{{\boldsymbol{u}_ - ^\dag {\boldsymbol{u}_ + }}}{{u_o^2}} + {\rm O}\left( {{z^{ - 1}}} \right),{\kern 3pt}{b_{22}} = \dfrac{{\boldsymbol{u}_ + ^\dag {\boldsymbol{u}_ - }}}{{u_o^2}} + {\rm O}\left( {{z^{ - 1}}} \right),{\kern 7pt}z \to \infty ,{\kern 15pt}{s_{22}} = \dfrac{{\boldsymbol{u}_ - ^\dag {\boldsymbol{u}_ + }}}{{u_o^2}} + {\rm O}\left( z \right),{\kern 3pt}{b_{22}} = \dfrac{{\boldsymbol{u}_ + ^\dag {\boldsymbol{u}_ - }}}{{u_o^2}} + {\rm O}\left( z \right),{\kern 3pt}z \to 0,\\
{s_{33}} = \dfrac{{\boldsymbol{u}_ + ^\dag {\boldsymbol{u}_ - }}}{{u_o^2}} + {\rm O}\left( {{z^{ - 1}}} \right),{\kern 3pt}{b_{33}} = \dfrac{{\boldsymbol{u}_ - ^\dag {\boldsymbol{u}_ + }}}{{u_o^2}} + {\rm O}\left( {{z^{ - 1}}} \right),{\kern 7pt}z \to \infty ,{\kern 15pt}{s_{33}} = 1 + {\rm O}\left( z \right),{\kern 21pt}{b_{33}} = 1 + {\rm O}\left( z \right),{\kern 22pt}z \to 0.
\end{array}$$

\section{Inverse scattering problem}

\subsection{Riemann-Hilbert problem}

The task of this part is to formulate the inverse scattering problem as a matrix Riemann-Hilbert problem for an appropriate set of piecewise meromorphic functions in the complex $z$-plane, with assigned jumps across $\Sigma$. To achieve this, we use the scattering relation (\ref{eq20}) and Eqs.(\ref{eq53}) to construct the corresponding piecewise meromorphic matrices.

\newtheorem{lemma}{Lemma}
\begin{lemma}\label{lem:1}
For $z \in {\Xi _j}$, defining the piecewise meromorphic matrices $M\left( {z,x,t} \right) = {M_j}\left( {z,x,t} \right){\kern 2pt}(j = 1, \mkern-2mu\cdots\mkern-2mu ,4)$ that are
\renewcommand{\arraystretch}{2}
$$\begin{array}{l}
{\boldsymbol{M}_1}\left( {z,x,t} \right) = {\boldsymbol{P}_1}{e^{ - i\boldsymbol{H}}}{\left[ {\mathrm{diag}\left( {{s_{11}},1,{b_{22}}} \right)} \right]^{ - 1}} = \left( {\dfrac{{{\boldsymbol{\phi} _{ - ,1}}}}{{{s_{11}}}},{\boldsymbol{\phi} _{ + ,2}},\dfrac{{{\boldsymbol{\chi} _1}}}{{{b_{22}}}}} \right),{\kern 5pt}z \in {\Xi _1},\\
{\boldsymbol{M}_2}\left( {z,x,t} \right) = {\boldsymbol{P}_2}{e^{ - i\boldsymbol{H}}}{\left[ {\mathrm{diag}\left( {1,{s_{22}},{b_{11}}} \right)} \right]^{ - 1}} = \left( {{\boldsymbol{\phi} _{ + ,1}},\dfrac{{{\boldsymbol{\phi} _{ - ,2}}}}{{{s_{22}}}},\dfrac{{{\boldsymbol{\chi} _2}}}{{{b_{11}}}}} \right),{\kern 5pt}z \in {\Xi _2},\\
{\boldsymbol{M}_3}\left( {z,x,t} \right) = {\boldsymbol{P}_3}{e^{ - i\boldsymbol{H}}}{\left[ {\mathrm{diag}\left( {{b_{33}},{s_{22}},1} \right)} \right]^{ - 1}} = \left( {\dfrac{{{\boldsymbol{\chi} _3}}}{{{b_{33}}}},\dfrac{{{\boldsymbol{\phi} _{ - ,2}}}}{{{s_{22}}}},{\boldsymbol{\phi} _{ + ,3}}} \right),{\kern 5pt}z \in {\Xi _3},\\
{\boldsymbol{M}_4}\left( {z,x,t} \right) = {\boldsymbol{P}_4}{e^{ - i\boldsymbol{H}}}{\left[ {\mathrm{diag}\left( {{b_{22}},1,{s_{33}}} \right)} \right]^{ - 1}} = \left( {\dfrac{{{\boldsymbol{\chi} _4}}}{{{b_{22}}}},{\boldsymbol{\phi} _{ + ,2}},\dfrac{{{\boldsymbol{\phi} _{ - ,3}}}}{{{s_{33}}}}} \right),{\kern 5pt}z \in {\Xi _4},
\end{array}$$
and the corresponding jump conditions are
\begin{equation}\label{eq74}
{\boldsymbol{M}^ + }\left( z \right) = {\boldsymbol{M}^ - }\left( z \right)\left[ {\boldsymbol{I} - {e^{i\boldsymbol{H}\left( z \right)}}\boldsymbol{V}\left( z \right){e^{ - i\boldsymbol{H}\left( z \right)}}} \right],{\kern 5pt}z \in \Sigma ,
\end{equation}
where $\boldsymbol{M} = {\boldsymbol{M}^ - }$ for $z \in {\Xi ^ - } = {\Xi _2} \cup {\Xi _4}$ and $\boldsymbol{M} = {\boldsymbol{M}^ + }$ for $z \in {\Xi ^ + } = {\Xi _1} \cup {\Xi _3}$ (i.e. ${\boldsymbol{M}^ + } = {\boldsymbol{M}_1}$ for $z \in {\Xi _1}$, ${\boldsymbol{M}^ - } = {\boldsymbol{M}_2}$ for $z \in {\Xi _2}$, ${\boldsymbol{M}^ + } = {\boldsymbol{M}_3}$ for $z \in {\Xi _3}$, ${\boldsymbol{M}^ - } = {\boldsymbol{M}_4}$ for $z \in {\Xi _4}$). Here, the matrix $\boldsymbol{V}$ is given on every segment ${\Sigma _j}$ on the contour as

\renewcommand{\arraystretch}{1.2}
$$\begin{array}{l}
{\boldsymbol{V}}\left( z \right) = \left( {\begin{array}{*{20}{c}}
{{\Lambda _1}}\\
{ - {\beta _1}}\\
{{\beta _1}{\beta _3} - {\beta _2}}
\end{array}\begin{array}{*{20}{c}}
{{\Lambda _2}}\\
0\\
{{\beta _3}}
\end{array}\begin{array}{*{20}{c}}
{{\Lambda _3}}\\
{\mu {{\mathord{\buildrel{\lower3pt\hbox{$\scriptscriptstyle\frown$}}
\over \beta } }_3}}\\
{ - \mu {{\mathord{\buildrel{\lower3pt\hbox{$\scriptscriptstyle\frown$}}
\over \beta } }_3}{\beta _3}}
\end{array}} \right),{\kern 5pt}z \in {\Sigma _1},{\kern 5pt}
{\boldsymbol{V}}\left( z \right) = \left( {\begin{array}{*{20}{c}}
{{{\mathord{\buildrel{\lower3pt\hbox{$\scriptscriptstyle\smile$}}
\over \beta } }_2}{{\mathord{\buildrel{\lower3pt\hbox{$\scriptscriptstyle\frown$}}
\over \beta } }_2}}\\
0\\
{{{\mathord{\buildrel{\lower3pt\hbox{$\scriptscriptstyle\smile$}}
\over \beta } }_2}}
\end{array}\begin{array}{*{20}{c}}
0\\
0\\
0
\end{array}\begin{array}{*{20}{c}}
-{{{\mathord{\buildrel{\lower3pt\hbox{$\scriptscriptstyle\frown$}}
\over \beta } }_2}}\\
0\\
0
\end{array}} \right),{\kern 5pt}z \in {\Sigma _2},\\
{\boldsymbol{V}}\left( z \right) = \left( {\begin{array}{*{20}{c}}
{ - {{\mathord{\buildrel{\lower3pt\hbox{$\scriptscriptstyle\smile$}}
\over \beta } }_2}{{\hat \beta }_2}}\\
{{\Lambda _5}}\\
{{{\mathord{\buildrel{\lower3pt\hbox{$\scriptscriptstyle\smile$}}
\over \beta } }_2}}
\end{array}\begin{array}{*{20}{c}}
{{\Lambda _4}}\\
{{\Lambda _6}}\\
{ - {\beta _3}}
\end{array}\begin{array}{*{20}{c}}
{{{\hat \beta }_2}}\\
{{\Lambda _7}}\\
0
\end{array}} \right),{\kern 5pt}z \in {\Sigma _3},{\kern 33pt}
{\boldsymbol{V}}\left( z \right) = \left( {\begin{array}{*{20}{c}}
{{\beta _2}{{\hat \beta }_2}}\\
{{\Lambda _8}}\\
{ - {\beta _2}}
\end{array}\begin{array}{*{20}{c}}
0\\
0\\
0
\end{array}\begin{array}{*{20}{c}}
{{{\hat \beta }_2}}\\
{{\Lambda _9}}\\
0
\end{array}} \right),{\kern 5pt}z \in {\Sigma _4},
\end{array}$$
with
$$\begin{array}{l}
{\Lambda _1} = {\beta _1}{{\mathord{\buildrel{\lower3pt\hbox{$\scriptscriptstyle\frown$}}\over \beta } }_2}{\beta _3} - {\beta _2}{{\mathord{\buildrel{\lower3pt\hbox{$\scriptscriptstyle\frown$}}\over \beta } }_2} - \dfrac{{iz}}{{{u_o}}}{\beta _1}{{\hat \beta }_3},{\kern 36pt}
{\Lambda _2} = {\beta _3}{{\mathord{\buildrel{\lower3pt\hbox{$\scriptscriptstyle\frown$}}\over \beta } }_2} - \dfrac{{iz}}{{{u_o}}}{{\hat \beta }_3},{\kern 19pt}
{\Lambda _3} =  - {{\mathord{\buildrel{\lower3pt\hbox{$\scriptscriptstyle\frown$}}\over \beta } }_2} - \mu {{\mathord{\buildrel{\lower3pt\hbox{$\scriptscriptstyle\frown$}}\over \beta } }_2}{\beta _3}{{\mathord{\buildrel{\lower3pt\hbox{$\scriptscriptstyle\frown$}}\over \beta } }_3} + \dfrac{{iz}}{{{u_o}}}\mu {{\mathord{\buildrel{\lower3pt\hbox{$\scriptscriptstyle\frown$}}
\over \beta } }_3}{{\hat \beta }_3},\\
{\Lambda _5} = \dfrac{{{u_o}}}{{iz}}{{\mathord{\buildrel{\lower3pt\hbox{$\scriptscriptstyle\smile$}}\over \beta } }_2}{{\hat \beta }_1} + \dfrac{{z\mu }}{{i{u_o}}}{{\mathord{\buildrel{\lower3pt\hbox{$\scriptscriptstyle\smile$}}
\over \beta } }_3} + \dfrac{{z\mu }}{{i{u_o}}}{{\mathord{\buildrel{\lower3pt\hbox{$\scriptscriptstyle\smile$}}\over \beta } }_2}{{\hat \beta }_2}{{\mathord{\buildrel{\lower3pt\hbox{$\scriptscriptstyle\smile$}}
\over \beta } }_3},{\kern 17pt}
{\Lambda _4} = {\beta _3}{{\hat \beta }_2} - \dfrac{z}{{i{u_o}}}{{\hat \beta }_3},{\kern 16pt}
{\Lambda _6} = \dfrac{{{u_o}}}{{iz}}{\beta _3}{{\hat \beta }_1} + \dfrac{{{z^2}\mu }}{{u_o^2}}{{\hat \beta }_3}{{\mathord{\buildrel{\lower3pt\hbox{$\scriptscriptstyle\smile$}}\over \beta } }_3} + \dfrac{{z\mu }}{{i{u_o}}}{\beta _3}{{\hat \beta }_2}{{\hat \beta }_3},\\
{\Lambda _8} = \dfrac{{z\mu }}{{i{u_o}}}{{\mathord{\buildrel{\lower3pt\hbox{$\scriptscriptstyle\smile$}}\over \beta } }_3} - \dfrac{{z\mu }}{{i{u_o}}}{\beta _2}{{\hat \beta }_2} - {\beta _1} + \dfrac{{i{u_o}}}{z}{\beta _2}{{\hat \beta }_1},{\kern 3pt}
{\Lambda _7} = \dfrac{{iz\mu }}{{{u_o}}}{{\hat \beta }_2}{{\mathord{\buildrel{\lower3pt\hbox{$\scriptscriptstyle\smile$}}\over \beta } }_3} + \dfrac{{i{u_o}}}{z}{{\hat \beta }_1},{\kern 2pt}
{\Lambda _9} = \dfrac{{iz\mu }}{{{u_o}}}{{\hat \beta }_2}{{\hat \beta }_3} + \dfrac{{i{u_o}}}{z}{{\hat \beta }_1} + \mu {{\mathord{\buildrel{\lower3pt\hbox{$\scriptscriptstyle\frown$}}
\over \beta } }_3},
\end{array}$$
where ${\Sigma _j}$ is the boundary of ${{\overline \Xi }_j} \cap {{\overline \Xi }_{j + 1\bmod 4}}$, ${\overline{\Xi}_j}$ denotes the closure of ${\Xi_j}$, and ${\beta _j} = {\beta _j}\left( z \right)$, ${{\hat \beta }_j} = {\beta _j}\left( { - u_o^2/z} \right)$, ${{\mathord{\buildrel{\lower3pt\hbox{$\scriptscriptstyle\frown$}}
\over \beta } }_j} = \beta _j^*\left( {{z^*}} \right)$, ${{\mathord{\buildrel{\lower3pt\hbox{$\scriptscriptstyle\smile$}}
\over \beta } }_j} = \beta _j^*\left( { - u_o^2/{z^*}} \right)$.
\end{lemma}

Combined with the asymptotic behavior in Section 3, it can be derived that
\begin{equation}\label{eq76}
\boldsymbol{M}\left( z \right) = {\boldsymbol{M}_\infty } + {\rm O}\left( {{z^{ - 1}}} \right),{\kern 5pt}z \to \infty ,{\kern 15pt}\boldsymbol{M}\left( z \right) = \frac{i}{z}{\boldsymbol{M}_o} + {\rm O}\left( {{z^{ - 1}}} \right),{\kern 5pt}z \to 0,
\end{equation}
where
$${\boldsymbol{M}_\infty } = \left( {\begin{array}{*{20}{c}}
1\\
0
\end{array}\begin{array}{*{20}{c}}
0\\
{\boldsymbol{u}_ + ^ \bot /{u_o}}
\end{array}\begin{array}{*{20}{c}}
0\\
{{\boldsymbol{u}_ + }/{u_o}}
\end{array}} \right),{\kern 5pt}{\boldsymbol{M}_0} = \left( {\begin{array}{*{20}{c}}
0\\
{{\boldsymbol{u}_ + }}
\end{array}\begin{array}{*{20}{c}}
0\\
0
\end{array}\begin{array}{*{20}{c}}
{{u_o}}\\
0
\end{array}} \right),$$
which regularize the Riemann-Hilbert problem (\ref{eq74}) and ensure that it has a unique solution. In addition to the above asymptotic behavior, we need to compute the residue conditions to fully specified the Riemann-Hilbert problem (\ref{eq74}). For simplicity, define a symbol $\boldsymbol{M}_{res,\omega }^ \pm $ to denote the residue of ${\boldsymbol{M}^ \pm }$ at $z = \omega $ and ${\boldsymbol{M}^ \pm } = \left( {m_1^ \pm ,m_2^ \pm ,m_3^ \pm } \right)$. Direct computation yields the residue conditions for the piecewise meromorphic matrices ${\boldsymbol{M}^ \pm }$ as
\renewcommand{\arraystretch}{1.8}
\begin{equation}\label{eq78}
\begin{array}{l}
\boldsymbol{M}_{res,{z_n}}^ + \left( {x,t} \right) = {A_n}\left[ {m_3^ + \left( {{z_n}} \right),0,0} \right],{\kern 7pt}\boldsymbol{M}_{res, - u_o^2/z_n^*}^ + \left( {x,t} \right) = \dfrac{{z_n^*}}{{i{u_o}}}{\check{A}_n}\left[ {m_1^ - \left( {z_n^*} \right),0,0} \right],{\kern 8pt}\boldsymbol{M}_{res, - u_o^2/{z_n}}^ - \left( {x,t} \right) = \dfrac{{{z_n}}}{{i{u_o}}}{{\bar A}_n}\left[ {0,0,m_3^ + \left( {{z_n}} \right)} \right],\\
\boldsymbol{M}_{res,z_n^*}^ - \left( {x,t} \right) = {{\hat A}_n}\left[ {0,0,m_1^ - \left( {z_n^*} \right)} \right],{\kern 9pt}\boldsymbol{M}_{res, - u_o^2/{\omega _n}}^ - \left( {x,t} \right) = {{\bar C}_n}\left[ {m_2^ + \left( {{\omega _n}} \right),0,0} \right],{\kern 16pt}\boldsymbol{M}_{res,\omega _n^*}^ - \left( {x,t} \right) = {{\hat C}_n}\left[ {0,m_3^ - \left( {\omega _n^*} \right),0} \right],\\
\boldsymbol{M}_{res,{\omega _n}}^ + \left( {x,t} \right) = {C_n}\left[ {0,0,m_2^ + \left( {{\omega _n}} \right)} \right],{\kern 4pt}\boldsymbol{M}_{res, - u_o^2/\omega _n^*}^ + \left( {x,t} \right) = \dfrac{{\omega _n^*}}{{i{u_o}}}{\check{C}_n}\left[ {0,m_3^ - \left( {\omega _n^*} \right),0} \right],{\kern 3pt}\boldsymbol{M}_{res,\xi _n^*}^ - \left( {x,t} \right) = {{\hat D}_n}\left[ {0,m_1^ - \left( {\xi _n^*} \right),0} \right],\\
\boldsymbol{M}_{res,{\xi _n}}^ + \left( {x,t} \right) = {D_n}\left[ {m_2^ + \left( {{\xi _n}} \right),0,0} \right],{\kern 5pt}\boldsymbol{M}_{res, - u_o^2/\xi _n^*}^ + \left( {x,t} \right) = \dfrac{{\xi _n^*}}{{i{u_o}}}{\check{D}_n}\left[ {0,m_1^ - \left( {\xi _n^*} \right),0} \right],{\kern 6pt}\boldsymbol{M}_{res, - u_o^2/{\xi _n}}^ - \left( {x,t} \right) = {{\bar D}_n}\left[ {0,0,m_2^ + \left( {{\xi _n}} \right)} \right],
\end{array}
\end{equation}
with norming constants
\renewcommand{\arraystretch}{2.2}
$$\begin{array}{l}
{A_n} = \dfrac{{{a_n}}}{{{s'_{11}}\left( {{z_n}} \right)}}{e^{ - 2i{h_1}\left( {{z_n}} \right)}},{\kern 52pt}{\check{A}_n} = \dfrac{{{\check{a}_n}}}{{{b'_{33}}\left( { - u_o^2/z_n^*} \right)}}{e^{2i{h_1}\left( {z_n^*} \right)}},{\kern 31pt}{{\hat A}_n} = \dfrac{{{{\hat a}_n}}}{{{b'_{11}}\left( {z_n^*} \right)}}{e^{2i{h_1}\left( {z_n^*} \right)}},\\
{{\bar A}_n} = \dfrac{{{{\bar a}_n}{b_{22}}\left( {{z_n}} \right)}}{{{s'_{33}}\left( { - u_o^2/{z_n}} \right)}}{e^{ - 2i{h_1}\left( {{z_n}} \right)}},{\kern 32pt}{C_n} = \dfrac{{{c_n}}}{{{b'_{22}}\left( {{\omega _n}} \right)}}{e^{i\left[ {{h_1}\left( {{\omega _n}} \right) + {h_2}\left( {{\omega _n}} \right)} \right]}},{\kern 23pt}{\check{C}_n} = \dfrac{{{\check{c}_n}{b_{11}}\left( {\omega _n^*} \right)}}{{{s'_{22}}\left( { - u_o^2/\omega _n^*} \right)}}{e^{ - i\left[ {{h_1}\left( {\omega _n^*} \right) + {h_2}\left( {\omega _n^*} \right)} \right]}},\\
{{\bar C}_n} = \dfrac{{{{\bar c}_n}}}{{{b'_{22}}\left( { - u_o^2/{\omega_n}} \right)}}{e^{i\left[ {{h_1}\left( {{\omega _n}} \right) + {h_2}\left( {{\omega _n}} \right)} \right]}},{\kern 6pt}{{\hat C}_n} = \dfrac{{{{\hat c}_n}{b_{11}}\left( {\omega _n^*} \right)}}{{{s'_{22}}\left( {\omega _n^*} \right)}}{e^{ - i\left[ {{h_1}\left( {\omega _n^*} \right) + {h_2}\left( {\omega _n^*} \right)} \right]}},{\kern 4pt}{D_n} = \dfrac{{{d_n}}}{{{s'_{11}}\left( {{\xi _n}} \right)}}{e^{i\left[ {{h_2}\left( {{\xi _n}} \right) - {h_1}\left( {{\xi _n}} \right)} \right]}},\\
{\check{D}_n} = \dfrac{{{\check{d}_n}}}{{{s'_{22}}\left( { - u_o^2/\xi _n^*} \right)}}{e^{i\left[ {{h_1}\left( {\xi _n^*} \right) - {h_2}\left( {\xi _n^*} \right)} \right]}},{\kern 5pt}{{\hat D}_n} = \dfrac{{{{\hat d}_n}}}{{{s'_{22}}\left( {\xi _n^*} \right)}}{e^{i\left[ {{h_1}\left( {\xi _n^*} \right) - {h_2}\left( {\xi _n^*} \right)} \right]}},{\kern 23pt}{{\bar D}_n} = \dfrac{{{{\bar d}_n}}}{{{s'_{33}}\left( { - u_o^2/{\xi _n}} \right)}}{e^{i\left[ {{h_2}\left( {{\xi _n}} \right) - {h_1}\left( {{\xi _n}} \right)} \right]}},
\end{array}$$
where ${z_n}{\kern 2pt}(n = 1, \mkern-2mu\cdots\mkern-2mu ,{N_1}),{\omega _n}{\kern 2pt}(n = 1, \mkern-2mu\cdots\mkern-2mu ,{N_2})$ and ${\xi _n}{\kern 2pt}(n = 1, \mkern-2mu\cdots\mkern-2mu ,{N_3})$. The symmetries of the norming constants are derived from symmetry relations of the scattering coefficients:

$$\begin{array}{l}
{{\hat A}_n} =  - A_n^*,{\kern 5pt}{\check{A}_n} =  - \dfrac{{u_o^2}}{{{{\left( {z_n^*} \right)}^2}}}A_n^*,{\kern 5pt}{{\bar A}_n} = \dfrac{{u_o^2}}{{z_n^2}}{A_n},{\kern 5pt}{{\hat C}_n} =  - \dfrac{{C_n^*}}{{\mu \left( {\omega _n^*} \right)}},{\kern 5pt}{\check{C}_n} = \dfrac{{u_o^3}}{{i{{\left( {\omega _n^*} \right)}^3}}}\dfrac{{C_n^*}}{{\mu \left( {\omega _n^*} \right)}},\\
{{\bar C}_n} = \dfrac{{{u_o}}}{{i{\omega _n}}}{C_n},{\kern 23pt}{{\bar D}_n} = \dfrac{{{u_o}}}{{i{\xi _n}}}{D_n},{\kern 23pt}{{\hat D}_n} =  - \dfrac{{D_n^*}}{{\mu \left( {\xi _n^*} \right)}},{\kern 24pt}{\check{D}_n} = \dfrac{{u_o^3}}{{i{{\left( {\xi _n^*} \right)}^3}}}\dfrac{{D_n^*}}{{\mu \left( {\xi _n^*} \right)}}.
\end{array}$$

\subsection{Reconstruction formula}

The Riemann-Hilbert problem (\ref{eq74}) in Lemma \ref{lem:1} is regularized by eliminating the asymptotic behavior at infinity and subtracting the contributions of all poles within the discrete spectrum. Following this, the Cauchy projection is applied, leading to the proposition presented below.

\begin{proposition}
The solution to the Riemann-Hilbert problem (\ref{eq74}) is given by a matrix algebraic-integral equation
\begin{equation}\label{eq81}
\boldsymbol{M}\left( {z,x,t} \right) = {\boldsymbol{G}_ + }\left( z \right) + \sum\limits_{n = 1}^N {\left( {\frac{{\boldsymbol{M}_{res,{\tau _n}}^ + }}{{z - {\tau _n}}} + \frac{{\boldsymbol{M}_{res,\tau _n^*}^ - }}{{z - \tau _n^*}} + \frac{{\boldsymbol{M}_{res, - u_o^2/\tau _n^*}^ + }}{{z + \left( {u_o^2/\tau _n^*} \right)}} + \frac{{\boldsymbol{M}_{res, - u_o^2/{\tau _n}}^ - }}{{z + \left( {u_o^2/{\tau _n}} \right)}}} \right)}  - \frac{1}{{2i\pi }}\int_\Sigma  {\frac{{{\boldsymbol{M}^ - }\left( \gamma  \right)}}{{\gamma  - z}}\bar {\boldsymbol{V}}\left( \gamma  \right)\mathrm{d}\gamma },
\end{equation}
where $\bar {\boldsymbol{V}} = {e^{i\boldsymbol{H}}}\boldsymbol{V}{e^{ - i\boldsymbol{H}}}$,$\boldsymbol{M} = {\boldsymbol{M}^ \pm }$ for $z \in {\Xi ^ \pm }$ and $\left\{ {{\tau _n}} \right\}_{n = 1}^N$ represents the set of discrete eigenvalues. Furthermore, the eigenfunctions involved in the residue conditions (\ref{eq78}) are determined by
$$m_2^ + \left( z \right) = \left( {\begin{array}{*{20}{c}}
0\\
{\dfrac{{\boldsymbol{u}_ + ^ \bot }}{{{u_o}}}}
\end{array}} \right) + \sum\limits_{n = 1}^{{N_2}} {\left( {\frac{{{{\hat C}_n}}m_3^ - \left( {\omega _n^*} \right)}{{z - \omega _n^*}} - \frac{{i\omega_n^*}}{{{u_o}}}\frac{{{\check{C}_n}}m_3^ - \left( {\omega _n^*} \right)}{{z + \left( {u_o^2/\omega _n^*} \right)}}} \right)}  + \sum\limits_{n = 1}^{{N_3}} {\left( {\frac{{{{\hat D}_n}}m_1^ - \left( {\xi _n^*} \right)}{{z - \xi _n^*}} - \frac{{i\xi _n^*}}{{{u_o}}}\frac{{{\check{D}_n}}m_1^ - \left( {\xi _n^*} \right)}{{z + \left( {u_o^2/\xi _n^*} \right)}}} \right)} - \frac{1}{{2i\pi }}\int_\Sigma  {\frac{{{{\left( {{\boldsymbol{M}^ - }\bar {\boldsymbol{V}}} \right)}_2}\left( \gamma  \right)}}{{\gamma  - z}}\mathrm{d}\gamma },{\kern 5pt}z = {\omega _{i'}},{\xi _{l'}}, $$
$$m_3^ - \left( z \right) = \left( {\begin{array}{*{20}{c}}
{\dfrac{{i{u_o}}}{z}}\\
{\dfrac{{{\boldsymbol{u}_ + }}}{{{u_o}}}}
\end{array}} \right) + \sum\limits_{n = 1}^{{N_1}} {\left( {\frac{{{{\hat A}_n}m_1^ - \left( {z_n^*} \right)}}{{z - z_n^*}} - \frac{{i{z_n}}}{{{u_o}}}\frac{{{{\bar A}_n}m_3^ + \left( {{z_n}} \right)}}{{z + \left( {u_o^2/{z_n}} \right)}}} \right)}  + \sum\limits_{n = 1}^{{N_2}} {\frac{{{C_n}m_2^ + \left( {{\omega _n}} \right)}}{{z - {\omega _n}}}}  + \sum\limits_{n = 1}^{{N_3}} {\frac{{{{\bar D}_n}m_2^ + \left( {{\xi _n}} \right)}}{{z + \left( {u_o^2/{\xi _n}} \right)}}}  - \frac{1}{{2i\pi }}\int_\Sigma  {\frac{{{{\left( {{\boldsymbol{M}^ - }\bar {\boldsymbol{V}}} \right)}_3}\left( \gamma  \right)}}{{\gamma  - z}}\mathrm{d}\gamma }, {\kern 5pt}z = \omega _{i'}^*,$$
$$m_1^ - \left( z \right) = \left( {\begin{array}{*{20}{c}}
1\\
{\dfrac{{i{\boldsymbol{u}_ + }}}{z}}
\end{array}} \right) + \sum\limits_{n = 1}^{{N_1}} {\left( {\frac{{{A_n}m_3^ + \left( {{z_n}} \right)}}{{z - {z_n}}} - \frac{{iz_n^*}}{{{u_o}}}\frac{{{\check{A}_n}m_1^ - \left( {z_n^*} \right)}}{{z + \left( {u_o^2/z_n} \right)}}} \right)}  + \sum\limits_{n = 1}^{{N_2}} {\frac{{{{\bar C}_n}m_2^ + \left( {{\omega _n}} \right)}}{{z + \left( {u_o^2/{\omega _n}} \right)}}}  + \sum\limits_{n = 1}^{{N_3}} {\frac{{{D_n}m_2^ + \left( {{\xi _n}} \right)}}{{z - {\xi _n}}}}  - \frac{1}{{2i\pi }}\int_\Sigma  {\frac{{{{\left( {{\boldsymbol{M}^ - }\bar {\boldsymbol{V}}} \right)}_1}\left( \gamma  \right)}}{{\gamma  - z}}\mathrm{d}\gamma } ,{\kern 5pt}z = \xi _{l'}^*,z_{j'}^*,$$
$$m_3^ + \left( z \right) = \left( {\begin{array}{*{20}{c}}
{\dfrac{{i{u_o}}}{z}}\\
{\dfrac{{{\boldsymbol{u}_ + }}}{{{u_o}}}}
\end{array}} \right) + \sum\limits_{n = 1}^{{N_1}} {\left( {\frac{{{{\hat A}_n}m_1^ - \left( {z_n^*} \right)}}{{z - z_n^*}} - \frac{{i{z_n}}}{{{u_o}}}\frac{{{{\bar A}_n}m_3^ + \left( {{z_n}} \right)}}{{z + \left( {u_o^2/z_n^*} \right)}}} \right)}  + \sum\limits_{n = 1}^{{N_2}} {\frac{{{C_n}m_2^ + \left( {{\omega _n}} \right)}}{{z - {\omega _n}}}}  + \sum\limits_{n = 1}^{{N_3}} {\frac{{{{\bar D}_n}m_2^ + \left( {{\xi _n}} \right)}}{{z + \left( {u_o^2/{\xi _n}} \right)}}}  - \frac{1}{{2i\pi }}\int_\Sigma  {\frac{{{{\left( {{\boldsymbol{M}^ - }\bar {\boldsymbol{V}}} \right)}_3}\left( \gamma  \right)}}{{\gamma  - z}}\mathrm{d}\gamma } ,{\kern 5pt}z = {z_{j'}},$$
where $i' = 1, \mkern-2mu\cdots\mkern-2mu ,{N_1},j' = 1, \mkern-2mu\cdots\mkern-2mu ,{N_2}$ and $l' = 1, \mkern-2mu\cdots\mkern-2mu ,{N_3}$.
\label{prop:11}
\end{proposition}

With the help of Eq.(\ref{eq81}), the potential function of the focusing two-component Hirota equation (\ref{eq2}) can be reconstructed from the scattering coefficients and norming constants. Using the asymptotic behavior (\ref{eq63}) of the ${\boldsymbol{\phi} _{ \pm ,1}}$, we obtain
$${u_k}\left( {x,t} \right) =  - i\mathop {\lim }\limits_{z \to \infty } \left[ {z{\boldsymbol{\phi} _{ \pm ,\left( {k + 1} \right)1}}\left( {z,x,t} \right)} \right],{\kern 5pt}k = 1,2.$$
Combining the above equation and Proposition \ref{prop:11}, the reconstruction formula for the focusing two-component Hirota equation is summarized in the following theorem.

\newtheorem{theorem}{Theorem}
\begin{theorem}\label{thm:2}
The solution of Eq.(\ref{eq2}) is
$$\begin{array}{l}
{{u}_k}\left( {x,t} \right) = {u_{ + ,k}} -i\mathop{\displaystyle\sum}\limits_{n=1}^{N_1} {\left[ {{A_n}m_{\left( {k + 1} \right)3}^ + \left( {{z_n}} \right) + {\check{A}_n}m_{\left( {k + 1} \right)3}^ + \left( { - u_o^2/z_n^*} \right)} \right]}  \\
\qquad {\kern 11pt}  - i\displaystyle\sum\limits_{n = 1}^{{N_2}} {{{\bar C}_n}m_{\left( {k + 1} \right)2}^ + \left( { - u_o^2/{\omega _n}} \right)}
  - i\displaystyle\sum\limits_{n = 1}^{{N_3}} {{D_n}m_{\left( {k + 1} \right)2}^ + \left( {{\xi _n}} \right)}
 - \dfrac{1}{{2\pi }}\displaystyle\sum\limits_{n = 1}^4 {\int_{{\Sigma _n}} {{{\left( {{M^ - }{{\bar V}_n}} \right)}_{\left( {k + 1} \right)1}}\left( \gamma  \right)\mathrm{d}\gamma } } ,{\kern 5pt}k = 1,2.
\end{array}$$
\end{theorem}

\subsection{Trace formulae}

The trace formula serves as a critical tool for elucidating the connection between scattering data and the potential function, facilitating the resolution of the inverse scattering problem. By using the same approach as in \cite{N31,N40},  we can derive the explicit expressions for the analytic scattering coefficients.

\begin{proposition}
The expressions for the analytic scattering coefficients are
\begin{equation}\label{eq98}
\begin{array}{l}
{s_{11}}\left( z \right) = \mathrm{exp}\left( {\dfrac{1}{{2i\pi }}\mathop{\displaystyle\int_\Sigma}  {\dfrac{{K\left( \gamma  \right)}}{{\gamma  - z}}\mathrm{d}\gamma } } \right)\mathop{\displaystyle\prod}\limits_{n = 1}^{{N_1}} {\dfrac{{z - {z_n}}}{{z - z_n^*}}\dfrac{{z + \left( {u_o^2/z_n^*} \right)}}{{z + \left( {u_o^2/{z_n}} \right)}}} \mathop{\displaystyle\prod}\limits_{n = 1}^{{N_2}} {\dfrac{{z + \left( {u_o^2/{\omega _n}} \right)}}{{z + \left( {u_o^2/\omega _n^*} \right)}}} \mathop{\displaystyle\prod}\limits_{n = 1}^{{N_3}} {\dfrac{{z - {\xi _n}}}{{z - \xi _n^*}}} ,\\
{b_{22}}\left( z \right) = \mathrm{exp}\left( { - i\Delta h - \dfrac{1}{{2i\pi }}\mathop{\displaystyle\int_R} {\dfrac{{{K_o}\left( \gamma  \right)}}{{\gamma  - z}}\mathrm{d}\gamma } } \right)\mathop{\displaystyle\prod}\limits_{n = 1}^{{N_2}} {\dfrac{{z - {\omega _n}}}{{z - \omega _n^*}}\dfrac{{z + \left( {u_o^2/{\omega _n}} \right)}}{{z + \left( {u_o^2/\omega _n^*} \right)}}} \mathop{\displaystyle\prod}\limits_{n = 1}^{{N_3}} {\dfrac{{z - {\xi _n}}}{{z - \xi _n^*}}\dfrac{{z + \left( {u_o^2/{\xi _n}} \right)}}{{z + \left( {u_o^2/\xi _n^*} \right)}}} .
\end{array}
\end{equation}
where the jumps conditions $K\left( \gamma  \right)={K_j}\left( \gamma  \right){\kern 2pt}(j = 1, \mkern-2mu\cdots\mkern-2mu ,4)$ are
$$\begin{array}{l}
{K_1}\left( z \right) =  - \ln \left[ {1  + {\beta _2}\left( z \right)\beta _2^*\left( {{z^*}} \right)}+ {\mu ^{ - 1}}{\beta _1}\left( z \right)\beta _1^*\left( {{z^*}} \right) \right],{\kern 5pt}z \in {\Sigma _1},\\
{K_2}\left( z \right) = \dfrac{1}{{2i\pi }}\mathop{\displaystyle\int_R} {\dfrac{{{K_o}\left( z \right)}}{{\gamma  - z}}\mathrm{d}\gamma }  - \ln \left[ {1 - \beta _2^*\left( {{z^*}} \right)\beta _2^*\left( { - \dfrac{{u_o^2}}{{{z^*}}}} \right)} \right],{\kern 5pt}z \in {\Sigma _2},\\
{K_3}\left( z \right) =  - \ln \left[ {1 + {\beta _2}\left( { - \dfrac{{u_o^2}}{z}} \right)\beta _2^*\left( { - \dfrac{{u_o^2}}{{{z^*}}}} \right) + {{\dfrac{{u_o^2}}{{{z^2}\mu \left( z \right)}}}}{\beta _1}\left( { - \dfrac{{u_o^2}}{z}} \right)\beta _1^*\left( { - \dfrac{{u_o^2}}{{{z^*}}}} \right)} \right],{\kern 5pt}z \in {\Sigma _3},\\
{K_4}\left( z \right) = \dfrac{i}{{2\pi }}\mathop{\displaystyle\int_R} {\dfrac{{{K_o}\left( z \right)}}{{\gamma  - z}}\mathrm{d}\gamma }  - \ln \left[ {1 - \beta _2^*\left( {{z^*}} \right)\beta _2^*\left( { - \dfrac{{u_o^2}}{{{z^*}}}} \right)} \right],{\kern 5pt}z \in {\Sigma _4},
\end{array}$$
with
$${K_o}\left( z \right) = \ln \left[ {1 + \mu \left( \gamma  \right){\beta _3}\left( \gamma  \right)\beta _3^*\left( {{\gamma ^*}} \right) + {\gamma ^2}u_o^{ - 2}\mu \left( \gamma  \right){\beta _3}\left( { - u_o^2/\gamma } \right)\beta _3^*\left( { - u_o^2/{\gamma ^*}} \right)} \right].$$
\label{prop:13}
\end{proposition}

It is worth noting that other trace formulas for the analytic scattering coefficients can be derived using the symmetry relations (\ref{eq43}) and (\ref{eq50}).  In addition, the trace formulas for the focusing case are more complex than those for the defocusing case, owing to the richer discrete spectrum in the focusing case. Furthermore, applying the trace formula for $s_{11}$ as $z \to 0$, one can derive the asymptotic phase difference $\Delta h = {h_ + } - {h_ - }$, which is given by
$$\Delta h =  - 4\sum\limits_{n = 1}^{{N_1}} {\arg \left( {{z_n}} \right)}  + 2\sum\limits_{n = 1}^{{N_2}} {\arg \left( {{\omega _n}} \right)}  - 2\sum\limits_{n = 1}^{{N_3}} {\arg \left( {{\xi _n}} \right)}  + \frac{1}{{2\pi }}\int_\Sigma  {{\gamma ^{ - 1}}K\left( \gamma  \right)\mathrm{d}\gamma }. $$

\section{Pure soliton solutions}

\begin{theorem}\label{thm:1}
The $N$-soliton solution of Eq.(\ref{eq2}) is
\renewcommand{\arraystretch}{1.2}
\begin{equation}\label{eq100}
\boldsymbol{u}\left( {x,t} \right) = \frac{1}{{\det {\boldsymbol{Q}}}}\left( {\begin{array}{*{20}{c}}
{\det {\boldsymbol{Q}}_1^{aug}}\\
{\det {\boldsymbol{Q}}_2^{aug}}
\end{array}} \right),{\kern 5pt}\boldsymbol{Q}_k^{aug} = \left( {\begin{array}{*{20}{c}}
{{u_{ + ,k}}}\\
{{\boldsymbol{\Theta} _k}}
\end{array}\begin{array}{*{20}{c}}
{{\boldsymbol{E}^T}}\\
\boldsymbol{Q}
\end{array}} \right),{\kern 5pt}k = 1,2,
\end{equation}
where the vectors $\boldsymbol{E}$, $\boldsymbol{\Theta} _k$ and matrix $\boldsymbol{Q}$ are represented as
$$\boldsymbol{E} = {\left( {{\eta _1}, \cdots ,{\eta _{2{N_1} + {N_2} + {N_3}}}} \right)^T},{\kern 5pt}{\boldsymbol{\Theta} _k} = {\left( {{\theta _{k1}}, \cdots ,{\theta _{k(2{N_1} + {N_2} + {N_3})}}} \right)^T},{\kern 5pt}\boldsymbol{Q} = \boldsymbol{I} - \boldsymbol{W},$$
respectively. Here
\begin{equation}\label{eq102}
{\eta _l} = \left\{ \begin{array}{l}
i{A_l},{\kern 34pt}l = 1, \mkern-2mu\cdots\mkern-2mu ,{N_1},\\
\dfrac{{z_{l - {N_1}}^*}}{{{u_o}}}{\check{A}_{l - {N_1}}},{\kern 4pt}l = {N_1} + 1, \mkern-2mu\cdots\mkern-2mu ,2{N_1},\\
i{{\bar C}_{l - 2{N_1}}},{\kern 18pt}l = 2{N_1} + 1, \mkern-2mu\cdots\mkern-2mu ,2{N_1} + {N_2},\\
i{D_{l  - {N_2}- 2{N_1}}},{\kern 5pt}l = 2{N_1} + {N_2} + 1, \mkern-2mu\cdots\mkern-2mu ,2{N_1} + {N_2} + {N_3},
\end{array} \right.
\end{equation}
and
$${\theta _{kl}} = \left\{ \begin{array}{l}
{u_{ + ,k}}/{u_o},{\kern 237pt}l = 1, \mkern-2mu\cdots\mkern-2mu ,{N_1},\\
i{u_{ + ,k}}/z_{l - {N_1}}^*,{\kern 225pt}l = {N_1} + 1, \mkern-2mu\cdots\mkern-2mu ,2{N_1},\\
{\left( { - 1} \right)^{k + 1}}\dfrac{{u_{ + ,\bar k}^*}}{{{u_o}}} + \dfrac{{{u_{ + ,k}}}}{{{u_o}}}\mathop{\displaystyle\sum}\limits_{n = 1}^{{N_2}} {Y_n^{\left( 1 \right)}\left( {{\omega _{l - 2{N_1}}}} \right)}  + i{u_{ + ,k}}\mathop{\displaystyle\sum}\limits_{n = 1}^{{N_3}} {\dfrac{{Y_n^{\left( 2 \right)}\left( {{\omega _{l - 2{N_1}}}} \right)}}{{\xi _n^*}}} ,{\kern 27pt}l = 2{N_1} + 1, \mkern-2mu\cdots\mkern-2mu ,2{N_1} + {N_2},\\
{\left( { - 1} \right)^{k + 1}}\dfrac{{u_{ + ,\bar k}^*}}{{{u_o}}} + \dfrac{{{u_{ + ,k}}}}{{{u_o}}}\mathop{\displaystyle\sum}\limits_{n = 1}^{{N_2}} {Y_n^{\left( 1 \right)}\left( {{\xi _{l  - {N_2}- 2{N_1}}}} \right)}  + i{u_{ + ,k}}\mathop{\displaystyle\sum}\limits_{n = 1}^{{N_3}} {\dfrac{{Y_n^{\left( 2 \right)}\left( {{\xi _{l  - {N_2}- 2{N_1}}}} \right)}}{{\xi _n^*}}} ,{\kern 5pt}l = 2{N_1} + {N_2} + 1, \mkern-2mu\cdots\mkern-2mu ,2{N_1} + {N_2} + {N_3},
\end{array} \right.$$
with $\bar{k}=3-k$. The detailed expressions for $\boldsymbol{W} = {W_{\varepsilon l}}\left( {x,t} \right)$ as follows. When $\varepsilon ,l = 1, \mkern-2mu\cdots\mkern-2mu ,{N_1}$,
$${W_{\varepsilon l}} = Y_l^{\left( 4 \right)}\left( {{z_\varepsilon }} \right).$$
When $\varepsilon  = 1, \mkern-2mu\cdots\mkern-2mu ,{N_1}$ and $l = {N_1} + 1, \mkern-2mu\cdots\mkern-2mu ,2{N_1},$
$${W_{\varepsilon l}} = Y_{l - {N_1}}^{\left( 3 \right)}\left( {{z_\varepsilon }} \right).$$
When $\varepsilon  = 1, \mkern-2mu\cdots\mkern-2mu ,{N_1}$ and $l = 2{N_1} + 1, \mkern-2mu\cdots\mkern-2mu ,2{N_1} + {N_2},$
$${W_{\varepsilon l}} = Y_{l - 2{N_1}}^{\left( 5 \right)}\left( {{z_\varepsilon }} \right).$$
When $\varepsilon  = 1, \mkern-2mu\cdots\mkern-2mu ,{N_1}$ and $l = 2{N_1} + {N_2} + 1, \mkern-2mu\cdots\mkern-2mu ,2{N_1} + {N_2} + {N_3},$
$${W_{\varepsilon l}} = Y_{l  - {N_2}- 2{N_1}}^{\left( 6 \right)}\left( {{z_\varepsilon }} \right).$$
When $\varepsilon  = {N_1} + 1, \mkern-2mu\cdots\mkern-2mu ,2{N_1}$ and $l = 1, \mkern-2mu\cdots\mkern-2mu ,{N_1},$
$${W_{\varepsilon l}} = Y_l^{\left( 7 \right)}\left( {z_{\varepsilon  - {N_1}}^*} \right).$$
When $\varepsilon, l  = {N_1} + 1, \mkern-2mu\cdots\mkern-2mu ,2{N_1}$,
$${W_{\varepsilon l}} = Y_{l - {N_1}}^{\left( 8 \right)}\left( {z_{\varepsilon  - {N_1}}^*} \right).$$
When $\varepsilon  = {N_1} + 1, \mkern-2mu\cdots\mkern-2mu ,2{N_1}$ and $l = 2{N_1} + 1, \mkern-2mu\cdots\mkern-2mu ,2{N_1} + {N_2},$
$${W_{\varepsilon l}} = Y_{l - 2{N_1}}^{\left( 9 \right)}\left( {z_{\varepsilon  - {N_1}}^*} \right).$$
When $\varepsilon  = {N_1} + 1, \mkern-2mu\cdots\mkern-2mu ,2{N_1}$ and $l = 2{N_1} + {N_2} + 1, \mkern-2mu\cdots\mkern-2mu ,2{N_1} + {N_2} + {N_3},$
$${W_{\varepsilon l}} = Y_{l  - {N_2}- 2{N_1}}^{\left( {10} \right)}\left( {z_{\varepsilon  - {N_1}}^*} \right).$$
When $\varepsilon  = 2{N_1} + 1, \mkern-2mu\cdots\mkern-2mu ,2{N_1} + {N_2}$ and $l = 1, \mkern-2mu\cdots\mkern-2mu ,{N_1},$
$${W_{\varepsilon l}} = \sum\limits_{n = 1}^{{N_2}} {Y_n^{\left( 1 \right)}\left( {{\omega _{\varepsilon  - 2{N_1}}}} \right)Y_l^{\left( 4 \right)}\left( {\omega _n^*} \right)}  + \sum\limits_{n = 1}^{{N_3}} {Y_n^{\left( 2 \right)}\left( {{\omega _{\varepsilon  - 2{N_1}}}} \right)Y_l^{\left( 7 \right)}\left( {\xi _n^*} \right)} .$$
When $\varepsilon  = 2{N_1} + 1, \mkern-2mu\cdots\mkern-2mu ,2{N_1} + {N_2}$ and $l = {N_1} + 1, \mkern-2mu\cdots\mkern-2mu ,2{N_1},$
$${W_{\varepsilon l}} = \sum\limits_{n = 1}^{{N_2}} {Y_n^{\left( 1 \right)}\left( {{\omega _{\varepsilon  - 2{N_1}}}} \right)Y_{l - {N_1}}^{\left( 3 \right)}\left( {\omega _n^*} \right)}  + \sum\limits_{n = 1}^{{N_3}} {Y_n^{\left( 2 \right)}\left( {{\omega _{\varepsilon  - 2{N_1}}}} \right)Y_{l - {N_1}}^{\left( 8 \right)}\left( {\xi _n^*} \right)} .$$
When $\varepsilon  = 2{N_1} + 1, \mkern-2mu\cdots\mkern-2mu ,2{N_1} + {N_2}$ and $l = 2{N_1} + 1, \mkern-2mu\cdots\mkern-2mu ,2{N_1} + {N_2},$
$${W_{\varepsilon l}} = \sum\limits_{n = 1}^{{N_2}} {Y_n^{\left( 1 \right)}\left( {{\omega _{\varepsilon  - 2{N_1}}}} \right)Y_{l - 2{N_1}}^{\left( 5 \right)}\left( {\omega _n^*} \right)}  + \sum\limits_{n = 1}^{{N_3}} {Y_n^{\left( 2 \right)}\left( {{\omega _{\varepsilon  - 2{N_1}}}} \right)Y_{l - 2{N_1}}^{\left( 9 \right)}\left( {\xi _n^*} \right)} .$$
When $\varepsilon  = 2{N_1} + 1, \mkern-2mu\cdots\mkern-2mu ,2{N_1} + {N_2}$ and $l = 2{N_1} + {N_2} + 1, \mkern-2mu\cdots\mkern-2mu ,2{N_1} + {N_2} + {N_3},$
$${W_{\varepsilon l}} = \sum\limits_{n = 1}^{{N_2}} {Y_n^{\left( 1 \right)}\left( {{\omega _{\varepsilon  - 2{N_1}}}} \right)Y_{l - {N_2} - 2{N_1}}^{\left( 6 \right)}\left( {\omega _n^*} \right)}  + \sum\limits_{n = 1}^{{N_3}} {Y_n^{\left( 2 \right)}\left( {{\omega _{\varepsilon  - 2{N_1}}}} \right)Y_{l  - {N_2}- 2{N_1}}^{\left( {10} \right)}\left( {\xi _n^*} \right)} .$$
When $\varepsilon  = 2{N_1} + {N_2} + 1, \mkern-2mu\cdots\mkern-2mu ,2{N_1} + {N_2} + {N_3}$ and $l = 1, \mkern-2mu\cdots\mkern-2mu ,{N_1},$
$${W_{\varepsilon l}} = \sum\limits_{n = 1}^{{N_2}} {Y_n^{\left( 1 \right)}\left( {{\xi _{\varepsilon   - {N_2}- 2{N_1}}}} \right)Y_l^{\left( 4 \right)}\left( {\omega _n^*} \right)}  + \sum\limits_{n = 1}^{{N_3}} {Y_n^{\left( 2 \right)}\left( {{\xi _{\varepsilon - {N_2} - 2{N_1} }}} \right)Y_l^{\left( 7 \right)}\left( {\xi _n^*} \right)} .$$
When $\varepsilon  = 2{N_1} + {N_2} + 1, \mkern-2mu\cdots\mkern-2mu ,2{N_1} + {N_2} + {N_3}$ and $l = {N_1} + 1, \mkern-2mu\cdots\mkern-2mu ,2{N_1},$
$${W_{\varepsilon l}} = \sum\limits_{n = 1}^{{N_2}} {Y_n^{\left( 1 \right)}\left( {{\xi _{\varepsilon  - {N_2}- 2{N_1} }}} \right)Y_{l - {N_1}}^{\left( 3 \right)}\left( {\omega _n^*} \right)}  + \sum\limits_{n = 1}^{{N_3}} {Y_n^{\left( 2 \right)}\left( {{\xi _{\varepsilon - {N_2} - 2{N_1} }}} \right)Y_{l - {N_1}}^{\left( 8 \right)}\left( {\xi _n^*} \right)} .$$
When $\varepsilon  = 2{N_1} + {N_2} + 1, \mkern-2mu\cdots\mkern-2mu ,2{N_1} + {N_2} + {N_3}$ and $l = 2{N_1} + 1, \mkern-2mu\cdots\mkern-2mu ,2{N_1} + {N_2},$
$${W_{\varepsilon l}} = \sum\limits_{n = 1}^{{N_2}} {Y_n^{\left( 1 \right)}\left( {{\xi _{\varepsilon - {N_2} - 2{N_1} }}} \right)Y_{l - 2{N_1}}^{\left( 5 \right)}\left( {\omega _n^*} \right)}  + \sum\limits_{n = 1}^{{N_3}} {Y_n^{\left( 2 \right)}\left( {{\xi _{\varepsilon  - {N_2}- 2{N_1} }}} \right)Y_{l - 2{N_1}}^{\left( 9 \right)}\left( {\xi _n^*} \right)} .$$
When $\varepsilon ,l = 2{N_1} + {N_2} + 1, \mkern-2mu\cdots\mkern-2mu ,2{N_1} + {N_2} + {N_3},$
$${W_{\varepsilon l}} = \sum\limits_{n = 1}^{{N_2}} {Y_n^{\left( 1 \right)}\left( {{\xi _{\varepsilon  - {N_2}- 2{N_1} }}} \right)Y_{l - {N_2}- 2{N_1} }^{\left( 6 \right)}\left( {\omega _n^*} \right)}  + \sum\limits_{n = 1}^{{N_3}} {Y_n^{\left( 2 \right)}\left( {{\xi _{\varepsilon  - {N_2}- 2{N_1} }}} \right)Y_{l - {N_2}- 2{N_1} }^{\left( 10 \right)}\left( {\xi _n^*} \right)} .$$
The symbols $Y_n^{\left( i \right)}{\kern 2pt}(i = 1, \mkern-2mu\cdots\mkern-2mu ,10)$ stand for
$$\begin{array}{l}
Y_n^{\left( 1 \right)}\left( z \right) = \dfrac{{\omega _n^*}}{{{iu_o}}}\dfrac{{{\check{C}_n}}}{{z + \left( {u_o^2/\omega _n^*} \right)}}+\dfrac{{{{\hat C}_n}}}{{z - \omega _n^*}} ,{\kern 5pt}Y_n^{\left( 4 \right)}\left( z \right) = \dfrac{{z_n}}{{i{u_o}}}\dfrac{{{{\bar A}_n}}}{{z + \left( {u_o^2/{z_n}} \right)}},{\kern 5pt}Y_n^{\left( 5 \right)}\left( z \right) = \dfrac{{{C_n}}}{{z - {\omega _n}}},\\
Y_n^{\left( 2 \right)}\left( z \right) = \dfrac{{{{\hat D}_n}}}{{z - \xi _n^*}} - \dfrac{{i\xi _n^*}}{{{u_o}}}\dfrac{{{\check{D}_n}}}{{z + \left( {u_o^2/\xi _n^*} \right)}},{\kern 12pt}Y_n^{\left( 8 \right)}\left( z \right) = \dfrac{{z_n^*}}{{i{u_o}}}\dfrac{{{\check{A}_n}}}{{z + \left( {u_o^2/z_n^*} \right)}},{\kern 5pt}Y_n^{\left(9\right)}\left( z \right) = \dfrac{{{{\bar C}_n}}}{{z + \left( {u_o^2/{\omega _n}} \right)}},\\
Y_n^{\left( 3 \right)}\left( z \right) = \dfrac{{{{\hat A}_n}}}{{z - z_n^*}},{\kern 10pt}Y_n^{\left( 6 \right)}\left( z \right) = \dfrac{{{{\bar D}_n}}}{{z + \left( {u_o^2/{\xi _n}} \right)}},{\kern 10pt}Y_n^{\left( 7 \right)}\left( z \right) = \dfrac{{{A_n}}}{{z - {z_n}}},{\kern 10pt}Y_n^{\left( {10} \right)}\left( z \right) = \dfrac{{{D_n}}}{{z - {\xi _n}}}.
\end{array}$$
\end{theorem}

\noindent \textbf{Proof.} Assuming the reflectionless case, using Proposition \ref{prop:11}, we obtain
$$m_{\left( {k + 1} \right)2}^ + \left( {{\omega _{i'}}} \right) =  \sum\limits_{n = 1}^{{N_2}} {Y_n^{\left( 1 \right)}\left( {{\omega _{i'}}} \right)m_{\left( {k + 1} \right)3}^ - \left( {\omega _n^*} \right)}  + \sum\limits_{n = 1}^{{N_3}} {Y_n^{\left( 2 \right)}\left( {{\omega _{i'}}} \right)m_{\left( {k + 1} \right)1}^ - \left( {\xi _n^*} \right)}+{\left( { - 1} \right)^{k + 1}}\frac{{u_{ + ,\bar k}^*}}{{{u_o}}} ,$$
$$m_{\left( {k + 1} \right)2}^ + \left( {{\xi _{l'}}} \right) = \sum\limits_{n = 1}^{{N_2}} {Y_n^{\left( 1 \right)}\left( {{\xi _{l'}}} \right)m_{\left( {k + 1} \right)3}^ - \left( {\omega _n^*} \right)}  + \sum\limits_{n = 1}^{{N_3}} {Y_n^{\left( 2 \right)}\left( {{\xi _{l'}}} \right)m_{\left( {k + 1} \right)1}^ - \left( {\xi _n^*} \right)}+{\left( { - 1} \right)^{k + 1}}\frac{{u_{ + ,\bar k}^*}}{{{u_o}}},$$
$$m_{\left( {k + 1} \right)3}^ - \left( {\omega _{i'}^*} \right) =\sum\limits_{n = 1}^{{N_1}} {\left[ {Y_n^{\left( 4 \right)}\left( {\omega _{i'}^*} \right)m_{\left( {k + 1} \right)3}^ + \left( {{z_n}} \right)+Y_n^{\left( 3 \right)}\left( {\omega _{i'}^*} \right)m_{\left( {k + 1} \right)1}^ - \left( {z_n^*} \right)} \right]} {\kern 1pt}  + \sum\limits_{n = 1}^{{N_2}} {Y_n^{\left( 5 \right)}\left( {\omega _{i'}^*} \right)m_{\left( {k + 1} \right)2}^ + \left( {{\omega _n}} \right)}  + \sum\limits_{n = 1}^{{N_3}} {Y_n^{\left( 6 \right)}\left( {\omega _{i'}^*} \right)m_{\left( {k + 1} \right)2}^ + \left( {{\xi _n}} \right)}+\frac{{{u_{ + ,k}}}}{{{u_o}}},$$
$$m_{\left( {k + 1} \right)1}^ - \left( {z_{j'}^*} \right) = \sum\limits_{n = 1}^{{N_1}} {\left[ {Y_n^{\left( 7 \right)}\left( {z_{j'}^*} \right)m_{\left( {k + 1} \right)3}^ + \left( {{z_n}} \right) + Y_n^{\left( 8 \right)}\left( {z_{j'}^*} \right)m_{\left( {k + 1} \right)1}^ - \left( {z_n^*} \right)} \right]}  + \sum\limits_{n = 1}^{{N_2}} {Y_n^{\left( 9 \right)}\left( {z_{j'}^*} \right)m_{\left( {k + 1} \right)2}^ + \left( {{\omega _n}} \right)}  + \sum\limits_{n = 1}^{{N_3}} {Y_n^{\left( {10} \right)}\left( {z_{j'}^*} \right)m_{\left( {k + 1} \right)2}^ + \left( {{\xi _n}} \right)}+\frac{{i{u_{ + ,k}}}}{{z_{j'}^*}},$$
$$m_{\left( {k + 1} \right)1}^ - \left( {\xi _{l'}^*} \right) = \sum\limits_{n = 1}^{{N_1}} {\left[ {Y_n^{\left( 7 \right)}\left( {\xi _{l'}^*} \right)m_{\left( {k + 1} \right)3}^ + \left( {{z_n}} \right) + Y_n^{\left( 8 \right)}\left( {\xi _{l'}^*} \right)m_{\left( {k + 1} \right)1}^ - \left( {z_n^*} \right)} \right]}  + \sum\limits_{n = 1}^{{N_2}} {Y_n^{\left( 9 \right)}\left( {\xi _{l'}^*} \right)m_{\left( {k + 1} \right)2}^ + \left( {{\omega _n}} \right)}  + \sum\limits_{n = 1}^{{N_3}} {Y_n^{\left( {10} \right)}\left( {\xi _{l'}^*} \right)m_{\left( {k + 1} \right)2}^ + \left( {{\xi _n}} \right)}+\frac{{i{u_{ + ,k}}}}{{\xi _{l'}^*}},$$
$$m_{\left( {k + 1} \right)3}^ + \left( {{z_{j'}}} \right) = \sum\limits_{n = 1}^{{N_1}} {\left[ { Y_n^{\left( 4 \right)}\left( {{z_{j'}}} \right)m_{\left( {k + 1} \right)3}^ + \left( {{z_n}} \right)+Y_n^{\left( 3 \right)}\left( {{z_{j'}}} \right)m_{\left( {k + 1} \right)1}^ - \left( {z_n^*} \right) } \right]}  + \sum\limits_{n = 1}^{{N_2}} {Y_n^{\left( 5 \right)}\left( {{z_{j'}}} \right)m_{\left( {k + 1} \right)2}^ + \left( {{\omega _n}} \right)}  + \sum\limits_{n = 1}^{{N_3}} {Y_n^{\left( 6 \right)}\left( {{z_{j'}}} \right)m_{\left( {k + 1} \right)2}^ + \left( {{\xi _n}} \right)}+\frac{{{u_{ + ,k}}}}{{{u_o}}}.$$

From the above results, at $z = {\omega _{i'}}$ and $z = {\xi _{l'}}$, we derive the following equations:

\begin{equation}\label{eq128}
\begin{array}{l}
m_{\left( {k + 1} \right)2}^ + \left( z \right) =  i{u_{ + ,k}}\mathop{\displaystyle\sum}\limits_{n = 1}^{{N_3}} {\dfrac{{Y_n^{\left( 2 \right)}\left( z \right)}}{{\xi _n^*}}}  + \mathop{\displaystyle\sum}\limits_{n = 1}^{{N_2}} {\mathop{\displaystyle\sum}\limits_{n' = 1}^{{N_3}} {Y_n^{\left( 1 \right)}\left( z \right)Y_{n'}^{\left( 6 \right)}\left( {\omega _n^*} \right)m_{\left( {k + 1} \right)2}^ + \left( {{\xi _{n'}}} \right)} }+{\left( { - 1} \right)^{k + 1}}\dfrac{{u_{ + ,\bar k}^*}}{{{u_o}}} + \dfrac{{{u_{ + ,k}}}}{{{u_o}}}\mathop{\displaystyle\sum}\limits_{n = 1}^{{N_2}} {Y_n^{\left( 1 \right)}\left( z \right)}\\
{\kern 40pt}  + \mathop{\displaystyle\sum}\limits_{n = 1}^{{N_3}} {\mathop{\displaystyle\sum}\limits_{n' = 1}^{{N_3}} {Y_n^{\left( 2 \right)}\left( z \right)Y_{n'}^{\left( {10} \right)}\left( {\xi _n^*} \right)m_{\left( {k + 1} \right)2}^ + \left( {{\xi _{n'}}} \right)} }+ \mathop{\displaystyle\sum}\limits_{n = 1}^{{N_2}} {\mathop{\displaystyle\sum}\limits_{n' = 1}^{{N_1}} {Y_n^{\left( 1 \right)}\left( z \right)\left[ { Y_{n'}^{\left( 4 \right)}\left( {\omega _n^*} \right)m_{\left( {k + 1} \right)3}^ + \left( {{z_{n'}}} \right)+Y_{n'}^{\left( 3 \right)}\left( {\omega_n^*} \right)m_{\left( {k + 1} \right)1}^ - \left( {z_{n'}^*} \right) } \right]} }  \\
{\kern 40pt} + \mathop{\displaystyle\sum}\limits_{n = 1}^{{N_3}} {\mathop{\displaystyle\sum}\limits_{n' = 1}^{{N_2}} {Y_n^{\left( 2 \right)}\left( z \right)Y_{n'}^{\left( 9 \right)}\left( {\xi _n^*} \right)m_{\left( {k + 1} \right)2}^ + \left( {{\omega _{n'}}} \right)} }+ \mathop{\displaystyle\sum}\limits_{n = 1}^{{N_3}} {\mathop{\displaystyle\sum}\limits_{n' = 1}^{{N_1}} {Y_n^{\left( 2 \right)}\left( z \right)\left[ {Y_{n'}^{\left( 8 \right)}\left( {\xi _n^*} \right)m_{\left( {k + 1} \right)1}^ - \left( {z_{n'}^*} \right)+Y_{n'}^{\left( 7 \right)}\left( {\xi _n^*} \right)m_{\left( {k + 1} \right)3}^ + \left( {{z_{n'}}} \right)  } \right]} }   \\
{\kern 40pt} + \mathop{\displaystyle\sum}\limits_{n = 1}^{{N_2}} {\mathop{\displaystyle\sum}\limits_{n' = 1}^{{N_2}} {Y_n^{\left( 1 \right)}\left( z \right)Y_{n'}^{\left( 5 \right)}\left( {\omega _n^*} \right)m_{\left( {k + 1} \right)2}^ + \left( {{\omega _{n'}}} \right)} } .
\end{array}
\end{equation}

Let ${\boldsymbol{J}_k} = {\left( {{J_{k1}}, \cdots ,{J_{k\left( {2{N_1} + {N_2} + {N_3}} \right)}}} \right)^T}$, where
$${J_{kl}} = \left\{ \begin{array}{l}
m_{\left( {k + 1} \right)3}^ + \left( {{z_l}} \right),{\kern 28pt}l = 1, \mkern-2mu\cdots\mkern-2mu ,{N_1},\\
m_{\left( {k + 1} \right)1}^ - \left( {z_{l - {N_1}}^*} \right),{\kern 16pt}l = {N_1} + 1, \mkern-2mu\cdots\mkern-2mu ,2{N_1},\\
m_{\left( {k + 1} \right)2}^ + \left( {{\omega _{l - 2{N_1}}}} \right),{\kern 10pt}l = 2{N_1} + 1, \mkern-2mu\cdots\mkern-2mu ,2{N_1} + {N_2},\\
m_{\left( {k + 1} \right)2}^ + \left( {{\xi _{l - {N_2}- 2{N_1} }}} \right),l = 2{N_1} + {N_2} + 1, \mkern-2mu\cdots\mkern-2mu ,2{N_1} + {N_2} + {N_3}.
\end{array} \right.$$
Then Eqs.(\ref{eq128}) are equivalent to $\left( {\boldsymbol{I} - \boldsymbol{W}} \right){\boldsymbol{J}_k} = {\boldsymbol{\Theta} _k}$. By solving this equation using Cramer's rule and noting that $\boldsymbol{I} - \boldsymbol{W}=\boldsymbol{Q}$, we get
$${J_{kl}} = \frac{{\det \boldsymbol{Q}_{kl}^{aug}}}{{\det {\boldsymbol{Q}}}},{\kern 5pt}k = 1,2,{\kern 5pt}l = 1, \mkern-2mu\cdots\mkern-2mu ,2{N_1} + {N_2} + {N_3},$$
where $\boldsymbol{Q}_{kl}^{aug} = \left( {{\boldsymbol{Q}_1}, \cdots ,{\boldsymbol{Q}_{l - 1}},{\boldsymbol{\Theta} _k},{\boldsymbol{Q}_{l + 1}}, \cdots ,{\boldsymbol{Q}_{2{N_1} + {N_2} + {N_3}}}} \right).$ Bringing the above results into Theorem \ref{thm:2} and taking into account Eqs.(\ref{eq102}), we can obtain Eqs.(\ref{eq100}).   $\Box$

Under the assumption of reflectionless, Eqs.(\ref{eq98}) degenerate into
$$\begin{array}{l}
{s_{11}}\left( z \right) = \mathop{\displaystyle \prod} \limits_{n = 1}^{{N_1}} \dfrac{{z + \left( {u_o^2/z_n^*} \right)}}{{z + \left( {u_o^2/{z_n}} \right)}}\dfrac{{z - {z_n}}}{{z - z_n^*}}\mathop{\displaystyle\prod} \limits_{n = 1}^{{N_2}} \dfrac{{z + \left( {u_o^2/{\omega _n}} \right)}}{{z + \left( {u_o^2/\omega _n^*} \right)}}\mathop{\displaystyle \prod} \limits_{n = 1}^{{N_3}} \dfrac{{z - {\xi _n}}}{{z - \xi _n^*}},\\
{b_{22}}\left( z \right) = {e^{ - i\Delta h}}\mathop{\displaystyle \prod} \limits_{n = 1}^{{N_2}} \dfrac{{z - {\omega _n}}}{{z - \omega _n^*}}\dfrac{{z + \left( {u_o^2/{\omega _n}} \right)}}{{z + \left( {u_o^2/\omega _n^*} \right)}}\mathop{\displaystyle\prod} \limits_{n = 1}^{{N_3}} \dfrac{{z - {\xi _n}}}{{z - \xi _n^*}}\dfrac{{z + \left( {u_o^2/{\xi _n}} \right)}}{{z + \left( {u_o^2/\xi _n^*} \right)}},
\end{array}$$
where ${N_i}{\kern 2pt}(i = 1, \mkern-2mu\cdots\mkern-2mu ,3)$ indicate the quantity of the three types of eigenvalues, respectively. Distinct types of eigenvalues and their combinations lead to various soliton solutions, as discussed in detail below.

\subsection{Diversity of soliton solutions.}

In this subsection, we use Theorem \ref{thm:1} to construct different types of soliton solutions for the focusing two-component Hirota equation (\ref{eq2}) with NZBCs. To facilitate the construction of normalization constants and eigenvalues, we define the parameters ${q_i},{g_i}, \in \mathbb{R}$, ${\delta _i} > \left| {{u_o}} \right|$ and $0 < {\kappa _i} < 1$, for $i = 1, \mkern-2mu\cdots\mkern-2mu ,6$.

First, we consider the existence of eigenvalues of the first type alone. By setting ${N_1} = 1$, ${N_2} = {N_3} = 0$, and denoting the normalization constants as ${a_1} = {e^{{q_1} + i{g_1}}},{z_1} = {\delta _1}{e^{i{\kappa _1}\pi }},$ applying Eqs.({\ref{eq100}}) to construct one breather-breather soliton solution for the two components $u_1$ and $u_2$. This breather soliton consists of two downward valleys and one upward peak, as shown in Fig.\ref{fig:1}: (a) and (b).

Next, by setting ${N_2} = 1,{N_1} = {N_3} = 0$, and defining ${c_1} = {e^{{q_2} + i{g_2}}},{\omega _1} = {\delta _2}{e^{i{\kappa _2}\pi }},$ we focus solely on the eigenvalues of the second type. In this scenario, three distinct solitons arise: ($i$). When ${u_{+,1}}{u_{+,2}} = 0$ and ${u_{+,k}} \in \mathbb{R}$ for $k = 1,2$, one dark-bright soliton solution is obtained, as shown in Fig.\ref{fig:1}: (c) and (d). ($ii$). If ${u_{+,1}}{u_{+,2}} = 0$ and ${u_{+,k}} \in \mathbb{C}$, one bright-kink soliton solution emerges, as depicted in Fig.\ref{fig:1}: (e) and (f). ($iii$). Finally, when ${u_{+,1}}{u_{+,2}} \ne 0$, one breather-breather soliton solution is generated, similar to the one in Fig.\ref{fig:1}: (a) and (b), though its corresponding images are omitted here.

Further, when setting ${N_3} = 1,{N_1} = {N_2} = 0$, and choosing ${d_1} = {e^{{q_3} + i{g_3}}},{\xi _1} = {\delta _3}{e^{i{\kappa _3}\pi }}$, two distinct solitons emerge: ($i$). If ${u_{+,1}}{u_{+,2}} = 0$, the solution manifests as one dark-bright soliton, similar to Fig.\ref{fig:1}: (c) and (d). ($ii$). When ${u_{+,1}}{u_{+,2}} \ne 0$, it resulting solution exhibits one $W$-$M$ shaped soliton, as depicted in Fig.\ref{fig:1}: (g) and (h).

\begin{figure}[!htbp]\label{f1}
\centering
\subfigure[]{\includegraphics[height=1.3in,width=1.69in]{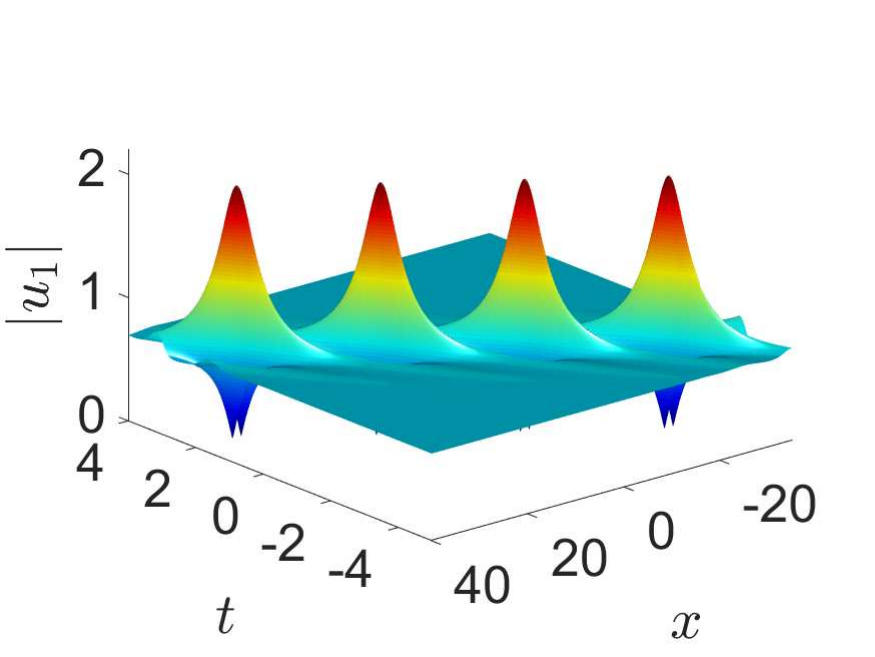}}\hspace{0.1cm}
\subfigure[]{\includegraphics[height=1.3in,width=1.69in]{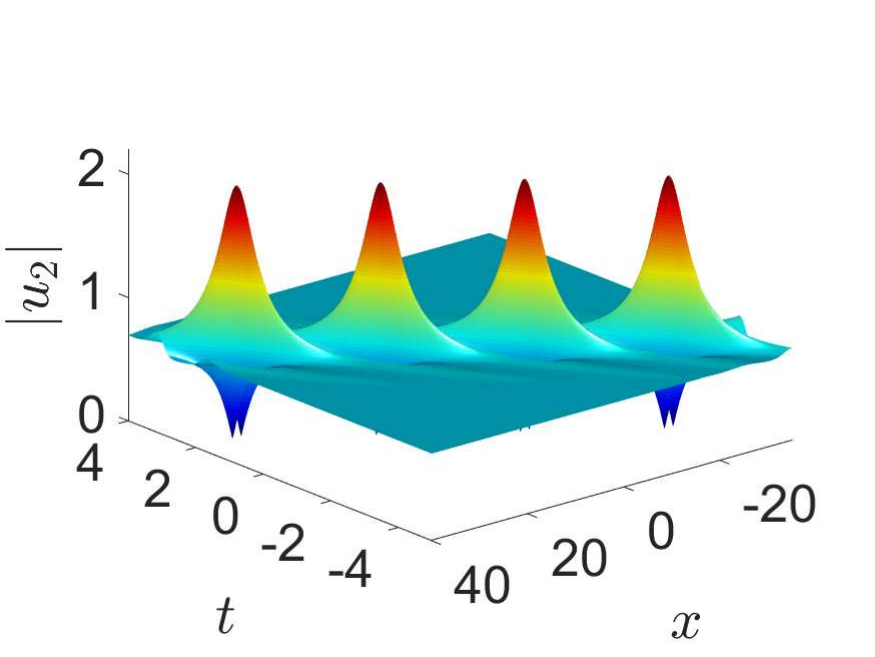}}\hspace{0.1cm}
\subfigure[]{\includegraphics[height=1.3in,width=1.69in]{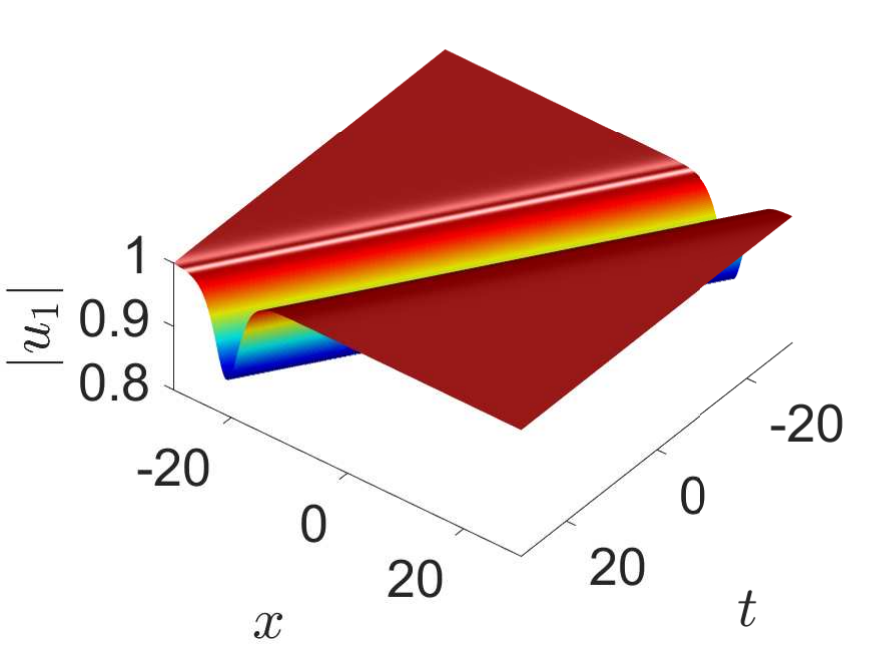}}\hspace{0.1cm}
\subfigure[]{\includegraphics[height=1.3in,width=1.69in]{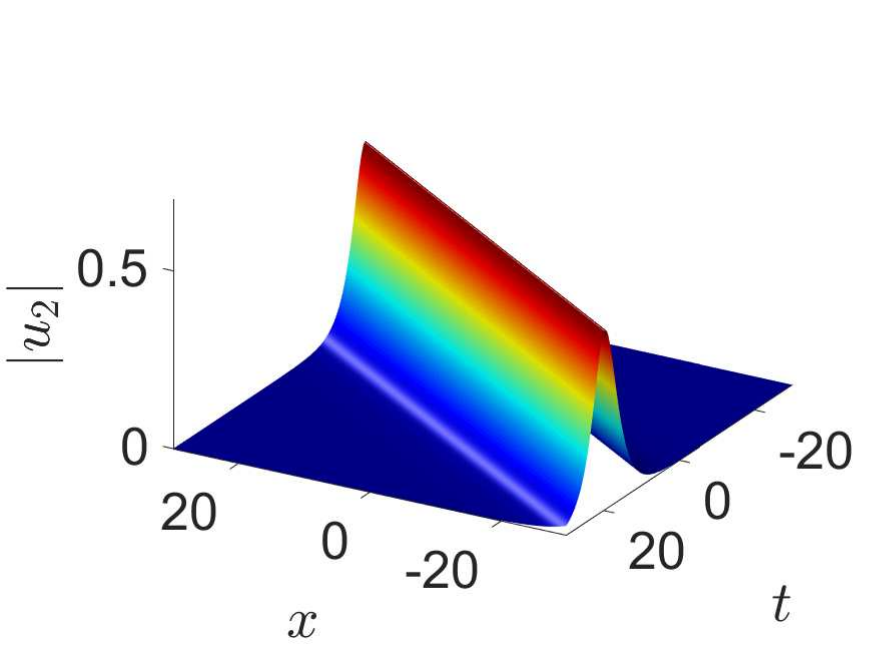}}\hspace{0.1cm}
\subfigure[]{\includegraphics[height=1.3in,width=1.69in]{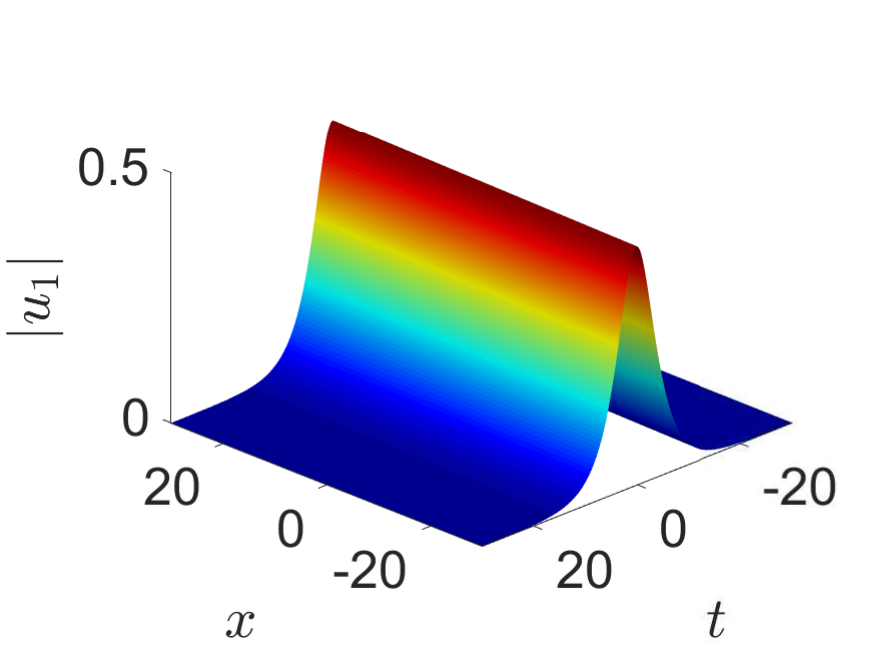}}\hspace{0.1cm}
\subfigure[]{\includegraphics[height=1.3in,width=1.69in]{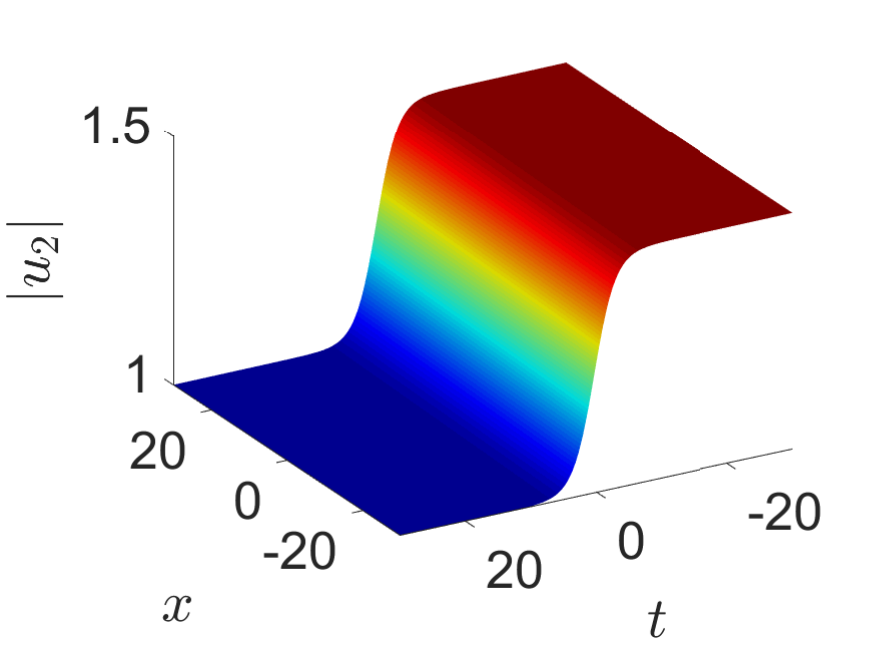}}\hspace{0.1cm}
\subfigure[]{\includegraphics[height=1.3in,width=1.69in]{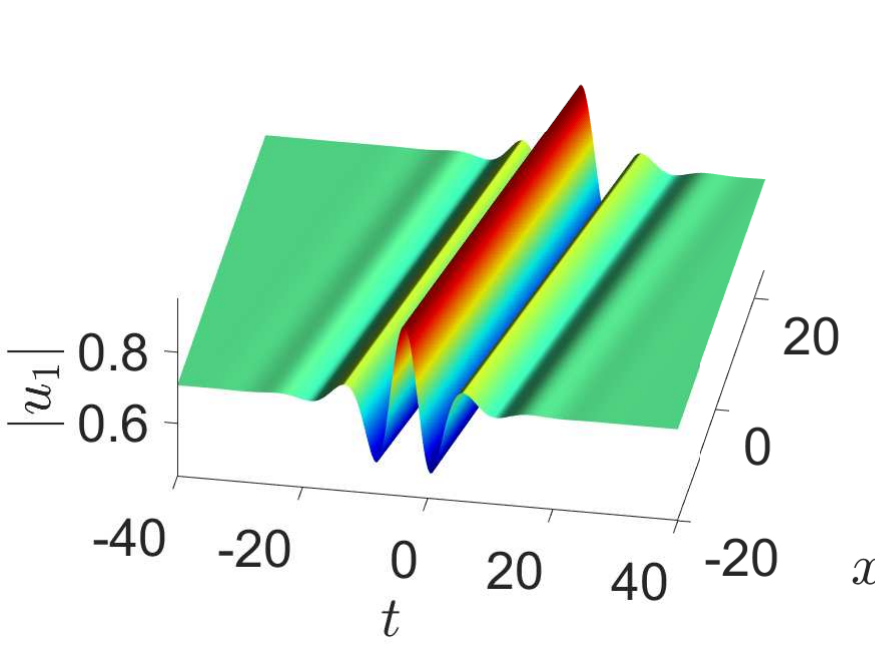}}\hspace{0.1cm}
\subfigure[]{\includegraphics[height=1.3in,width=1.69in]{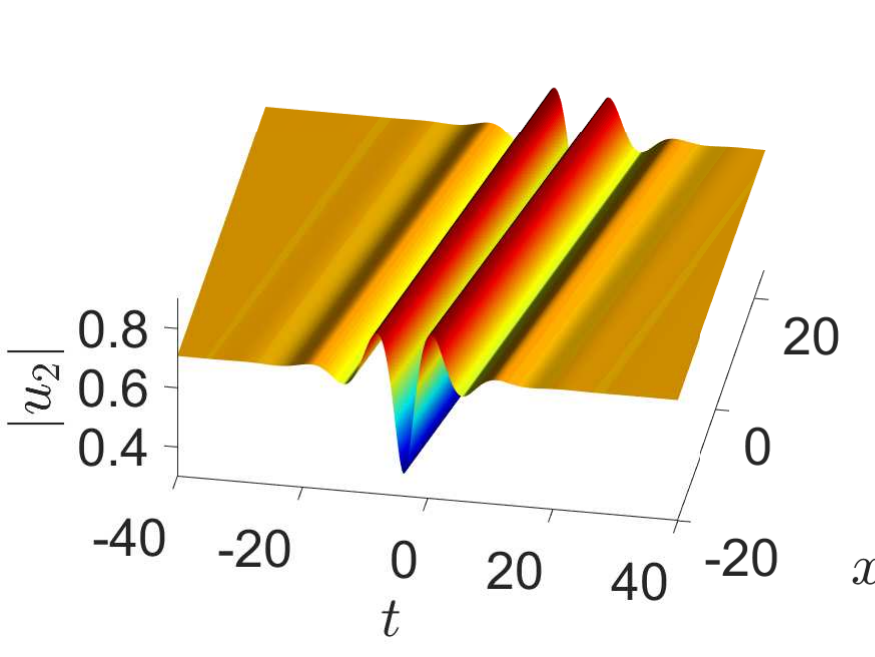}}\hspace{0.1cm}
\caption{(a) and (b): One breather-breather soliton  solution with parameters are chosen as ${N_1} = 1$, ${q_1} = 2$, ${g_1} = 1$, ${\delta_1} = 1.1$, ${\kappa_1} = 0.4$, $\boldsymbol{u}_+ = \left( {\sqrt 2}/{2}, -{\sqrt 2}/2 \right)^T$, ${\alpha _1} = {\alpha _2} = 1$. (c) and (d): One dark-bright soliton solution with parameters are chosen as ${N_2} = 1$, ${q_2} = 0$, ${g_2} = 5$, ${\delta_2} = 1.1$, ${\kappa_2} = \frac{1}{8}$, ${\boldsymbol{u}_ + } = \left( {1,0} \right)^T$, ${\alpha _1} = {\alpha _2} = 1$. (e) and (f): One bright-kink soliton solution with parameters are chosen as ${N_2} = 1$, ${q_2} = 0$, ${g_2} = 5$, ${\delta_2} = 1.1$, ${\kappa_2} = \frac{1}{8}$, ${\boldsymbol{u}_ + } = \left( {0,i} \right)^T$, ${\alpha _1} = {\alpha _2} = 1$. (g) and (h): One $W$-$M$ shaped soliton solution with parameters are chosen as ${N_3} = 1$, ${q_3} = 1$, ${g_3} = 5$, ${\delta_3} = 1.2$, ${\kappa_3} = \frac{1}{8}$, ${\boldsymbol{u}_ + } = {\left( {\sqrt 2 /2, - \sqrt 2 /2} \right)^T}$, ${\alpha _1} = {\alpha _2} = 1$.}
\label{fig:1}
\end{figure}

Next, we analyze the scenarios where eigenvalues appear in pairs. Specifically, let ${N_1} = 2,{N_2} = {N_3} = 0$, and define $a_1 = {e^{{q_1} + i{g_1}}},{a_2} = {e^{{q_4} + i{g_4}}},{z_1} = {\delta _1}{e^{i{\kappa _1}\pi }}$ and ${z_2} = {\delta _4}{e^{i{\kappa _4}\pi }}$. Under these conditions, the solution corresponds to two-breather solitons, where each breather wave exhibits two downward valleys and one upward peak, as depicted in Fig.\ref{fig:2}: (a) and (b).

With ${N_3} = 2,{N_1} = {N_2} = 0$, the associated normalization constants are ${d_1} = {e^{{q_3} + i{g_3}}}$ and ${d_2} = {e^{{q_5} + i{g_5}}}$, while the eigenvalues are specified as ${\xi _1} = {\delta _3}{e^{i{\kappa _3}\pi }}$ and ${\xi _2} = {\delta _5}{e^{i{\kappa _5}\pi }}$. Under these conditions, two distinct types of solitons can be identified. ($i$). For ${u_{+,1}}{u_{+,2}} = 0$, the solution comprise two bright-dark solitons. ($ii$). Alternatively, when ${u_{+,1}}{u_{+,2}} \ne 0$, the configuration results in one dark-bright and one breather-breather solitons. The corresponding visual representations are provided in Fig.\ref{fig:2}: (c)-(d) and (e)-(f), respectively.

Considering a pair of second class eigenvalues, let ${N_2} = 2,{N_1} = {N_3} = 0$, with the normalization constants ${c_1} = {e^{{q_2} + i{g_2}}},{c_2} = {e^{{q_6} + i{g_6}}}$ and eigenvalues ${\omega _1} = {\delta _2}{e^{i{\kappa _2}\pi }},{\omega _2} = {\delta _6}{e^{i{\kappa _6}\pi }}$. Distinct from previous scenarios, when ${u_{+,1}}{u_{+,2}} \ne 0$, the solution yields two breather-breather solitons illustrated in Fig.\ref{fig:2}: (g) and (h). The figures reveal that the two breather solitons exhibit contrasting structures: for $u_1$, each breather wave consists of two upward peaks and one downward valley, whereas for $u_2$, the wave is characterized by two downward valleys and one upward peak.  Conversely, when ${u_{+,1}}{u_{+,2}} = 0$, the solution reduces to two bright-dark solitons, similar to Fig.\ref{fig:2}: (c) and (d).

\begin{figure}[!htbp]\label{f2}
\centering
\subfigure[]{\includegraphics[height=1.3in,width=1.69in]{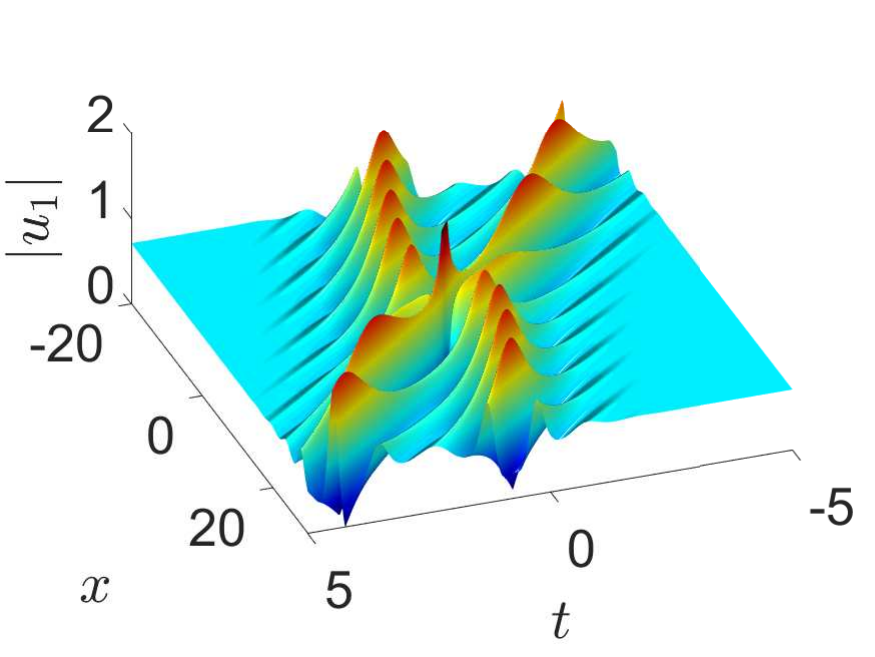}}\hspace{0.1cm}
\subfigure[]{\includegraphics[height=1.3in,width=1.69in]{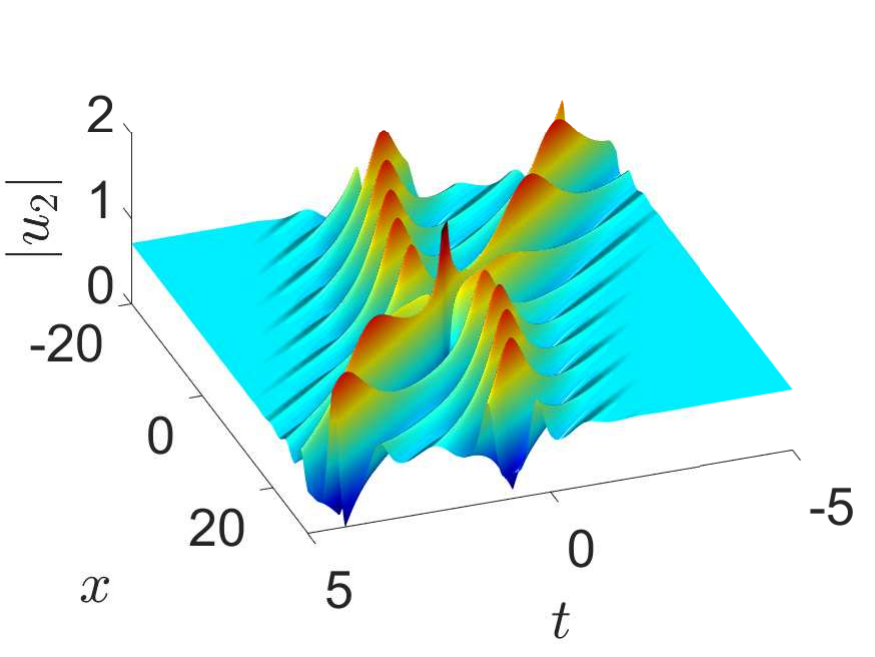}}\hspace{0.1cm}
\subfigure[]{\includegraphics[height=1.3in,width=1.69in]{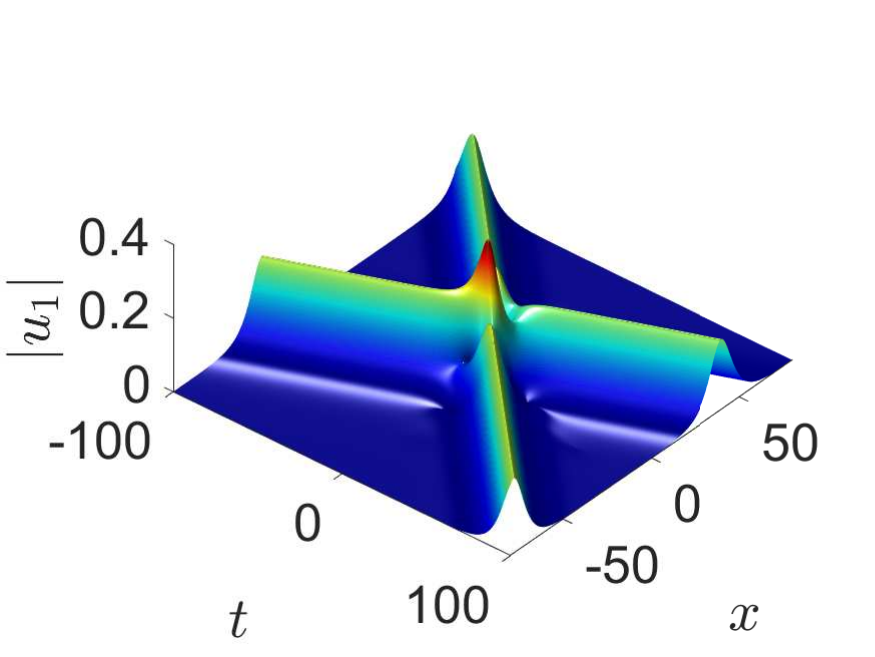}}\hspace{0.1cm}
\subfigure[]{\includegraphics[height=1.3in,width=1.69in]{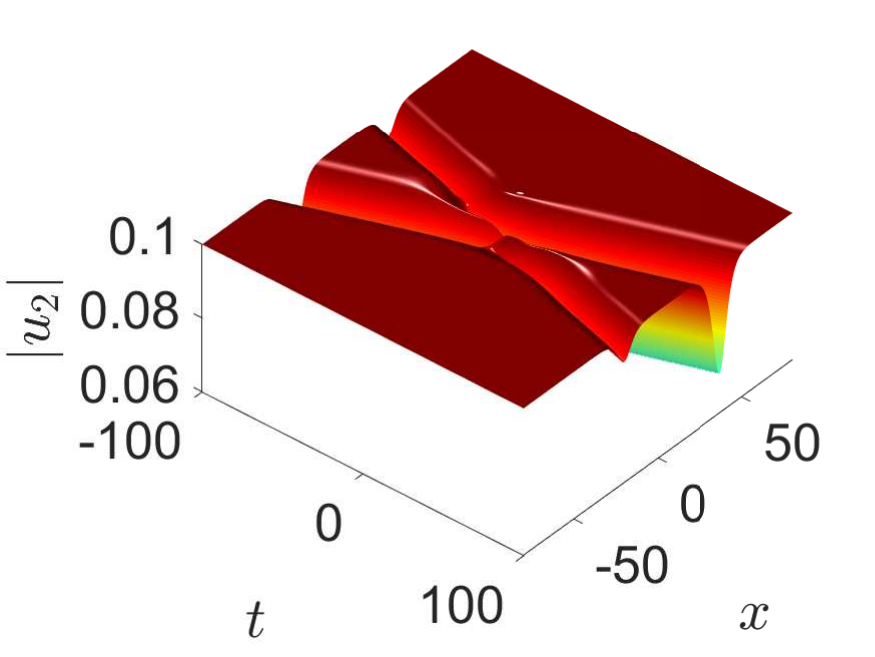}}\hspace{0.1cm}
\subfigure[]{\includegraphics[height=1.3in,width=1.69in]{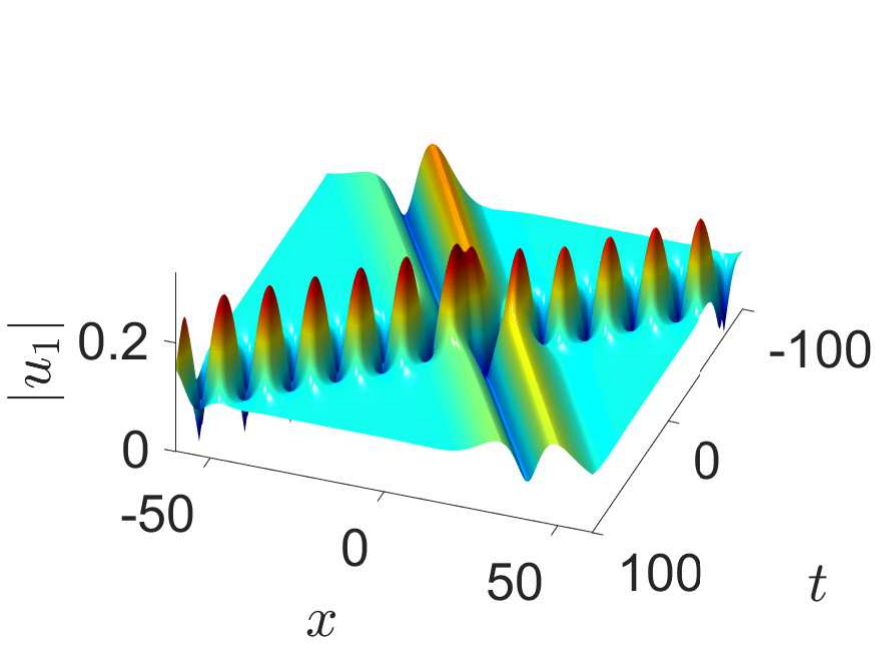}}\hspace{0.1cm}
\subfigure[]{\includegraphics[height=1.3in,width=1.69in]{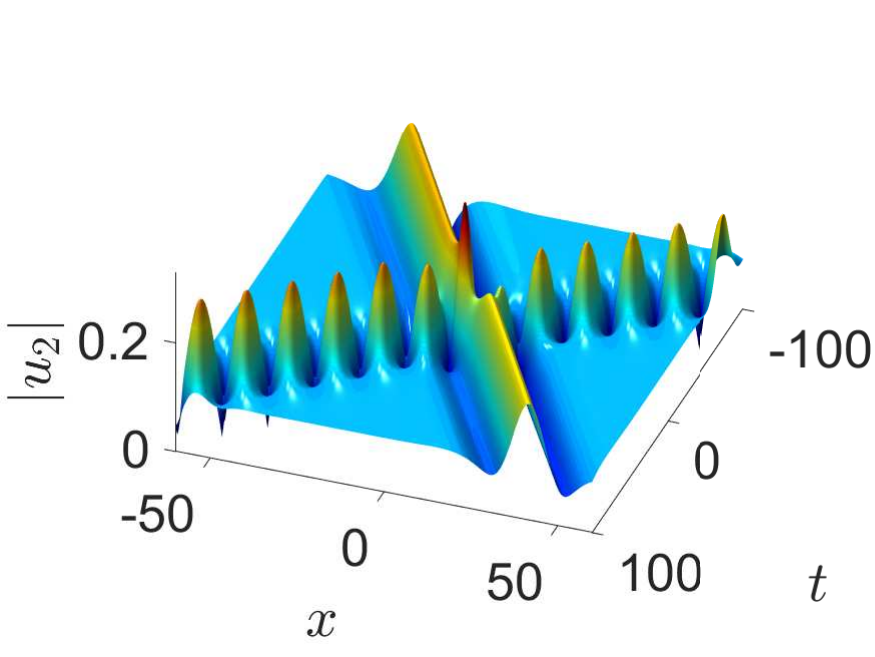}}\hspace{0.1cm}
\subfigure[]{\includegraphics[height=1.3in,width=1.69in]{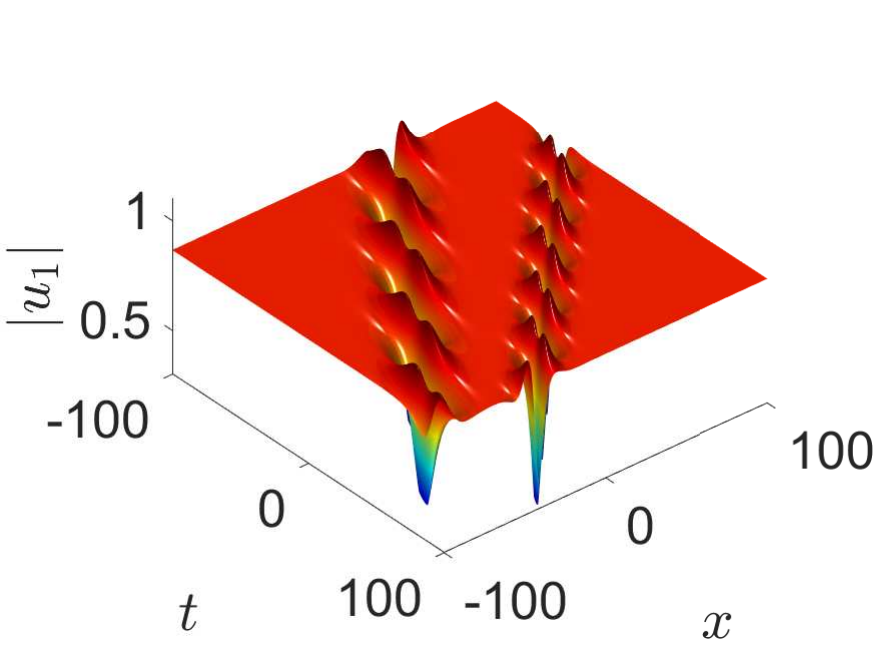}}\hspace{0.1cm}
\subfigure[]{\includegraphics[height=1.3in,width=1.69in]{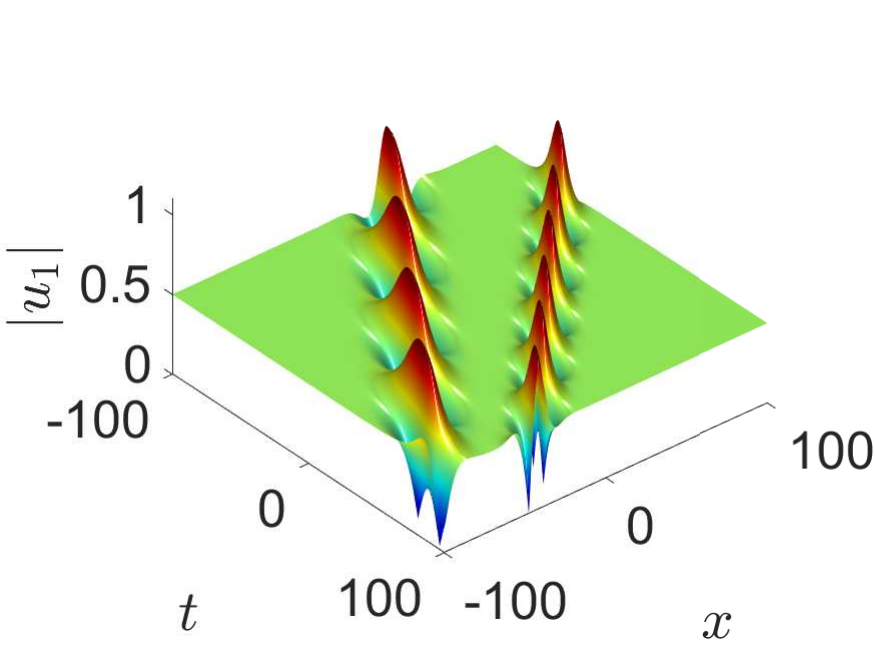}}\hspace{0.1cm}
\caption{(a) and (b): Two breather-breather solitons solution with parameters are chosen as ${N_1} = 2$, ${q_1} = 0$, ${g_1}={q_4}={g_4}=1$, ${\delta_1}=1.1$, ${\delta_2}=1.2$, ${\kappa_1}={\kappa_2}=\frac{1}{4}$, ${\alpha _1} ={\alpha_2}= 1$, ${{\boldsymbol{u}}_ + } = {\left( {\sqrt 2 /2, - \sqrt 2 /2} \right)^T}$. (c) and (d): Two bright-dark solitons solution with parameters are chosen as ${N_3} = 2$, ${q_3} = {g_3} = {q_5} = {g_{5}} = 1$, ${\delta_3} =1.5$, ${\delta_5} = \frac{5}{4}$, ${\kappa_3} = \frac{1}{8}$, ${\kappa_{5}} = \frac{1}{4}$, ${\alpha _1} = 1$, ${\alpha _2} = 2$, ${\boldsymbol{u}_ + } = {\left( {0,1} \right)^T}$. (e) and (f): One bright-bright and one breather-breather solitons solution with parameters are chosen as ${N_3} = 2$, ${q_3} = {g_3} = {q_5} = {g_{5}} = 1$, ${\delta_3} =1.5$, ${\delta_5} = \frac{5}{4}$, ${\kappa_3} = \frac{1}{8}$, ${\kappa_{5}} = \frac{1}{4}$, ${\alpha _1} = 1$, ${\alpha _2} = 3$, ${\boldsymbol{u}_ + } =  {\left( {\sqrt 2 /2, - \sqrt 2 /2} \right)^T}$. (g) and (h): Two breather-breather solitons solution with parameters are chosen as ${N_2} = 2$, ${q_2} = 10$, ${g_2} = 1$, ${q_{6}} =  - 5$, ${g_{6}} = 1$, ${\delta_2} = 3$, ${\delta_6} =4$, $\kappa_2=\kappa_6 = \frac{1}{4}$, ${\alpha _1} = 2$, ${\alpha _2} = {10^{ - 3}}$, ${\boldsymbol{u}_ + } = {\left( {\sqrt 3 /2,1/2} \right)^T}$.  }
\label{fig:2}
\end{figure}

Finally, after analyzing the cases where each type of eigenvalue exists independently, we now turn to scenarios where the three types of eigenvalues are mixed. First, consider the coexistence of the first and second types,where ${N_1} = {N_2} = 1,{N_3} = 0$, with normalization constants ${a_1} = {e^{{q_1} + i{g_1}}},{c_1} = {e^{{q_2} + i{g_2}}}$, and eigenvalues ${z_1} = {\delta _1}{e^{i{\kappa _1}\pi }},{\omega _1} = {\delta _2}{e^{i{\kappa _2}\pi }}$. This combination yields two distinct types of solitons. ($i$). When ${u_{ + ,1}}=1$ and ${u_{ + ,2}}= 0$, the solution represents an interaction of a breather and a dark solitons in $u_1$, while $u_2$ corresponds to a bright soliton, as depicted in Fig.\ref{fig:3}: (a) and (b). ($ii$). When ${u_{+,1}}{u_{+,2}} \ne 0$, the solution exhibits two breather-breather solitons, as shown in Fig.\ref{fig:3}: (c) and (d).

Moreover, when the first and third types of eigenvalues coexist, we set ${N_1} = {N_3} = 1,{N_2} = 0,$ with the normalization constants ${a_1} = {e^{{q_1} + i{g_1}}},{d_1} = {e^{{q_3} + i{g_3}}}$ and eigenvalues ${z_1} = {\delta _1}{e^{i{\kappa _1}\pi }},{\xi _1} = {\delta _3}{e^{i{\kappa _3}\pi }}$. This configuration results in one breather-breather and $M$-$W$ shaped solitons, as illustrated in Fig.\ref{fig:3}: (e) and (f). Finally, for the scenario where the second and third types of eigenvalues appear together, we take ${N_2} = {N_3} = 1,{N_1} = 0$, with ${c_1} = {e^{{q_2} + i{g_2}}},{d_1} = {e^{{q_3} + i{g_3}}}$, and eigenvalues ${\omega _1} = {\delta _2}{e^{i{\kappa _2}\pi }},{\xi _1} = {\delta _3}{e^{i{\kappa _3}\pi }}$. This configuration yields one bright-bright and one dark-bright solitons. It is worth noting that varying $N_1$, $N_2$ and $N_3$ and exploring different combinations can generate an even wider variety of soliton solutions.

\begin{figure}[!htbp]\label{f3}
\centering
\subfigure[]{\includegraphics[height=1.3in,width=1.69in]{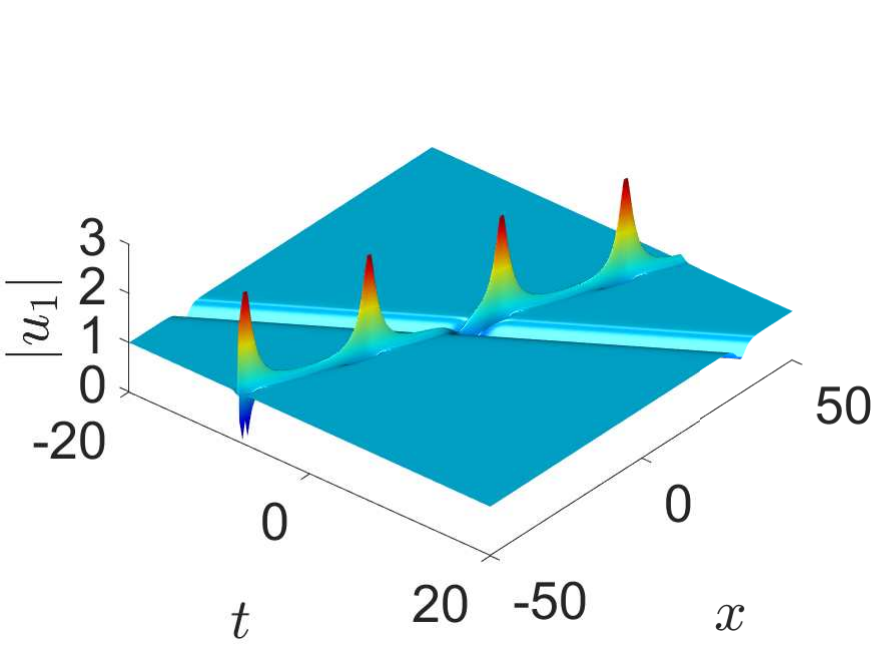}}\hspace{0.1cm}
\subfigure[]{\includegraphics[height=1.3in,width=1.69in]{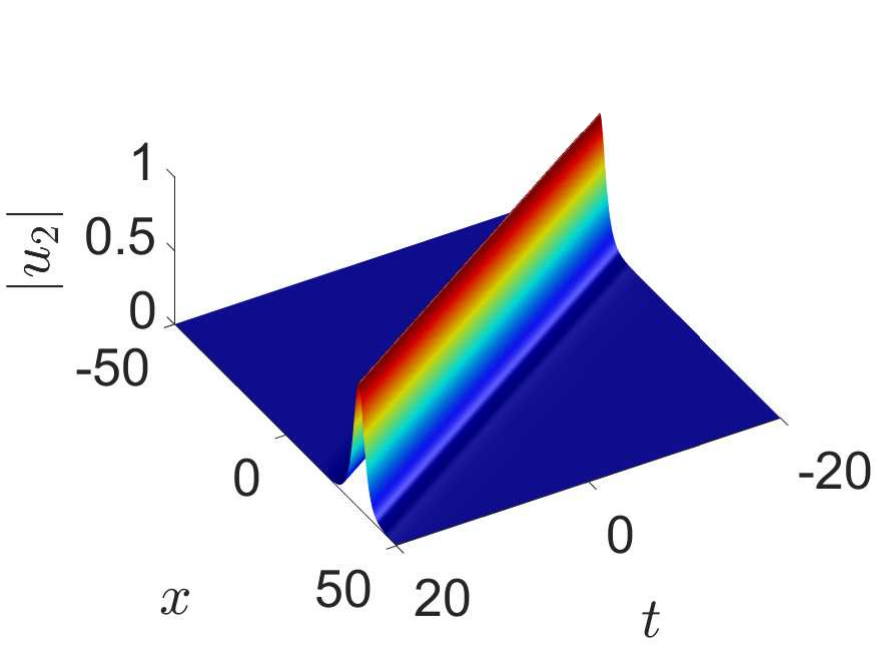}}\hspace{0.1cm}
\subfigure[]{\includegraphics[height=1.3in,width=1.69in]{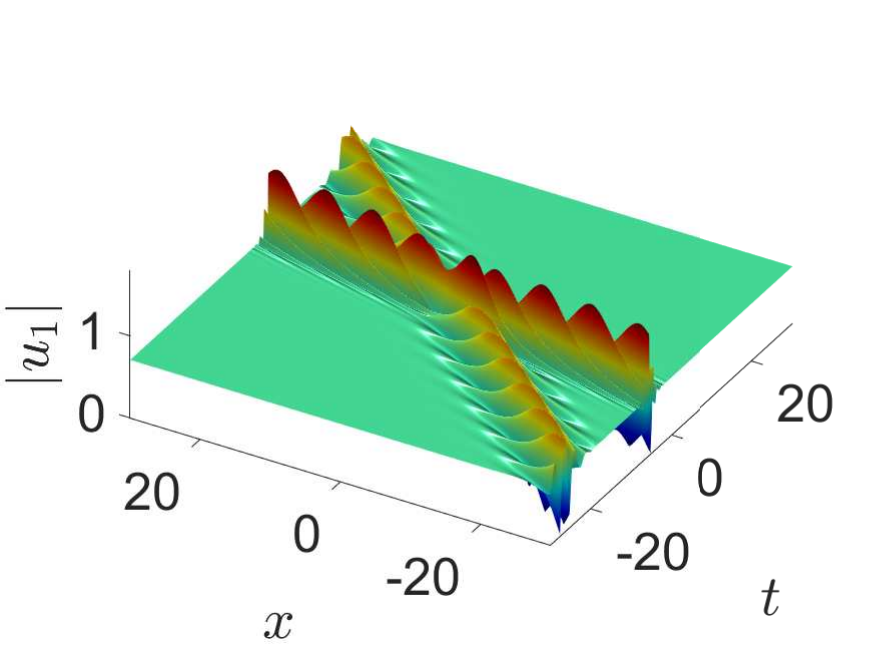}}\hspace{0.1cm}
\subfigure[]{\includegraphics[height=1.3in,width=1.69in]{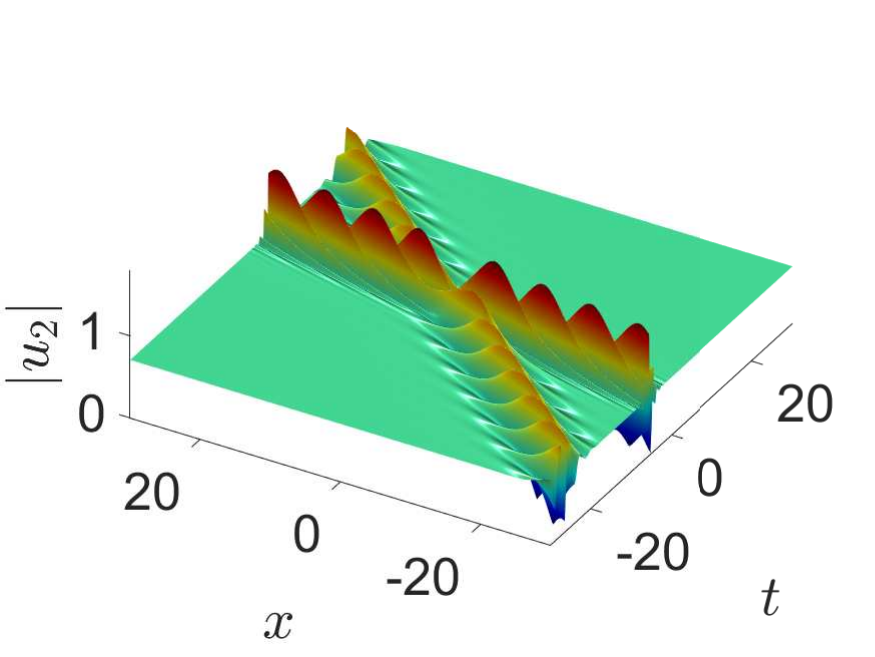}}\hspace{0.1cm}
\subfigure[]{\includegraphics[height=1.3in,width=1.69in]{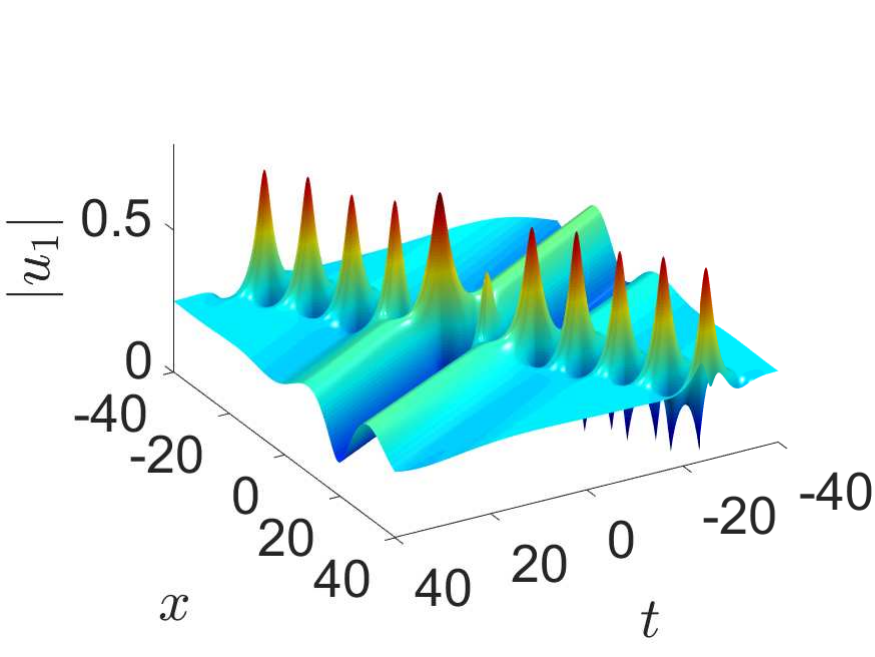}}\hspace{0.1cm}
\subfigure[]{\includegraphics[height=1.3in,width=1.69in]{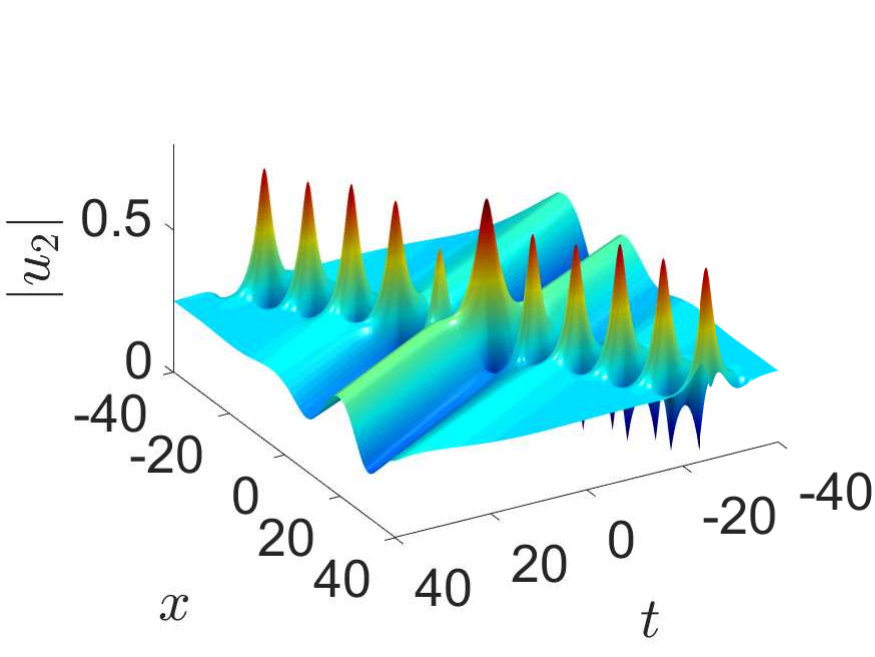}}\hspace{0.1cm}
\subfigure[]{\includegraphics[height=1.3in,width=1.69in]{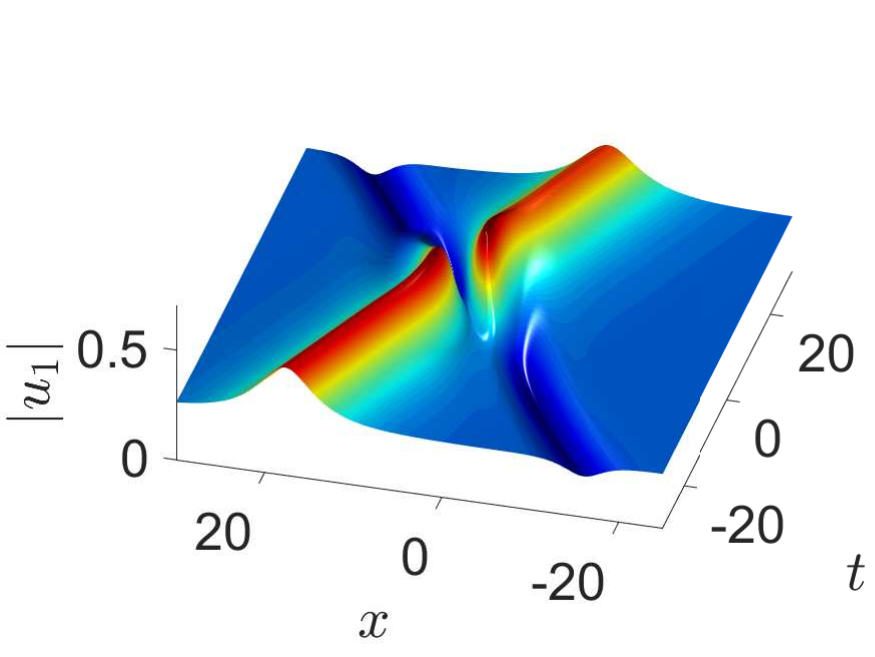}}\hspace{0.1cm}
\subfigure[]{\includegraphics[height=1.3in,width=1.69in]{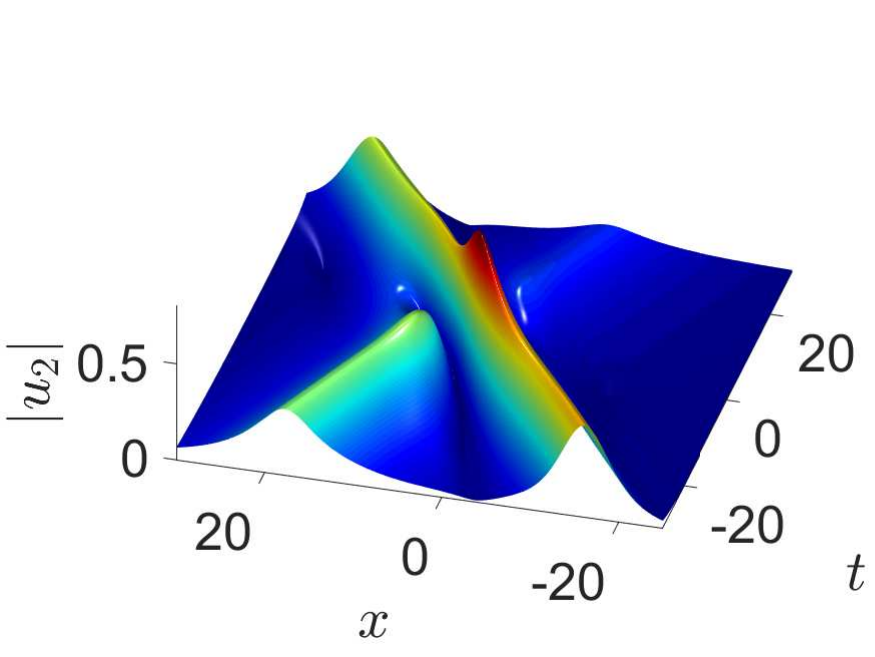}}\hspace{0.1cm}
\caption{(a) and (b): One dark-bright and one breather solitons solution with parameters are chosen as $N_1=N_2=1$, $q_1=q_2=0$, $g_1=g_2=1$, $\delta_1=1.4$, $\delta_2=1.5$, $\kappa_1=0.5$, $\kappa_2=0.25$, $\alpha_1=\alpha2=1$, ${\boldsymbol{u}_ + } = {\left( {1,0} \right)^T}$. (c) and (d): Two breather-breather solitons solution with parameters are chosen as $N_1=N_2=1$, $q_1=q_2=0$, $g_1=g_2=1$, $\delta_1=\delta_2=1.2$, $\kappa_1=0.2$, $\kappa_2=0.25$, $\alpha_1=\alpha2=1$, ${\boldsymbol{u}_ + } ={\left( {\sqrt {2} /2, - \sqrt 2 /2} \right)^T}$. (e) and (f): One breather-breather and one $M$-$W$ shaped solitons solution with parameters are chosen as $N_1=N_3=1$, $q_1=q_3=0$, $g_1=g_3=1$, $\delta_1=1.2$, $\delta_2=\frac{5}{4}$, $\kappa_1=\frac{1}{4}$, $\kappa_2=\frac{1}{8}$, $\alpha_1=\alpha2=1$, ${\boldsymbol{u}_ + } = {\left( {\sqrt 2 /2, - \sqrt 2 /2} \right)^T}$. (g) and (h): One bright-bright and one dark-bright solitons solutions with parameters are chosen as $N_2=N_3=1$, $q_2=0$, $g_2=q_3=g_3=1$, $\delta_2=\frac{5}{4}$, $\delta_3=1.5$, $\kappa_2=\frac{1}{6}$, $\kappa_3=0.25$, $\alpha_1=\alpha2=1$, ${\boldsymbol{u}_ + } = {\left( 1,0 \right)^T}$.   }
\label{fig:3}
\end{figure}

\section{Conclusions}

We construct the IST for the focusing two-component Hirota equation with NZBCs, successfully obtaining various types of soliton solutions. The study provides a detailed analysis of the analyticity of the scattering coefficients and Jost eigenfunctions, as well as two symmetry relations, offering a comprehensive characterization of the discrete spectrum. The results show that the discrete spectrum generates three kinds of discrete eigenvalues. Different types of eigenvalues and their combinations lead to the formation of distinct soliton types. We also illustrate the interaction characteristics of these solitons with graphical representations.

In contrast to the defocusing case, which involves only two fundamental regions, the focusing case contains four.  This difference complicates the analyticity properties and symmetry relations of the eigenfunctions and scattering coefficients. Furthermore, when formulating the inverse scattering problem via the Riemann-Hilbert problem, we define four piecewise meromorphic matrices, two more than in the defocusing case, which results in greater complexity in the construction of the Riemann-Hilbert problem. Moreover, in terms of soliton types, one bright-kink soliton (Fig.\ref{fig:1}: (e) and (f)) as well as one breather-breather and one $M$-$W$ shaped solitons (Fig.\ref{fig:3}: (e) and (f)) do not appear in the defocusing case, and the shapes of the two breather-breather solitons (Fig.\ref{fig:2}: (a) and (b)) are significantly different from those in \cite{N40}.

In this paper, we examine the focusing two-component Hirota equation with boundary conditions ${\boldsymbol{u}}\left( {x,t} \right) \to {{\boldsymbol{u}}_\pm}$ and $\left\| {{\boldsymbol{u}_+}} \right\| = \left\| {{\boldsymbol{u}_-}} \right\| = {{u}_o}$ as $x \to \pm \infty$, which are known as symmetric NZBCs. It is worth noting that in \cite{N43,N44,N45}, the IST for the scalar NLS equation with completely asymmetric NZBCs was studied. These conditions refer to the situation where both the asymptotic amplitude and phase shift are unequal. This extension has significant implications for physical applications, particularly in the theoretical study of perturbed soliton solutions and rogue waves in fiber optic systems, where different background amplitudes are applied at both ends. Future work will focus on constructing the IST for the focusing two-component Hirota equation with completely asymmetric NZBCs,  i.e., the case where $\left\| {{\boldsymbol{u}_ + }} \right\| \neq \left\| {{\boldsymbol{u}_ - }} \right\|$. These research efforts will contribute to the study of the long time asymptotic behavior of solutions with nontrivial boundary conditions.

\begin{flushleft}
\textbf{Acknowledgments}
\end{flushleft}

This work is supported by the National Natural Science Foundation of China (No. 12271488 and No. 11371326).

\begin{flushleft}
 \textbf{References}
\end{flushleft}

\small{

\end{document}